\documentclass[a4paper,11pt]{article}
\usepackage{amsmath}
\usepackage{jheppub} 

\usepackage[T1]{fontenc} 

\usepackage{cleveref}
\usepackage{braket}
\usepackage{amssymb}
\usepackage{subcaption}
\usepackage{xcolor}
\usepackage{axodraw2}
\crefname{section}{section}{sections}
\crefname{figure}{figure}{figures}

\newcommand{\eps}{\epsilon}
\newcommand{\al}{\alpha}
\newcommand{\eik}{\mathcal{J}}

\newcommand{\eikJ}{\mathcal{J}}
\newcommand{\formFactorArgs}{F\left(1,\al_s(Q^2),\eps\right)}
\newcommand{\jetFunctionArgs}{J_i\left(\frac{(2p_i\cdot n_i)^2}{n_i^2\mu^2},\al_s(\mu^2),\eps\right)}
\newcommand{\eikJetFunctionArgs}{\eikJ_i\left(\frac{2(\beta_i\cdot n_i)^2}{n_i^2},\al_s(\mu^2),\eps\right)}
\newcommand{\softFunctionArgs}{\mathcal{S}\left(\beta_1\cdot\beta_2,\al_s(\mu^2),\eps\right)}
\newcommand{\hardFunctionArgs}{H\left(\frac{Q^2}{\mu^2},\frac{(2p_i\cdot n_i)^2}{n_i^2\mu^2},\al_s(\mu^2)\right)}
\newcommand{\dDimAlpha}{\al_s\left(\lambda^2,\eps\right)}

\title{Relating amplitude and PDF factorisation through Wilson-line geometries}

\author{Giulio Falcioni,}
\author{Einan Gardi}
\author{and Calum Milloy}

\affiliation{Higgs Centre for Theoretical Physics, 
School of Physics and Astronomy, \\
The University of Edinburgh, Edinburgh EH9 3FD, Scotland, UK}
\emailAdd{Giulio.Falcioni@ed.ac.uk}
\emailAdd{Einan.Gardi@ed.ac.uk}
\emailAdd{Calum.Milloy@ed.ac.uk}

\abstract{We study long-distance singularities governing different physical quantities involving massless partons in perturbative QCD by using factorisation in terms of Wilson-line correlators. 
By isolating the process-independent hard-collinear singularities from quark and gluon form factors, and identifying these with the ones governing the elastic limit of the perturbative Parton Distribution Functions (PDFs) -- $\delta(1-x)$ in the large-$x$ limit of DGLAP splitting functions -- we extract the anomalous dimension controlling soft singularities of the PDFs, verifying that it admits Casimir scaling. We then perform an independent diagrammatic computation of the latter using its definition in terms of Wilson lines, confirming explicitly the above result through two loops. By comparing our eikonal PDF calculation to that of the eikonal form factor by Erdogan and Sterman and the classical computation of the closed parallelogram by Korchemsky and Korchemskaya, a consistent picture emerges whereby all singularities emerge in diagrammatic configurations localised at the cusps or along lightlike lines, but where distinct contributions to the anomalous dimensions are associated with finite (closed) lightlike segments as compared to infinite (open) ones. Both are relevant for resumming large logarithms in physical quantities, notably the anomalous dimension controlling Drell-Yan or Higgs production near threshold on the one hand, and the gluon Regge trajectory controlling the high-energy limit of partonic scattering on the other.}

\begin{document} 
\maketitle
\flushbottom

\section{Introduction}
\label{sec:intro}

It is well known that perturbative QCD at fixed order in $\al_s$, which is highly successful in describing hard processes at colliders, loses its predictive power in kinematic regions where there is a large hierarchy of scales. Familiar examples are Drell-Yan or Higgs production near threshold, see e.g.~\cite{Sterman:1986aj,Catani:1989ne,Korchemsky:1993uz,Contopanagos:1996nh,Laenen:2005uz,Becher:2007ty}, or at small transverse momentum, which are dominated by soft-gluon radiation. Another example is the high-energy limit of QCD scattering, where the centre-of-mass energy is much larger than the momentum transfer~\cite{Balitsky:1978ic,Kuraev:1976ge,Fadin:1975cb,Lipatov:1976zz,Korchemskaya:1994qp,Balitsky:1995ub,Caron-Huot:2013fea,Caron-Huot:2017fxr}. In each of these cases, and many others, factorisation techniques allow us to derive all-order resummation formulae, which extend the predictive power of QCD, leading to highly successful phenomenology in many cases.

\sloppy{The theory underlying factorisation relies on identifying the origin of any parametrically-enhanced corrections through operators, which capture the relevant divergences. Independently of whether one uses QCD fields~\cite{Collins:1989bt,Collins:2011zzd}, or Soft-Collinear Effective Theory~\cite{IntroToSCET} ones, the relevant operators involve Wilson lines, which follow the trajectory of fast-moving partons, and capture their interactions with soft gluons. These operators obey evolution equations, governed by corresponding anomalous dimensions, which are computable order by order in QCD perturbation theory. The most familiar amongst these is the (lightlike) cusp anomalous dimension~\cite{Korchemsky:1987wg}, $\gamma_{\rm cusp}$, which in particular describes double poles in the Sudakov form factor, originating in overlapping soft and collinear singularities. While the cusp anomalous dimension occurs universally, governing the leading singularities in any kinematic limit, single-logarithmic contributions characterising separately large-angle soft or hard-collinear or rapidity divergences, are somewhat less universal, and yet --- as we shall see --- recur in  a variety of physical quantities that are not a priori related.}
Resummation formulae are obtained upon solving the aforementioned evolution equations, leading to exponentiation. The anomalous dimensions therefore have a central role in the predictive power of QCD, and in certain cases their computation has been recently pushed to three-loop order, e.g.~\cite{Moch:2005ba,Li:2014afw,Grozin:2014hna,Grozin:2015kna,Almelid:2015jia,Almelid:2017qju}, with very recent progress towards four loops~\cite{Davies:2016jie,Henn:2016men,Moch:2017uml,Grozin:2017css,Moch:2018wjh,Lee:2019zop,Henn:2019rmi,Bruser:2019auj,vonManteuffel:2019wbj} (even more is known in maximally supersymmetric Yang-Mills theory, see e.g.~\cite{Beisert:2006ez,Boels:2017skl,Fioravanti:2009xt,Freyhult:2009my,Freyhult:2007pz,Dixon:2017nat}). Despite this impressive progress, there remains several unresolved questions regarding the anomalous dimensions governing single-logarithmic corrections and their universality, some of which we address below.

In the present paper we study two fundamental physical quantities, which are  recurrent ingredients in the factorisation of amplitudes and cross sections~\cite{Mueller:1989hs,Collins:2011zzd}. The first is
the massless on-shell form factor, associated e.g.\ with an electromagnetic vector current  in the case of quarks, or effective Higgs production vertex, $gg\to H$, in the case of gluons. The second is parton distribution functions (PDFs), or more precisely,  the large-$x$ limit of diagonal $qq$ and $gg$ Altarelli-Parisi splitting functions, governing the scale dependence of PDFs according to the Dokshitzer-Gribov-Lipatov-Altarelli-Parisi (DGLAP) evolution equation~\cite{Dokshitzer:1977sg,Gribov:1972ri,Altarelli:1977zs}. Each of these physical quantities is important in its own sake, and their infrared factorisation will be discussed in some detail in sections~\ref{sec:nEqual4} and~\ref{sec:pdfFactor}, respectively. The main motivation to our study comes from the relation between the two, namely a particular combination of single-pole anomalous dimensions, which respectively capture collinear singularities in these two quantities. The relation holds separately for quarks and for gluons:
\begin{align}
\gamma_G^q - 2\,B_{\delta}^q \equiv f_{\text{eik}}^q\,,\qquad \qquad
\gamma_G^g - 2\,B_{\delta}^g &\equiv f_{\text{eik}}^g,\label{MVVidentities}
\end{align}
where  $\gamma_G^q$ ($\gamma_G^g$) is defined by the function $G$ (see eq.~(\ref{def:gammaG})), which along with the cusp anomalous dimension, governs the infrared structure of the quark (gluon) form factor in eq.~(\ref{FFresum})  below; and $B_{\delta}^q$ ($B_{\delta}^g$) is the coefficient of the $\delta(1-x)$ term, in the large-$x$ limit of the quark-quark (gluon-gluon) splitting function, see eq.~(\ref{eq:splittinglarge-x}) below. It was observed long ago~\cite{Ravindran:2004mb,FormFactors} that while the separate perturbative results for $\gamma_G$ and $B_{\delta}$ are very different between quarks and gluons (this is expected: collinear singularities are known to depend on the parton's spin), the combination~(\ref{MVVidentities}) vanishes at one loop in both cases, and admits a Casimir-scaling relation\footnote{A Casimir-scaling relation similar to~(\ref{MVVidentities}) and~(\ref{eq:feik2l_into}) was deduced from factorisation already in~\cite{Korchemsky:1988si}; in this analysis single-pole collinear singularities are controlled by the anomalous dimension of the quark or gluon fields in axial gauge.} at two loops, namely
\begin{align}
  \label{eq:feik2l_into}
 \frac{ f_{\text{eik}}^q}{C_F}=\frac{f_{\text{eik}}^g}{C_A}
  &=\bigg(\frac{\alpha_s}{\pi}\bigg)^2  \bigg[C_A \left(\frac{101}{54}-\frac{11 }{24}\zeta_2-\frac{7}{4} \zeta_3\right)+T_f n_f\left(-\frac{14}{27}+\frac{1}{6}\zeta_2\right)\bigg]+{\cal{O}}\left(\alpha_s^3\right)\,.
\end{align}
The same Casimir-scaling property persists at three loops~\cite{FormFactors}.
This is a clear indication that $f_{\text{eik}}$ has an interpretation purely in terms of Wilson lines --- hence the name, an \emph{eikonal} function. A Wilson-line-based definition would explain why the result does not depend on the parton's spin, while it depends on its colour representation in proportion to the relevant quadratic Casimir through three loops. The question we would like to address is what is the Wilson-loop correlator corresponding to $f_{\text{eik}}$.

Before describing our approach to answer this question, let us note that the combination  in~(\ref{MVVidentities}) has a direct physical interpretation as the soft anomalous dimension associated with Drell-Yan production near partonic threshold~\cite{Sterman:1986aj,Catani:1989ne,Korchemsky:1993uz,Contopanagos:1996nh,Laenen:2005uz,Becher:2007ty}, namely $\gamma_G^q - 2\,B_{\delta}^q=\frac12 \Gamma_{\rm DY}$.
Similarly $\gamma_G^g - 2\,B_{\delta}^g$ is associated with Higgs production through gluon-gluon fusion near threshold.
The corresponding soft function is defined \emph{at cross-section level}, by replacing the energetic partons, which move in opposite lightlike directions (before annihilating at the hard interaction vertex), by Wilson lines that follow the same trajectory, in both the amplitude and its complex conjugate. The cusp where the complex-conjugate amplitude Wilson lines meet is displaced by a timelike distance with respect to the amplitude: this distance is the Fourier conjugate variable to the energy fraction carried by soft partons.\footnote{An additional displacement of the two cusps in transverse space can be used to resum transverse-momentum logarithms~\cite{Collins:1984kg}. The corresponding anomalous dimensions can be related to the DY soft function via a conformal transformation~\cite{Li:2016axz,Li:2016ctv,Vladimirov:2017ksc,Vladimirov:2016dll}.}
Final-state radiation, namely the set of soft particles connecting the amplitude side to the complex-conjugate amplitude side, are described by cut propagators.
This soft function admits an evolution equation governed by $\gamma_{\rm cusp}$ and $\Gamma_{\rm DY}$ (see e.g.\ eq. (9) in ref.~\cite{Korchemsky:1993uz}, or eqs.~(43--44) in ref.~\cite{Becher:2007ty}). The latter was computed through three loops directly based on the aforementioned Wilson-line definition~\cite{Belitsky:1998tc,Li:2014afw}, and the results agree with the combination of anomalous dimensions in~(\ref{MVVidentities}), which were extracted from independent QCD computations of the form factor~\cite{Harlander:2000mg,Ravindran:2004mb,FormFactors} and DGLAP splitting functions~\cite{Floratos:1977au,Floratos:1978ny,GonzalezArroyo:1979df,GonzalezArroyo:1979he,Curci:1980uw,Furmanski:1980cm,Floratos:1981hs,Moch:1999eb}. Thus, from this perspective, this physical quantity is well understood, and its Casimir-scaling property simply follows from the above-mentioned Wilson-line definition.

\begin{figure}
  \centering
  \begin{subfigure}[b]{0.3\textwidth}
    \centering
  \includegraphics[width=0.75\textwidth]{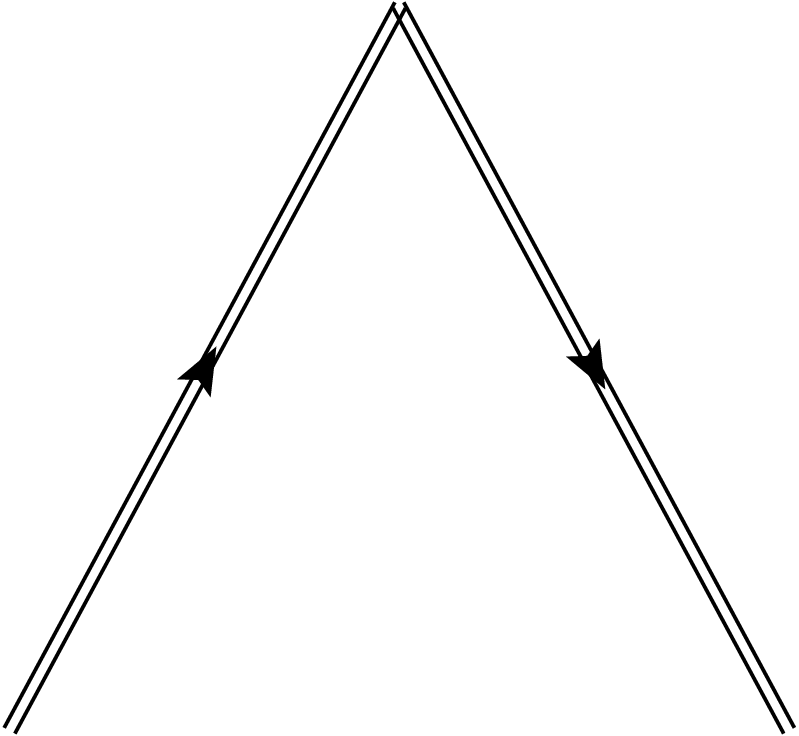}
  \caption{\label{fig:wedge} $\wedge$ geometry}
  \end{subfigure}
  \begin{subfigure}[b]{0.3\textwidth}
    \centering
  \includegraphics[width=0.75\textwidth]{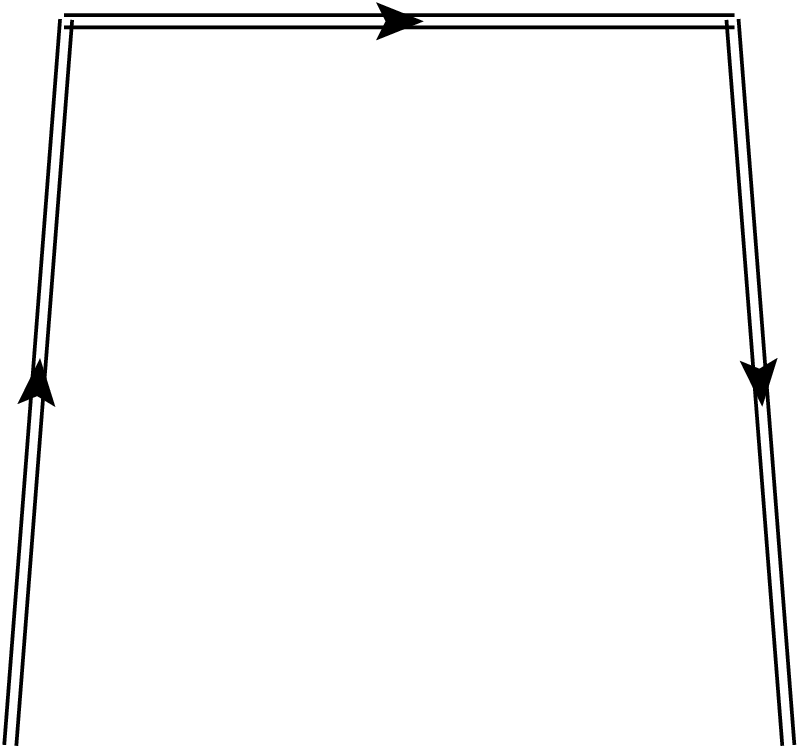}
  \caption{\label{fig:sqcap} $\sqcap$ geometry}
  \end{subfigure}
  \begin{subfigure}[b]{0.3\textwidth}
    \centering
  \includegraphics[width=0.75\textwidth]{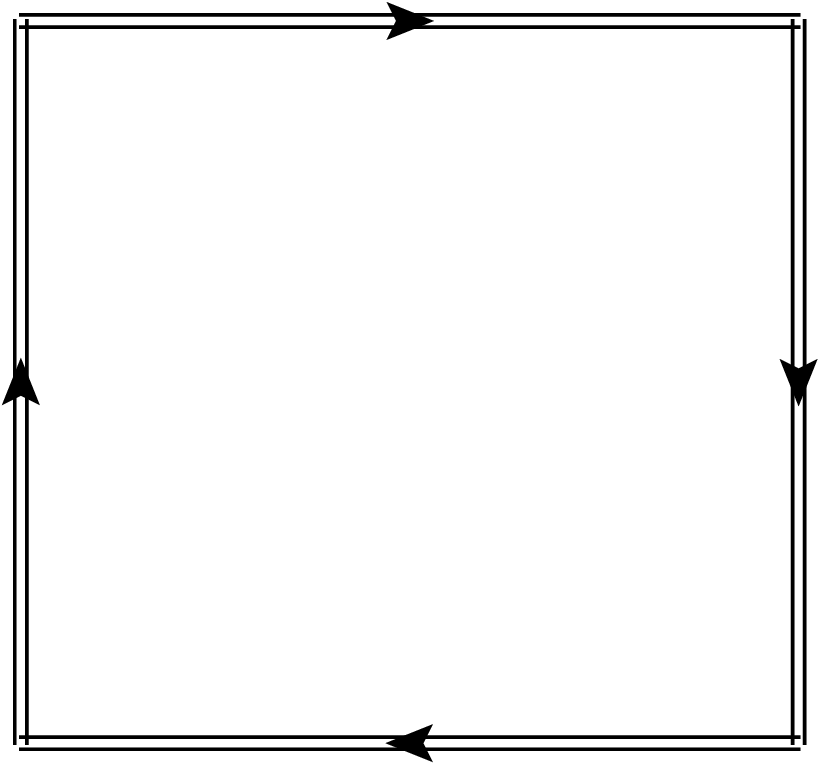}
  \caption{\label{fig:box} $\Box$ geometry}
\end{subfigure}
\caption{\label{fig:geometries} Contours of lightlike Wilson loops that contain semi-infinite Wilson lines, which arise in the factorisation of the form factor (a) and the parton distribution function (b). \, Contour (c), the parallelogram, which consists of four \emph{finite} lightlike segments, gives rise to the anomalous dimension on the right-hand-side of eq.~\eqref{box_proposition}.}
\end{figure}

Our own investigation starts with the simple observation that  the two-loop result for $\gamma_G - 2\,B_{\delta}$ in~(\ref{eq:feik2l_into}) \emph{also agrees}, up to an overall factor of 4, with the result for the parallelogram Wilson loop made of four lightlike segments (see figure~\ref{fig:box}), which was computed in 1992 by Korchemsky and Korchemskaya~\cite{Korchemsky}. It is a highly appealing proposition that\footnote{Note that we systematically omit the superscript $q/g$ in~(\ref{box_proposition}) and below, and specify the representation only when needed.}
\begin{equation}\label{box_proposition}
f_{\rm eik}\equiv\gamma_G - 2B_\delta = \frac{\Gamma_\Box}{4}\,,
\end{equation}
holds to all orders.\footnote{While the two-loop result for $\Gamma_\Box$ has been known for a while, we are not aware that the proposition~(\ref{box_proposition}) was made before. Unfortunately, there is no \emph{direct} three-loop computation of $\Gamma_\Box$ available at this~point.}
The parallelogram Wilson loop, is a very simple object: being compact it has no infrared divergences, so the singularities arise here from short distances, and the calculation can be done directly in dimensional regularisation. Importantly, in contrast to the Drell-Yan soft function described above, real corrections and cut propagators do not arise here. The natural questions to ask then are first, does the relation in~(\ref{box_proposition}) indeed hold to all orders, and second, can we see how a parallelogram Wilson loop arises from the definitions of the objects on the left-hand side of  eq.~(\ref{box_proposition}), the form factor and the PDF. Establishing this relation is one of the main goals of the present paper.

The infrared factorisation of the form factor is well understood~\cite{Mueller:1989hs,Collins:2011zzd,Dixon:2008gr}, and has been used as the starting point for the factorisation of massless amplitudes with any number of legs in general kinematics~\cite{Catani:1998bh,Sterman:2002qn,Aybat:2006wq,Aybat:2006mz,Becher:2009cu,Becher:2009qa,Gardi:2009qi,Gardi:2009zv}. The form-factor factorisation gives rise to a different Wilson-line configuration,
namely a couple of \emph{semi-infinite} lightlike Wilson lines (with different 4-velocities) meeting at the hard-interaction vertex, see figure~\ref{fig:wedge}. We shall refer to this contour as the $\wedge$ geometry.
We emphasise that in contrast  with the Drell-Yan soft function described above, where the cross section was considered~\cite{Belitsky:1998tc,Li:2014afw}, here the Wilson-line configuration is defined at amplitude level. At a difference with the parallelogram of~\cite{Korchemsky}, the $\wedge$ geometry is non-compact, and thus gives rise to infrared divergences, in addition to ultraviolet ones. We shall return to the $\wedge$ geometry and its properties below. At this point it suffices to say that considering the infrared factorisation of the form factor, the origin of the relation between $\gamma_G - 2B_\delta$ and the parallelogram geometry remains obscure: the $\wedge$ geometry has no finite segments while the parallelogram consists exclusively of such.

An important step in explaining the eikonal nature of $f_{\rm eik}$ in~(\ref{MVVidentities}), based on the infrared factorisation properties of the form factor and the PDF, was taken in 2008 in a paper by Dixon, Magnea and Sterman~\cite{Dixon:2008gr}. The fundamental explanation is that spin-dependent
hard-collinear contributions are common to both $\gamma_G$ and  $2\,B_{\delta}$ and drop in the difference, leaving behind a purely eikonal component. This is the premise we shall follow here as well.
However, ref.~\cite{Dixon:2008gr} relied on the \emph{assumption} that $B_{\delta}$, as the coefficient of $\delta(1-x)$, is a purely virtual quantity and
 hence the factorisation of the PDF could be done at ``amplitude level''. According to the factorisation outlined in~\cite{Dixon:2008gr}  the eikonal component of $B_{\delta}$ should correspond to Wilson lines with a $\wedge-$geometry, much like the form factor. Taking this at face value, if the eikonal components of  $\gamma_G$ and $B_\delta$ on the right-hand side of~(\ref{box_proposition}) indeed \emph{both} correspond to the $\wedge-$geometry, one concludes that the $\wedge$ and the $\Box$ anomalous dimensions must be proportional to each other, at least through two loops, or, put differently, one may deduce the anomalous dimension of the $\wedge-$geometry from~(\ref{eq:feik2l_into}).

The first direct two-loop computation of $\wedge-$geometry Wilson loop
 was performed only in 2015, by Erdogan and~Sterman~\cite{Sterman}.
This calculation is an important step forward also in the sense that it presents a new method for dealing directly with (semi)-infinite lightlike Wilson lines in configuration space (which a priori lead to scaleless integrals) without resorting to an extra regulator. This is done by cleverly using the exponentiation properties and isolating a well-defined integrand, before renormalising ultraviolet divergences by means of a suitable cutoff. We shall adopt and generalise this method in section~\ref{sec:GammaPiCalc} below.
The result of ref.~\cite{Sterman} is that the anomalous dimension corresponding to the $\wedge-$geometry Wilson loop is given by
\begin{equation}
\Gamma_\wedge = \left(\frac{\al_s}{\pi}\right)^2C_i\left[ C_A\left(\frac{101}{54}-\frac{11}{24}\zeta_2-\frac14\zeta_3\right)
+\left( - \frac{14}{27}+\frac{1}{6}\zeta_2
\right)T_fn_f \right]+ \mathcal{O}(\al_s^3),\label{GeikSE_intro}
\end{equation}
where $C_i=C_F$ for Wilson lines in the fundamental representation and $C_A$ for the adjoint.
As with $f_{\rm eik}$ and $\Gamma_\wedge$ above, we omit the superscript $q/g$ for $\Gamma_\wedge$  wherever it is not necessary.
While the result in~(\ref{GeikSE_intro}) bears a striking resemblance to $f_{\rm eik}$ in~(\ref{eq:feik2l_into}), it is evidently not identical; the coefficient of the $\zeta_3$ term is entirely different. The authors of ref.~\cite{Sterman} further provided a detailed diagrammatic analysis, comparing their calculation to that of the parallelogram in ref.~\cite{Korchemsky}, and explaining the origin of the difference in the coefficient of $\zeta_3$ as emanating from endpoint contributions that are present in finite lightlike segments, but are absent in infinite ones. This conclusion can be confirmed by a momentum-space computation.

It is useful to bear in mind that infinite and semi-infinite Wilson-line configurations (but not finite ones!) are of direct relevance to partonic scattering amplitudes in the high-energy limit (the Regge limit)~\cite{Korchemskaya:1994qp,Balitsky:1995ub,Caron-Huot:2013fea,Caron-Huot:2017fxr}.
Also, the explicit two-loop combination in~(\ref{GeikSE_intro}) appeared in the literature in that context long before the computation of the $\wedge$ configuration in ref.~\cite{Sterman}.
Specifically, considering $gg\to gg$,  $qq\to qq$ or $qg\to qg$ scattering
in the limit where the centre-of-mass energy is much larger than the momentum transfer, $s\gg-t$, the leading and next-to-leading logarithms in $s/(-t)$ in the (real part of the) amplitude exponentiate according to a simple replacement of the $t$-channel gluon propagator (dubbed gluon Reggeisation):
\begin{equation}
\frac{1}{t}\to\frac{1}{t}\left(\frac{s}{-t}\right)^{\al(t,\epsilon)}\,,
\end{equation}
where $\al(t,\epsilon)$ is the gluon Regge trajectory\footnote{See also a more recent observation in ref.~\cite{DelDuca:2017pmn} that the two-loop coefficient $\Gamma_\wedge^{g\,(2)}$ occurs also in the QCD impact factor.}~\cite{Fadin:1995xg,Fadin:1996tb,Fadin:1995km,Blumlein:1998ib,DelDuca:2001gu} given by:
\begin{equation}
\label{Regge}
\al(t,\epsilon) = \frac{\alpha_s}{\pi}\frac{\gamma_{\text{cusp}}^{g\, (1)}}{2\eps}+\left(\frac{\alpha_s}{\pi}\right)^2\frac{1}{4}\left(-\frac{C_A\hat{b}_0}{\eps^2}+\frac{\gamma_{\text{cusp}}^{g\,(2)}}{\eps}+2\Gamma_\wedge^{g\,(2)}+C_A\hat{b}_0\zeta_2\right)+\mathcal{O}(\al_s^3)
\end{equation}
where $\alpha_s= \al_s(-t,\eps)$, with $\epsilon=(4-d)/2$ the dimensional regularisation parameter, $\hat{b}_0$ is the one-loop QCD beta function of~(\ref{b0hat}), $\gamma_{\text{cusp}}^{g\,(n)}$ are the coefficient of   the cusp anomalous dimension of eq.~(\ref{twoLoopCusp}) for the gluon, and $\Gamma_\wedge^{g\,(2)}$ is the two-loop coefficient in eq.~(\ref{GeikSE_intro}), again with $C_i=C_A$.
We further recall that the overall similarity between the parallelogram Wilson loop in~\cite{Korchemsky} and the gluon Regge trajectory in~(\ref{Regge}), as well as the peculiar difference between them in the coefficient of $\zeta_3$, were already observed early on, in ref.~\cite{Korchemskaya:1996je}, where an evolution equation for the Regge trajectory was derived, considering the forward limit of crossed Wilson lines.
However, this raises no difficulty: as stressed above, it is the infinite Wilson-line geometry which is expected to be relevant for the factorisation of partonic scattering amplitudes, not the parallelogram.

A real puzzle arises, however, upon considering the explicit result for the $\wedge-$geometry anomalous dimension in eq.~(\ref{GeikSE_intro})
in view of  eq.~(\ref{eq:feik2l_into}), if the conclusion of ref.~\cite{Dixon:2008gr} is taken at face value. Given that the factorisation of the form factor is well understood, and the eikonal component of $\gamma_G$ is determined by the $\wedge-$geometry,  we are compelled to revisit the assumption of ref.~\cite{Dixon:2008gr} that $B_{\delta}$ is a purely virtual quantity, systematically establish the infrared factorisation of the PDFs at large $x$, and identify the eikonal component of $B_{\delta}$, which clearly \emph{must not} be proportional to $\Gamma_\wedge$.

We proceed as follows. In section~\ref{sec:nEqual4} we review the factorisation of long-distance singularities of the QCD form factor and identify the process-independent spin-dependent hard-collinear component of $\gamma_G$. In turn, in section~\ref{sec:pdfFactor} we discuss the factorisation of PDFs in the limit $x\rightarrow 1$. We show explicitly that the calculation of $B_\delta$ requires both real and virtual corrections. To this end we perform an explicit two-loop calculation of the splitting functions at large $x$ (the details are presented in appendix~\ref{subsec:pdf}). Next we identify the eikonal component of $B_\delta$ as the anomalous dimension associated with a $\sqcap$-shaped Wilson-line geometry, see figure~\ref{fig:sqcap}. By using the known value of $B_\delta$ along with the hard-collinear anomalous dimension extracted from the form factor, we then predict the $\Gamma_\sqcap$ anomalous dimension at two loops. Then, in section~\ref{sec:GammaPiCalc} we compute $\Gamma_\sqcap$ directly to this order, finding agreement with the extracted result of section~\ref{sec:pdfFactor}. In section~\ref{sec:GammaPiCalc} we also derive an evolution equation for the $\sqcap$-shaped Wilson-line and show that while in the ultraviolet it is characterised by double poles, as any other cusped Wilson loop, its infrared properties are different, displaying strictly single poles, in agreement with single-pole nature of PDFs themselves. In section~\ref{sec:finalRelation} we put together our results for the factorisation of the form factor and the PDF, and establish the relation of~(\ref{box_proposition})  with the parallelogram to all orders. We further summarise the state-of-the-art knowledge of higher-order corrections to $\Gamma_{\Box}$ in view of its relations with other physical quantities. We briefly summarise our conclusions in section~\ref{sec:conc}.


\section{Infrared factorisation of the on-shell form factor}
\label{sec:nEqual4}

Let us review the well-known factorisation of a colour-singlet on-shell form factor of coloured massless particles (quarks or gluons) in QCD~\cite{Mueller:1989hs,Collins:2011zzd,Collins:1989bt, Magnea:1990zb,FormFactors, Dixon:2008gr}.
We label the external momenta by $p_1$ (incoming) and $p_2$ (outgoing) with the momentum transfer $Q^2\equiv-(p_1-p_2)^2$, and, as usual, we renormalise all ultraviolet singularities in the $\overline{\rm MS}$ scheme, denoting the renormalisation scale by $\mu^2$.

The quark form factor is defined in terms of the electromagnetic vector current, proportional to $\bar{\psi} \gamma_{\mu}\psi$, which does not renormalise. The gluon form factor in turn is defined using an effective local interaction vertex with the Higgs field, $H G_{\mu\nu}^aG^{\mu\nu a}$, and it does renormalise, proportionally to the QCD beta function~\cite{Ravindran:2004mb}. The distinct ultraviolet properties of the quark and gluon form factors will be of little relevance for us: we focus instead on the infrared singularities of the form factor, which  have a rather similar structure for massless quarks and gluons.

For large $Q^2$ the form factor $F\left(Q^2/\mu^2,\al_s(\mu^2),\eps\right)$ features large logarithms in the ratio ${Q^2}/{\mu^2}$, and fixed-order perturbation theory breaks down. These large logarithms can be resummed using a renormalisation-group equation (see e.g.~\cite{Magnea:1990zb}), giving the following all-order formula for the form factor,
\begin{equation}\label{FFresum}
\formFactorArgs = \exp\bigg[\frac{1}{2}\int_0^{Q^2}\frac{d\lambda^2}{\lambda^2}\left(G(1,\al_s(\lambda^2,\eps),\eps)-\gamma_{\text{cusp}}(\al_s(\lambda^2,\eps))\log\frac{Q^2}{\lambda^2}\right)\bigg],
\end{equation}
where we set the renormalisation scale $\mu^2=Q^2$ for simplicity.
Note that we have absorbed into the function $G$ any operator renormalisation terms --- see~\cite{FormFactors,Ravindran:2004mb} for more details. Infrared singularities are generated in eq.~\eqref{FFresum} through an integration, from $\lambda^2=0$, over the $d=4-2\epsilon$ dimensional running coupling $\al_s(\mu^2,\epsilon)$, which obeys
\begin{align}
\label{eq:dCoupl}
\frac{d}{d\ln\mu^2}\left(\frac{\al_s(\mu^2,\eps)}{\pi}\right)&=-\epsilon \left(\frac{\al_s(\mu^2,\eps)}{\pi}\right)
-\sum_{n=0}^\infty\hat{b}_n\left(\frac{\al_s(\mu^2,\eps)}{\pi}\right)^{n+2}.
\end{align}
We report the coefficients $\hat{b}_0$, $\hat{b}_1$ and $\hat{b}_2$ of
the QCD beta function respectively at one~\cite{Vanyashin:1965ple,Khriplovich:1969aa,Gross:1973id,Politzer:1973fx},
two~\cite{Caswell:1974gg,Jones:1974mm,Tarasov:1976ef,Egorian:1978zx}
and three loops~\cite{Tarasov:1980au,Larin:1993tp}, because we will
use them in the rest of this paper\begin{subequations}\begin{align}
  \hat{b}_0=\,&\frac{11}{12}C_A-\frac{1}{3}T_fn_f ,\label{b0hat}\\
  \hat{b}_1=\,&\frac{17}{24}C_A^2-\frac{5}{12}C_AT_fn_f-\frac{1}{4}C_FT_fn_f,\\
  \hat{b}_2=\,&\frac{2857}{3456}C_A^3-\frac{1415}{1728}C_A^2T_fn_f-\frac{205}{576}C_AC_FT_fn_f+\frac{1}{32}C_F^2T_fn_f\\
  &+\frac{11}{144}C_FT_f^2n_f^2+\frac{79}{864}C_AT_f^2n_f^2,\nonumber
\end{align}
\end{subequations}
with the quadratic Casimir defined by ${\bf T}^a{\bf T}^a = C_R
\mathbf{1}$, with ${\bf T}$ being the ${\rm SU}(N_c)$ generator in the
representation $R$ and $C_A$ corresponds to the Casimir in the adjoint
representation; $n_f$ is the number of light quarks and the normalisation of
the generators $t^a$ in the fundamental representation, ${\rm
  Tr}(t^at^b)=T_f \delta_{ab}$, is conventionally set to $T_f=1/2$.

Equation~(\ref{FFresum}) applies for both quarks and gluons, but with distinct functions $\gamma_\text{cusp}(\al_s)$ and $G(Q^2/\mu^2,\al_s,\eps)$, which do depend on the type of particles (although this is suppressed in our notation).
The former, which captures all double poles, depends solely on the colour representation of the particles (fundamental and adjoint for quarks and gluons, respectively) while the latter, which controls single poles, depends also on their spin.
This distinction will be crucial in what follows and it is a direct consequence of the fact that $\gamma_\text{cusp}$ is an eikonal quantity, namely one that can be defined exclusively in terms of Wilson lines, while  $G(Q^2/\mu^2,\al_s,\eps)$ instead, contains  hard-collinear effects, which cannot fully be described by Wilson lines.
Specifically, $\gamma_\text{cusp}$ is the lightlike cusp anomalous dimension~\cite{Korchemsky:1987wg}, defined as the coefficient of the leading ultraviolet divergence occurring in a cusped Wilson loop, which evaluates to
\begin{align}\label{twoLoopCusp}
  \begin{split}
  \gamma_\text{cusp}(\alpha_s)=\,&\sum_{n=1}^{\infty} \bigg(\frac{\alpha_s}{\pi}\bigg)^n \,\gamma_\text{cusp}^{(n)}
\\=\,&
\frac{\alpha_s}{\pi}C_i+\bigg(\frac{\alpha_s}{\pi}\bigg)^2C_i\bigg[C_A\left(\frac{67}{36}-\frac{\zeta_2}{2}\right)-\frac{5}{9}n_fT_f\bigg]\\
  &+\bigg(\frac{\alpha_s}{\pi}\bigg)^3C_i\bigg[C_A^2\left(\frac{245}{96}-\frac{67}{36}\zeta_2+\frac{11}{24}\zeta_3+\frac{11}{8}\zeta_4\right)\\
&\hspace{3cm}+C_An_fT_f\left(-\frac{209}{216}+\frac{5}{9}\zeta_2-\frac{7}{6}\zeta_3\right)\\
    &\hspace{3cm}+C_Fn_fT_f\left(-\frac{55}{48}+\zeta_3\right)-\frac{(n_fT_f)^2}{27}\bigg]+{\cal{O}}\left(\alpha_s^4\right),
  \end{split}
\end{align}
where $C_i$, defined above, is the quadratic Casimir in the
fundamental or the adjoint representation for quarks and gluons,
respectively. The three-loop value of $\gamma_\text{cusp}$ was
computed in~\cite{splittingNonSinglet}, and recently there has been
significant progress towards a four-loop
determination~\cite{Moch:2018wjh,Lee:2019zop,Henn:2019rmi,Bruser:2019auj,vonManteuffel:2019wbj}. Through
three loops, the cusp anomalous dimension, much like other quantities
that are defined exclusively in terms of Wilson lines, depends on the
colour representation proportionally to the quadratic Casimir $C_i$,
as in~(\ref{twoLoopCusp}), adhering to the so-called \emph{Casimir
  scaling} property. Starting at four loops quartic Casimirs,
$d_{ij}^{(4)}\equiv d_i^{abcd}d_j^{abcd}$, appear as well, making the
dependence of the colour representation more involved. Differently
from $\gamma_\text{cusp}$, the function
$G(1,\al_s(\lambda^2,\eps),\epsilon)$ has an expansion both in $\al_s$
and $\epsilon$, as follows
\begin{equation}
\label{Gln}
  G\left(1,\alpha_s\left(\lambda^2,\eps\right),\eps\right) = \sum_{l=1}^\infty\sum_{n=0}^\infty G(l,n)\left(\frac{\alpha_s\left(\lambda^2,\eps\right)}{\pi}\right)^l\eps^n,
\end{equation}
therefore it generates both infrared poles and non-negative powers of
$\epsilon$ upon integrating over the scale $\lambda^2$ of the running
coupling as in eq.~(\ref{FFresum}). We isolate the divergent
contribution order-by-order in $\alpha_s$, by defining the anomalous dimension
$\gamma_G$ such that
\begin{equation}
  \int_0^{\mu^2}\frac{d\lambda^2}{\lambda^2}G(1,\al_s(\lambda^2,\eps),\epsilon)
  =
  \int_0^{\mu^2}\frac{d\lambda^2}{\lambda^2}\left[\gamma_G\left(\alpha_s\left(\lambda^2,\eps\right)\right)\right]
  + {\cal{O}}\left(\epsilon^0\right),
  \label{def:gammaG}
\end{equation}
 where $\gamma_G$ depends on $\eps$ only through the coupling. The coefficients $\gamma_G$ for the quark and for the gluon are well
 known in the literature; they are referred to sometimes as
``collinear anomalous dimensions'' and were denoted by $\widetilde{G}$ in~\cite{dy}, by $\mathcal{G}_0$ in~\cite{Dixon:2017nat} and by $\gamma^q$ and $\gamma^g$ in appendix I of~\cite{IntroToSCET}. The
 latter has a conventional factor of $-2$.  In practice, we derive
 here $\gamma_G$ to four loops by substituting the $d-$dimensional running coupling of eq.~(\ref{eq:dCoupl}) into
 eq.~(\ref{def:gammaG}) and then identifying the singularities arising on the two sides of equation~(\ref{def:gammaG}), getting
\begin{align}
  \begin{split}
    \gamma_G=\,&\frac{\alpha_s}{\pi}G(1,0)+\left(\frac{\alpha_s}{\pi}\right)^2\Big[G(2,0)-\hat{b}_0\,G(1,1)\Big]\\
& + \left(\frac{\alpha_s}{\pi}\right)^3\Big[G(3,0)-\hat{b}_0\,G(2,1)-\hat{b}_1\,G(1,1)+\hat{b}_0^2\,G(1,2)\Big]\\
&+\left(\frac{\alpha_s}{\pi}\right)^4\Big[G(4,0)-\hat{b}_0\,G(3,1)-\hat{b}_1\,G(2,1)-\hat{b}_2\,G(1,1)+\hat{b}_0^2\,G(2,2)\\
      &\hspace{2cm}+2\hat{b}_0 \hat{b}_1\,G(1,2)-\hat{b}_0^3G(1,3)\Big]+{\cal{O}}\left(\alpha_s^5\right)\,,
    \end{split}
\end{align}
where  $G(l,n)$ are defined in eq.~(\ref{Gln}) and their values can be extracted from refs.~\cite{Harlander:2000mg,Ravindran:2004mb,FormFactors} where the form factors have been computed to three loops.
For the purpose of this paper we only need explicit results for the collinear anomalous dimensions through two loops, which~read
\begin{align}
  \gamma_G^{q}=\left(\frac{\al_s}{\pi}\right)\frac{3C_F}{2} +\left(\frac{\al_s}{\pi}\right)^2\bigg\{&C_A C_F \left(\frac{11}{8} \zeta_2-\frac{13}{4}\zeta_3+\frac{961}{432}\right)\nonumber\\&+C_F^2 \left(-\frac{3}{2}\zeta_2+3 \zeta_3+\frac{3}{16}\right)-C_F T_f n_f\left(\frac{\zeta_2}{2}+\frac{65}{108}\right)\bigg\}+\mathcal{O}(\al_s^3)\nonumber\\
  \gamma_G^{g}=\left(\frac{\al_s}{\pi}\right)2\hat{b}_0 +\left(\frac{\al_s}{\pi}\right)^2\bigg\{&C_A^2\left(-\frac{11 \zeta_2}{24}-\frac{\zeta_3}{4}+\frac{173}{54}\right)\nonumber\\&+C_A T_fn_f \left(\frac{\zeta_2}{6}-\frac{32}{27}\right)- \frac{C_FT_f n_f}{2}\bigg\}+\mathcal{O}(\al_s^3)\,,
  \label{res:gammaG2}
\end{align}
where we added superscripts $i=q,g$ to distinguish between quarks
and gluons.

\subsection{Infrared factorisation}

At high energy ($Q^2\to\infty$) the infrared behaviour decouples from the hard scattering
\begin{align}\label{FFfactor}
  \begin{split}
    \formFactorArgs =\,& \hardFunctionArgs \prod_{i=1}^{2}\jetFunctionArgs\\
    &\times\left(\frac{\softFunctionArgs}{\prod_{k=1}^2\eikJetFunctionArgs}\right),
    \end{split}
\end{align}
where the jet function $J_i$, one for each external leg, captures the
collinear singularities, the soft function $\mathcal{S}$ contains the
contribution of any long-wavelength gluons exchanged between the external
particles and the eikonal jet function $\eikJ_i$ captures all the
singularities that are present both in $\mathcal{S}$ and in the
jet function $J_i$, which are associated with exchanges
that are both soft and collinear to the massless external
particles. Therefore, the ratio $\frac{\mathcal{S}}{\eikJ_1\eikJ_2}$ in
eq.~\eqref{FFfactor} includes only the divergences associated with soft
wide-angle emissions. $H$ is the hard function, found from
matching to the full form factor. Each other factor in eq.~\eqref{FFfactor} has an operator
definition which dictates their functional dependences in
eq.~(\ref{FFfactor}), involving the momenta $p_i$ of the external
particles and their lightlike velocities $\beta_i$, defined~by
\begin{equation}
  p_i^\mu = Q_0 \beta_i^\mu,
  \label{eq:betadef}
\end{equation}
where $Q_0$ is an arbitrary normalisation and would typically be of
the order of the hard scale of the process, $Q$.  The operator definitions of $\mathcal{S}$, $J_i$ and $\eikJ_i$ are
written in terms of the expectation values of Wilson lines, defined as
\begin{equation}
W_v(y,x)=\mathbf{P}\exp\left(ig_s\int_x^ydz
v_\mu A^\mu(z)\right),\label{wilsonLineDef}
\end{equation}
where $v$ is the direction of the line and $x$ and $y$ are its
endpoints. In general, the vector $v$ can be either lightlike $v^2=
0$ or non-lightlike $v^2\neq 0$. In the context of the on-shell massless form factor, lightlike kinematics for the external legs, $\beta_i^2=0$, is dictated by eq.~(\ref{eq:betadef}), and we define the functions entering the factorisation formula~(\ref{FFfactor}) by:
\begin{align}
  \eikJetFunctionArgs&=\braket{0|\mathrm{T}\Big[W_{n_i}(\infty,0)W_{\beta_i}(0,\infty)\Big]|0},\label{eikjetFunction}\\
  \softFunctionArgs&=\braket{0|\mathrm{T}\Big[W_{\beta_1}(\infty,0)W_{\beta_2}(0,\infty)\Big]|0},\label{softFunction}\phantom{\frac{1}{1}}\\
  u(p)\,\jetFunctionArgs&=\braket{0|\mathrm{T}\Big[W_{n_i}(\infty,0)\psi(0)\Big]|p}\label{jetFunction},
\end{align}
where $n_i$ is an auxiliary non-lightlike vector and the dependence on its choice must cancel in eq.~(\ref{FFfactor}). The contour defining $\mathcal{S}$ is shown in figure~\ref{fig:wedge}.
In eq.~(\ref{jetFunction})
we presented the jet function $J_i$ for fermion fields;
for a definition of the gluon jet function see refs.~\cite{Becher:2009th,
  Becher:2010pd, Magnea:2018ebr}.  The representation of the Wilson
lines in eq.~\eqref{softFunction} is the representation of the
corresponding external particle. Any function built solely from Wilson
lines, such as $\mathcal{S}$ and $\eikJ_i$, is called eikonal. As mentioned in the context of the cusp anomalous dimension, one of
the properties of eikonal quantities is that they admit Casimir scaling
up to three loops; this is a consequence of non-Abelian exponentiation.
Beyond three loops there are quartic (and
eventually higher order) Casimir contributions, but given that the
same Wilson-line diagrams contribute for quarks and gluons, differing
just by the representations of the Wilson lines, one expects a
relation between these quantities. Indeed, a conjectural relation was proposed in~\cite{Moch:2018wjh} based on partial four-loop computations; we shall return to this in section~\ref{sec:highOrder} below.

The individual functions in
eqs.~\eqref{eikjetFunction}--\eqref{jetFunction} are heavily constrained
by kinematic considerations, such as the dependence on the auxiliary
vectors $n_i$, and by renormalisation group evolution. These constraints can be solved to
give explicit formulae~\cite{Gardi:2009qi,Gardi:2009zv},
\begin{align}
  \mathcal{J}_i&=\exp\bigg\{-\frac{1}{4}\int_0^{\mu^2}\frac{d\lambda^2}{\lambda^2}\left(\Gamma_{\mathcal{J}}\left(\alpha_s(\lambda^2,\eps)\right)+\gamma_{\text{cusp}}(\alpha_s(\lambda^2,\eps))\log\frac{2(\beta_i\cdot n_i)^2\mu^2}{n_i^2\lambda^2}\right)\bigg\},\label{eikJFint}\\
  \mathcal{S}&=\exp\bigg\{-\frac{1}{2}\int_0^{\mu^2}\frac{d\lambda^2}{\lambda^2}\left(\Gamma_\wedge(\alpha(\lambda^2,\eps))+\gamma_{\text{cusp}}(\alpha_s(\lambda^2,\eps))\log\left(\frac{\beta_1\cdot\beta_2\mu^2}{\lambda^2}\right)\right)\bigg\},\label{softFint}
\end{align}
where $\Gamma_{\mathcal{J}}$ and $\Gamma_\wedge$ are constants to be
determined by direct calculation. Note that $\Gamma_\wedge$ was
denoted in the literature~\cite{Dixon:2008gr,Sterman} as
$-G_{\text{eik}}$. As in eq.~(\ref{FFresum}), the infrared
singularities of $\mathcal{J}_i$ and $\mathcal{S}$ are generated by
integrating over the $d$ dimensional running coupling
$\al_s(\lambda^2,\eps)$ from $\lambda^2=0$.
We notice that the soft function and the product of the eikonal jets share the same
dependence on $\gamma_{\text{cusp}}\ln \mu^2/\lambda^2$, which is associated with the
overlapping soft-collinear singularities of these two
quantities. This fact ensures that the ratio
$\frac{\mathcal{S}}{\eikJ_1\eikJ_2}$ is free of overlapping
divergences and depends only on the logarithm of the kinematic
variable
\begin{equation}
  \kappa=\frac{(\beta_1\cdot\beta_2)^2\,n_1^2n_2^2}{4(\beta_1\cdot n_1)^2(\beta_2\cdot n_2)^2},
  \label{eq:defCCR}
\end{equation}
which is insensitive to the normalisation of the vectors $\beta_i$ in the
definition eq.~\eqref{eq:betadef}.  Using the factorisation equation
eq.~(\ref{FFfactor}), we determine the partonic jet function by
dividing the form factor in eq.~(\ref{FFresum}) by the ratio
$\frac{\mathcal{S}}{\eikJ_1\eikJ_2}$, yielding
\begin{align}
    J_i&=\exp\bigg\{h_J(\al_s(p_n^2))+\frac{1}{2}\int_{\mu^2}^{p_n^2}\frac{d\lambda^2}{\lambda^2}\gamma_{i}(\alpha_s(\lambda^2))  + \frac{1}{4}\int_0^{p_n^2}\frac{d\lambda^2}{\lambda^2}\bigg(-\gamma_{\text{cusp}}(\alpha_s(\lambda^2,\eps))\log\left(\frac{p_n^2}{\lambda^2}\right)\nonumber\\
    &\qquad\qquad+\Gamma_\wedge(\alpha_s(\lambda^2,\eps))-\Gamma_{\mathcal{J}}(\alpha_s(\lambda^2,\eps))+G(1,\al_s(\lambda^2,\eps),\eps)\bigg)\bigg\},\label{jetFint}
\end{align}
where we introduced the variable $p_n^2=\frac{(2p\cdot
  n)^2}{n^2}$. The function $h_J(\al_s(p_n^2))$ is a matching
coefficient that captures the finite parts of the jet function and
$\gamma_i$, with $i=q$ for the quark and $i=g$ for the gluon, is the
anomalous dimension of the field $i$ in axial gauge. The latter is
only concerned with the ultraviolet behaviour of the jet function and indeed it
is not associated with any IR pole, because the contribution from the
IR region $\lambda^2\simeq 0$ is absent in the second term of
eq.~(\ref{jetFint}). All the IR poles of the form factor are generated
by the second integral in the equation above, involving the
anomalous dimensions $\gamma_{\text{cusp}}$, $\Gamma_\wedge$,
$\Gamma_\eikJ$ and the resummation function $G(1,\al_s,\epsilon)$. The
dependence on $\gamma_{\text{cusp}}$ is such that the combination with
$\frac{\mathcal{S}}{\eikJ_1\eikJ_2}$ reconstructs the kinematic
dependence of the form factor eq.~\eqref{FFresum}~through
\begin{equation}
  2\log\left(\frac{Q^2}{\lambda^2}\right)=\log(\kappa)+\log\left(\frac{p_{n_1}^2}{\lambda^2}\right)+\log\left(\frac{p_{n_2}^2}{\lambda^2}\right).
\end{equation}

\subsection{Isolating hard-collinear singularities}
\label{sec:FormFactorJJ}

The contribution of $\Gamma_\eikJ$ in eq.~(\ref{jetFint}) is associated
to the soft singularities of $J_i$, which cancel in the ratio of
$J_i$ and $\eikJ_i$ eq.~\eqref{FFfactor}. It is therefore convenient to
focus on the poles of pure hard-collinear origin, defined as
\begin{equation}
J_{i/\eikJ}\equiv \frac{J_i|_{\text{pole}}}{\eik_i},\label{JoverJ}
\end{equation}
where $J_i|_{\text{pole}}$ means only the poles of the jet function.
We extract the function $J_{i/\eikJ}$ for $i=q$ and $i=g$ from the
form factor of the quark and of the gluon, respectively, thus
providing the {\emph{process-independent}} components containing the
purely collinear singularities associated with massless external
partons.  In order to determine $J_{i/\eikJ}$, we isolate the pole
part of the jet function $J_i$, by replacing in eq.~(\ref{jetFint})
the function $G(1,\al_s,\eps)$ with $\gamma_G$, according to the
definition in eq.~(\ref{def:gammaG}), and we get the ratio
\begin{align}
  \log\left(\frac{J_i|_{\text{pole}}}{\eik_i}\right)
                                    &=\bigg[\frac{1}{4}\int_0^{\mu^2}\frac{d\lambda^2}{\lambda^2}\left(\gamma_G(\al_s(\lambda^2,\eps))+\Gamma_\wedge(\dDimAlpha)-\gamma_{\text{cusp}}\log\left(\frac{2(p_i\cdot n_i)^2}{(\beta_i\cdot n_i)^2\mu^2}\right)\right)\bigg]\nonumber\\
  &\equiv \frac{1}{2}\int_0^{\mu^2}\frac{d\lambda^2}{\lambda^2}\left[\gamma_{J/\eik}(\alpha_s(\lambda^2,\eps))-\frac{\gamma_{\text{cusp}}(\alpha_s(\lambda^2,\epsilon))}{2}\log\left(\frac{2(p_i\cdot n_i)^2}{(\beta_i\cdot n_i)^2\mu^2}\right)\right]\label{JJdef},
\end{align}
where on the last line we have defined the anomalous dimension
$\gamma_{J/\eik}$
\begin{align}
  2\gamma_{J/\eik}=\gamma_G + \Gamma_\wedge\label{gammaJJrel}.
\end{align}
As mentioned above, the collinear anomalous dimension $\gamma_G$ is known to three loops~\cite{Harlander:2000mg,Ravindran:2004mb,FormFactors} for both quarks and gluons, and we quoted the corresponding expressions through two loops in eq.~(\ref{res:gammaG2}).
The anomalous dimension $\Gamma_\wedge$, in turn, is derived from the
renormalisation of the soft function $\mathcal{S}$, that can be read
off eq.~\eqref{softFint}
\begin{equation}
  \mu\frac{d}{d\mu}\log\mathcal{S} = -\int_0^{\mu^2}\frac{d\lambda^2}{\lambda^2}\gamma_{\text{cusp}}\left(\al_s(\lambda^2)\right)-\left[\Gamma_\wedge\left(\al_s(\mu^2)\right)+\gamma_{\text{cusp}}(\alpha_s(\mu^2))\log(\beta_1\cdot\beta_2)\right].
\end{equation}
The equation above clarifies the meaning of the subscript $\wedge$,
which symbolises the contour of the lightlike Wilson loop in the
definition of the soft function in eq.~\eqref{softFunction} that defines
$\Gamma_\wedge$. This notation will be used throughout this paper and
it will be generalised for different contours. $\Gamma_\wedge$ is
known to two loops~\cite{Sterman} by direct computation of the
equation above
\begin{equation}
\Gamma_\wedge = \left(\frac{\al_s}{\pi}\right)^2\frac{C_i}{4}\left(-2\hat{b}_0\zeta_2 - \frac{56}{27}T_fn_f + C_A\left[\frac{202}{27}-\zeta_3\right]\right)+ \mathcal{O}(\al_s^3),\label{GeikSE}
\end{equation}
where $C_i$ is the quadratic Casimir dependent on the representation
of the Wilson lines in eq.~\eqref{softFunction}. Using the results in eqs.~\eqref{res:gammaG2}
and~\eqref{GeikSE} we determine $\gamma_{J/\eik}$ to two loops. First
for quarks we have
\begin{align}
  \gamma_{J/\eik}^{q}=\,&\left(\frac{\al_s}{\pi}\right)\frac{3C_F}{4}+\left(\frac{\al_s}{\pi}\right)^2\bigg\{C_A C_F \left(\frac{11 \zeta_2}{24}-\frac{7\zeta_3}{4}+\frac{1769}{864}\right)\nonumber\\&+C_F^2 \left(-\frac{3\zeta_2}{4}+\frac{3 \zeta_3}{2}+\frac{3}{32}\right)-C_FT_fn_f \left(\frac{\zeta_2}{6}+\frac{121}{216}\right)\bigg\}+\mathcal{O}(\al_s^3) .\label{QuarkJJ}
\end{align}
Then for gluons,
\begin{align}
  \gamma_{J/\eik}^{g}=\left(\frac{\al_s}{\pi}\right){\hat{b}_0}&+\left(\frac{\al_s}{\pi}\right)^2\bigg\{C_A^2 \left(-\frac{11 \zeta_2}{24}-\frac{\zeta_3}{4}+\frac{137}{54}\right)\nonumber\\&\hspace{2cm}+C_AT_fn_f \left(\frac{\zeta_2}{6}-\frac{23}{27}\right)-\frac{C_FT_fn_f}{4}\bigg\}+\mathcal{O}(\al_s^3).\label{GluonJJ}
\end{align}

We have thus isolated the hard-collinear singularities of the form factor and found the quantity $\gamma_{J/\eik}$ that governs this behaviour for quark and for gluons according to eq.~(\ref{JJdef}). We emphasises that in contrast to the conventional collinear anomalous dimension $\gamma_G$ given in eq.~(\ref{res:gammaG2}), which is specific to the form factor (recall eqs.~(\ref{def:gammaG}) and~(\ref{FFresum})),
the hard-collinear anomalous dimension $\gamma_{J/\eik}$ defined here is \emph{process independent}. This universality will now be put to use.
In the next section we will consider the factorisation of parton distribution functions (PDFs) at large $x$ where we will use the above two-loop results for $\gamma_{J/\eik}^{q}$ and $\gamma_{J/\eik}^{g}$ given in eqs.~(\ref{QuarkJJ}) and~(\ref{GluonJJ}) respectively, and ultimately identify the eikonal anomalous dimension relevant to the PDF evolution.


\section{Parton distribution functions at large \texorpdfstring{\boldmath $x$}{x}}
\label{sec:pdfFactor}

Parton distribution functions, $f_{AB}(x)$, describe the probability of finding parton $A$ with momentum fraction $x$ inside hadron (or parton) $B$. We will be interested here in PDF \emph{evolution}, which is the same for the partonic and for the hadronic quantities, and will therefore consider partonic PDFs. PDFs are inherently defined at \emph{cross-section} level with the need to combine real and virtual radiation to cancel soft singularities such that only pure collinear singularities associated with the massless initial-state parton are kept. We will see that in the elastic limit, $x\to 1$, the contributions from different regions factorise and claim that the hard-collinear behaviour of the initial-state partons is described by $\gamma_{J/\eik}$, the same anomalous dimension we identified in the factorisation of the form factor.

\subsection{Definition}
\label{subsec:PDFdef}

The light-cone PDF for a quark (gluon) in a parton $P$ of momentum $p$ with longitudinal momentum fraction $x$ is given by~\cite{pdf}
\begin{align}
  f_{qP}^{\text{bare}}\left(x,\eps\right)&=\frac{1}{2}\int\frac{dy}{2\pi}e^{-iy x p\cdot u}\braket{P|\bar{\psi}_q(y u)\gamma\cdot u W_u(y,0)\psi_q(0)|P}\label{quarkPDF}\\
  f_{gP}^{\text{bare}}\left(x,\eps\right)&=\frac{1}{x p\cdot u}\int\frac{dy}{2\pi}e^{-iy x p\cdot u}\braket{P|G_{\mu +}(y u)W_u(y,0)G^{+\mu}(0)|P}\label{gluonPDF}.
\end{align}
The Wilson-line operator $W_u$ is defined in eq.~\eqref{wilsonLineDef} and
$\ket{P}$ is either an on-shell quark or gluon, $P=q,g$. We take the
lightlike momentum $p$ to be in the ($+$) direction and then the velocity four-vector $u$ is in
the ($-$) direction. It is worthwhile noting here that the bare PDFs
$f_{j'j}^{\text{bare}}(x,\eps)$ are scaleless. This will be used later in the context of factorisation. They are renormalised through a
convolution,
\begin{equation}\label{eqn:Zfactor}
f_{jk}(x,\mu) = \sum_{j'}\int_x^1\frac{dz}{z}Z_{jj'}(z,\al_s,\eps)f_{j'k}^{\text{bare}}(x/z,\eps),
\end{equation}
where $Z_{jj'}$ is a renormalisation factor, removing the UV divergences from the bare PDF in the $\overline{\text{MS}}$ scheme and $f_{jk}$ is the renormalised PDF. From $Z_{jj'}(x,\al_s,\eps)$ we can get the splitting~functions,
\begin{equation}\label{eqn:SplitFunc}
 \frac{d}{d\log\mu} Z_{jk}(x,\al_s,\eps)=2\sum_{j'}\int_x^1\frac{dz}{z}P_{jj'}(z,\al_s)Z_{j'k}(x/z,\al_s,\eps).
\end{equation}

The RG evolution of the PDFs is governed by the DGLAP equations~\cite{Dokshitzer:1977sg,Gribov:1972ri,Altarelli:1977zs}:
\begin{equation}\label{eqn:splitting}
\frac{d}{d\log\mu} f_{jk}\left(x,\mu\right) = 2\sum_{j'}\int_x^1\frac{dz}{z}P_{jj'}(z,\al_s)f_{j'k}(x/z,\mu).
\end{equation}
The DGLAP splitting kernels $P_{jk}$ are known to three loops~\cite{Gross:1973ju,Georgi:1951sr,Altarelli:1977zs,Floratos:1977au,Floratos:1978ny,GonzalezArroyo:1979df,GonzalezArroyo:1979he,Curci:1980uw,Furmanski:1980cm,Floratos:1981hs,Hamberg:1991qt,Moch:1999eb,splittingNonSinglet,vogt} with some recent results at four loops~\cite{Davies:2016jie,Moch:2017uml,Moch:2018wjh}.

\subsection[Perturbative calculation at large \texorpdfstring{$x$}{x}]{\boldmath Perturbative calculation at large $x$}

In the limit $x\to 1$ the diagonal terms in the splitting functions, $P_{qq}$ and $P_{gg}$, feature divergent contributions~\cite{Korchemsky:1988si,Korchemsky:1992xv,Dokshitzer:2005bf,Belitsky:2008mg}, namely
\begin{equation}
 \label{eq:splittinglarge-x}
 P_{ii}=\frac{\gamma_{\text{cusp}}}{(1-x)_+} + B_\delta^{(i)}\,\delta(1-x)\,+\,\mathcal{O}\left(\log(1-x)\right),
\end{equation}
where the label $i=q,g$ indicates quarks and gluons, respectively, and the
plus distribution is defined as usual, see e.g.~\cite{Altarelli:1977zs}.

\begin{figure}
  \centering
  \includegraphics[width=.95\textwidth]{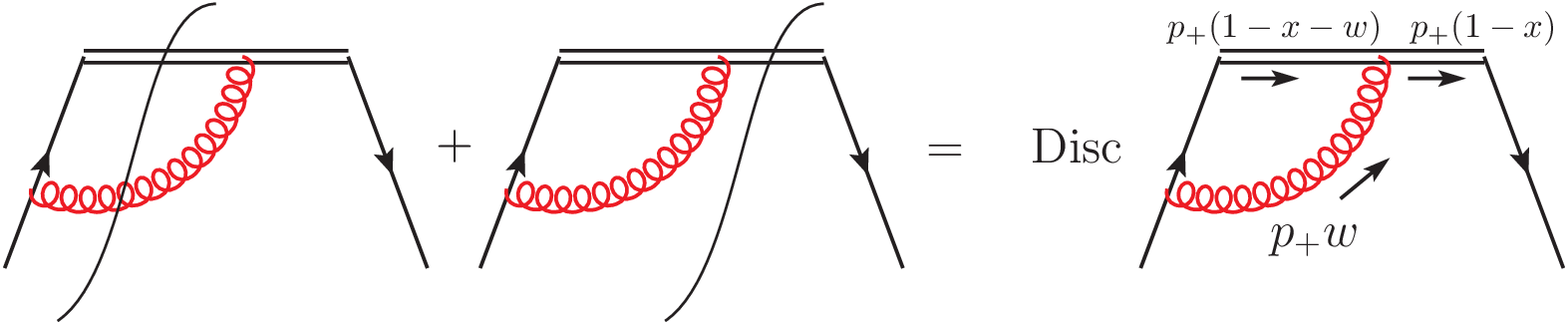}
  \caption{\label{fig:cutsToDisc}The vertex correction for the one-loop quark PDF. The left-hand side is the standard sum over cuts equating to the discontinuity of the amplitude. The double line is the Wilson line while the solid black line is a quark.}
\end{figure}

The splitting functions are determined from the UV singularities of
the PDFs defined in eqs.~\eqref{quarkPDF} and \eqref{gluonPDF}, which can be computed
perturbatively. We can relate these definitions to time-ordered
products by the discontinuity in $x$,
\begin{equation}
f_{qq}^{\text{bare}}(x,\eps) = \text{Disc}_x\frac{1}{2}\int\frac{dy}{2\pi}e^{-iy xp\cdot u}\braket{p|T\left[\bar{\psi}_q(y u)\gamma\cdot u W_u(y,0)\psi_q(0)\right]|p}.
\end{equation}
This relation, which is illustrated  diagrammatically in figure~\ref{fig:cutsToDisc}, can be derived as follows. One first splits the Wilson line in eq.~(\ref{quarkPDF}) into two Wilson lines that extend to infinity, $W_u(y,\infty)W_u(\infty,0)$, one then inserts a complete set of states between them and finally identifies the result as the discontinuity of the time-ordered product.
This relies on the fact that the condition $x\leq 1$ selects the cuts with positive energy~\cite{Collins:2011zzd,Schwartz:2013pla}. One can think that the coefficient $B_\delta$ in eq.~\eqref{eq:splittinglarge-x} is entirely determined by the contribution of the virtual diagrams, such as the second term in the left-handside in figure~\ref{fig:cutsToDisc}, however the explicit calculation will lead to a different conclusion.

At one loop, the relevant diagram is shown in the right-hand side of figure~\ref{fig:cutsToDisc}, which in Feynman gauge reads
\begin{equation}
f_{qq}^{\text{fig.1}}=\text{Disc}\,\,\frac{ g_s^2 }{\pi}C_F\int\frac{d^dq}{(2\pi)^d}\frac{p_+(p_+-q_+)}{\left(q^2+i0\right)\left((p-q)^2\!+\!i0\right)\left((p\!-\!k)\cdot u\!+\!i0\right)\left((p\!-\!k\!-\!q)\cdot u\!+\!i0\right)},
\end{equation}
where we used $p$ and $k$ respectively to denote the incoming and outgoing quark momenta, and $q$ the gluon momentum. For brevity, we also drop the superscript $\text{bare}$. It is straightforward to compute the integral over $q_-$ by complex analysis. This places a bound on $q_+$ i.e.\ $p_+>q_+>0$. The $q_T$ integral is scaleless but as we are interested only in the UV divergence it is simply a matter of replacing,
\begin{equation}
\int\frac{d^{d-2}q_T}{(2\pi)^{d-2}}\frac{1}{q_T^2}\to \frac{e^{\eps\gamma_E}}{(4\pi)^{1-\eps}}\frac{1}{\eps}.
\end{equation}
We then scale out $p_+$ by defining $q_+=p_+ w$ to produce an elegant integral representation,
\begin{equation}
f_{qq}^{\text{fig.1}}=\text{Disc}\,\,\frac{i\al_s}{\pi}C_F\frac{1}{4\pi}\frac{1}{\eps}\int_0^1dw\frac{1-w}{(1-x+i0)(1-x-w+i0)},\label{eq:niceIntegral}
\end{equation}
where we have absorbed the $(e^{\gamma_E}4\pi)^{\eps}$ factors in the $\overline{\text{MS}}$ coupling. The representation in eq.~\eqref{eq:niceIntegral} has the advantage of compactly displaying the sum over cuts: individual cuts can be isolated by computing the residues corresponding to each of the propagator poles. Using partial fractioning,
\begin{equation}
\frac{1}{1-x+i0}\frac{1}{1-x-w+i0} = \frac{1}{w}\left(\frac{1}{1-x-w+i0}-\frac{1}{1-x+i0}\right)
\label{eq:partfrac}
\end{equation}
so the full discontinuity of the integrand equals,
\begin{equation}
\text{Disc}\frac{1}{1-x+i0}\frac{1}{1-x-w+i0} = \frac{1}{w}(-2\pi i)\left(\delta(1-x-w)-\delta(1-x)\right)
\end{equation}
and we find
\begin{equation}
f_{qq}^{\text{fig.1}}=\frac{\al_s}{2\pi}\frac{C_F}{\eps}\left(\frac{x}{1-x}-\delta(1-x)\int_0^1dw\frac{1-w}{w}\right).
\end{equation}
Here the first term is a real emission cut, while the second, a virtual correction. As usual, the endpoint divergence in the first term is combined with the divergence as $w\to 0$ in the second to give,
\begin{equation}\label{onShellDiagram}
f_{qq}^{\,\,\text{fig. 1}}=\frac{\al_s}{4\pi}C_F\frac{1}{\eps}\left(\frac{2}{(1-x)_+}+2\delta(1-x)-2\right).
\end{equation}
We emphasise that it is ambiguous to determine which cuts have contributed to the $\delta(1-x)$ term, as its coefficient is only finite after the cancellation of the soft divergences between the real and the virtual cuts.

We combine eq.~\eqref{onShellDiagram} with the mirror diagram representing the correction of the right vertex, which yields an identical result, and with the box-type diagram, which does not contribute to divergent terms at large $x$. We complete the calculation of the (bare) PDF by including the two diagrams featuring radiative corrections on the external legs
\begin{equation}
f_{qq}^{\,\,\text{SE}} = (Z_2-1)\, \delta(1-x) = -\frac{\al_s}{4\pi\eps}C_F \,\delta(1-x),
\end{equation}
where we used the wavefunction renormalisation $Z_2$ at one loop. The expression of the UV singularities of the bare PDF at one loop reads
\begin{equation}
  f_{qq}^{\text{bare}} = \delta(1-x) + \frac{\al_s(\mu^2)}{4\pi\eps}C_F\left(\frac{4}{(1-x)_+}+3\delta(1-x)+\mathcal{O}((1-x)^0)\right)+ \mathcal{O}(\al_s^2).
\end{equation}
Following eq.~\eqref{eqn:Zfactor}, we derive the renormalisation factor $Z_{qq}$ that cancels the ultraviolet divergence in the equation above
\begin{equation}
  Z_{qq} = \delta(1-x) - \frac{\al_s(\mu^2)}{4\pi\eps}C_F\left(\frac{4}{(1-x)_+}+3\delta(1-x)+\mathcal{O}((1-x)^0)\right) + \mathcal{O}(\al_s^2).
\end{equation}
Finally, we obtain the splitting function by computing the derivative with respect to the renormalisation scale eq.~\eqref{eqn:splitting}, which yields the well-known result for the $qq$ splitting function
\begin{equation}
P_{qq}(x)=\frac{\al_s}{4\pi}C_F\left(\frac{4}{(1-x)_+}+3\delta(1-x)+\mathcal{O}((1-x)^0)\right) + \mathcal{O}(\al_s^2).
\end{equation}

\begin{figure}
  \centering
        \includegraphics[width=0.5\textwidth]{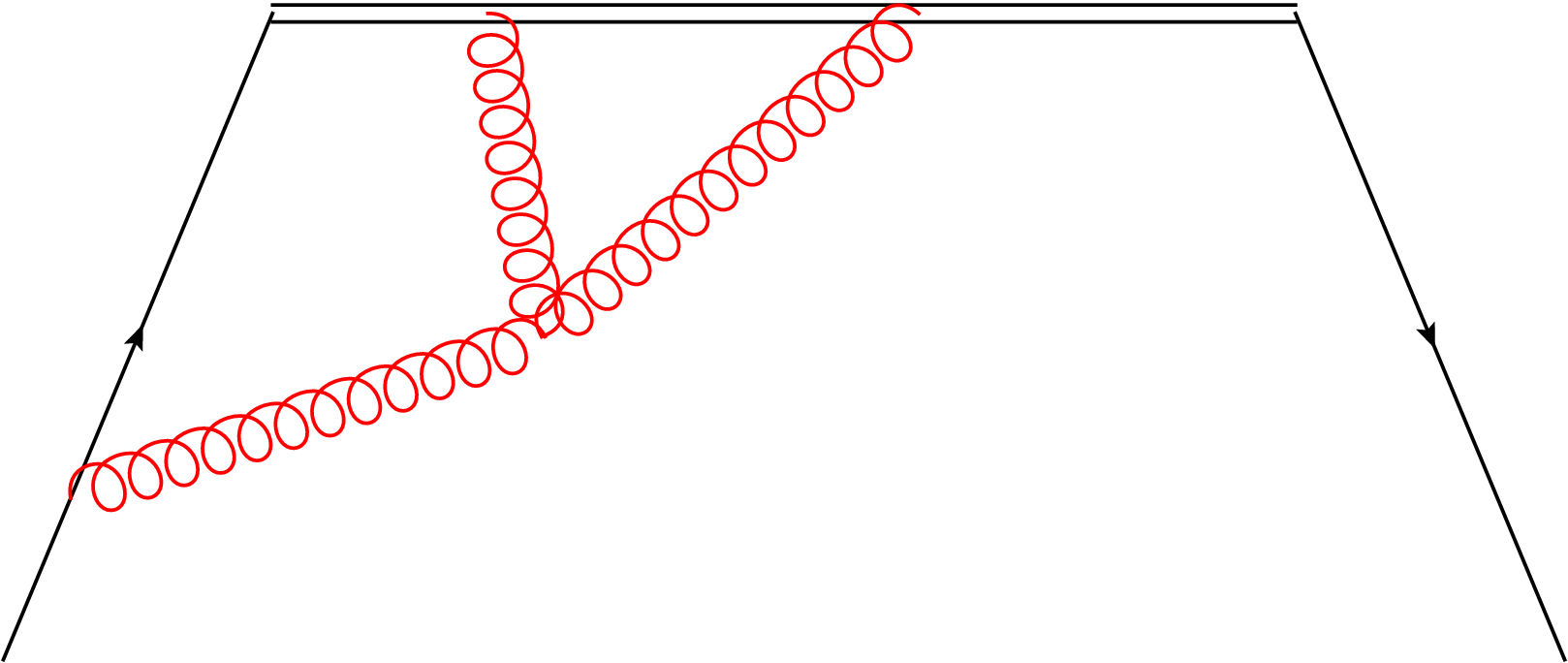}
        \caption{The diagram $f_{qq}^{(2),(e)}$.\label{fig:twoLoop12}}
\end{figure}

The one-loop calculation with on-shell states is straightforward but at two loops and beyond it becomes complicated to disentangle the UV from the IR in the transverse integrals. To regularise the IR we can take the initial states to be off-shell $p^2\neq 0$. The intermediate expressions become more verbose but introduce no major conceptual issues. As the states are now unphysical the correlators become gauge dependent. It means that the running of the gauge parameter, $\xi\to Z_A\xi$ has to be taken into account in $\mathcal{O}(\eps^0)$ finite terms. A similar observation was made in~\cite{Bruser:2019auj}.
Using this method we are able to arrive at the integral representation similar to eq.~\eqref{eq:niceIntegral} for each two-loop diagram. For two loops it is a two parameter integral with integrals over the plus component of the two loop momenta. As an example, the diagram in figure~\ref{fig:twoLoop12} can be represented as
\begin{align}
  \label{eq:fqq2e}
f_{qq}^{(2),(e)}=\text{Disc}\frac{i}{2\pi}C_AC_F\frac{\Gamma(\eps)\Gamma(2\eps)}{\Gamma(1+\eps)}\int_0^1&dydz\,y^{1-2\eps}(1-y)^{1-\eps}(1-z)^{-\eps}z^{-\eps}\nonumber\\&\times\frac{1-2z}{(1-x+i0)(1-x-y+i0)(1-x-yz+i0)}.
\end{align}
The three denominators correspond to the three Wilson-line propagators after integration over the $(-)$ and transverse components of the two loop momenta. We distinguish the contribution of the real emission and the ones of the virtual corrections by applying partial fractioning as in eq.~\eqref{eq:partfrac}. The discontinuity of the first propagator in eq.~\eqref{eq:fqq2e} is proportional to $\delta(1-x)$ and it determines the virtual contribution. The other two propagators in eq.~\eqref{eq:fqq2e} correspond to real emissions. Each term features infrared divergences, which cancel in the sum of all cuts. Furthermore, we notice that the real emission cuts yield UV poles that are proportional to $\delta(1-x)$ and therefore contribute to the $P_{qq}$ splitting function. This particular calculation is detailed in section~\ref{subsec:pdf}, where we also present the full two-loop results for quarks and gluons, diagram by diagram.

Our final result for the splitting functions eqs.~\eqref{app:Pqq} and~\eqref{app:Pgg} reproduces the known results~\cite{Gross:1973ju,Georgi:1951sr,Altarelli:1977zs,Floratos:1977au,Floratos:1978ny,GonzalezArroyo:1979df,GonzalezArroyo:1979he,Curci:1980uw,Furmanski:1980cm,Floratos:1981hs,Hamberg:1991qt,Moch:1999eb}. These previous splitting function calculations have been performed using different methods, including extracting them from corresponding deep inelastic structure-function calculations~\cite{Moch:1999eb}, by means of the operator product expansion~\cite{Gross:1973ju,Georgi:1951sr,Floratos:1977au,Floratos:1978ny,GonzalezArroyo:1979df,GonzalezArroyo:1979he,Floratos:1981hs,Hamberg:1991qt}, by means of light-cone axial gauge~\cite{Heinrich:1997kv,Bassetto:1998uv}, or by relating them to splitting amplitudes~\cite{Kosower:2003np}.
To our knowledge, our direct calculation is the first of its kind. This method has the advantage to show that not all the diagrams contribute to the singular behaviour of the splitting functions in eq.~\eqref{eq:splittinglarge-x} and that the coefficient $B_\delta$ includes both the virtual and the real corrections.

\subsection{Factorisation}
\label{subsec:PDFfactorise}

As $x\to 1$ the momentum of the final-state parton tends to the initial-state one, meaning that the contribution from soft gluon radiation dominates. It then implies a factorisation of the renormalised PDFs at large $x$, allowing us to separate the hard-collinear divergences from the soft divergences~\cite{Korchemsky:1988si,Berger:2002sv}. In the following we shall only consider diagonal splitting functions and since the formulae apply to both quarks and gluons we shall drop the subscript $jj$ on the partonic PDF and related quantities and only specialise when needed. To factorise the PDFs we shall transform into Mellin space,
\begin{equation}\label{MellinSpace}
\tilde{f}(N) = \int_0^1 dx x^{N-1}f(x),
\end{equation}
where convolutions become products. In this space the divergent terms become,
\begin{align}
  \delta(1-x) \to 1 &&  \frac{1}{(1-x)_+} \to -\log N-\gamma_E\,. \label{mellinT}
\end{align}
The large-$x$ limit corresponds to the large-$N$ limit. The factorisation works in much the same way as the form factor by defining two jet functions and two corresponding eikonal jet functions along with a soft function~\cite{Korchemsky:1988si,Berger:2002sv},
\begin{align}\label{PDFfactorise}
  \tilde{f}(N,\mu)&=H\left(\frac{(2p\cdot n)^2}{n^2\mu^2},\al_s(\mu^2)\right)\prod_{i=L,R}\frac{J_i\left(\frac{(2p\cdot n)^2}{n^2\mu^2},\al_s(\mu^2),\eps\right)}{\eik_i\left(\frac{(2\beta\cdot n)^2}{n^2},\al_s(\mu^2),\eps\right)}\nonumber\\ &\hspace{4cm}\qquad\qquad\times\tilde{S}_\sqcap\left(N,\frac{\beta\cdot u\mu}{p\cdot u},\al_s(\mu^2),\eps\right)+\mathcal{O}\left(\frac{\log N}{N}\right)
\end{align}
where the four-velocity $\beta$ is in the $p$ direction and
$L$ and $R$ indicate which side of the cut the jet functions are (see figure~\ref{fig:cutsToDisc}). The renormalised parton distribution functions are defined as pure counterterms in minimal subtraction schemes, because they can only depend on the factorisation scale. Since the hard function $H$ and the jet functions $J_i$ are the only functions with finite terms it must mean that their non-divergent terms cancel such that eq.~\eqref{PDFfactorise} contains only poles,
\begin{equation}\label{PDFfactorisePole}
\tilde{f}(N,\mu)=\left(\prod_{i=L,R}\frac{J_i\left(\frac{(2p\cdot n)^2}{n^2\mu^2},\al_s(\mu^2),\eps\right)\bigg|_{\text{pole}}}{\eik_i\left(\frac{(2\beta\cdot n)^2}{n^2},\al_s(\mu^2),\eps\right)}\right)\tilde{S}_\sqcap\left(N,\frac{\beta\cdot u\mu}{p\cdot u},\al_s(\mu^2),\eps\right)+\mathcal{O}\left(\frac{\log N}{N}\right)
\end{equation}
where $J|_{\text{pole}}$ has the same meaning as in eq.~\eqref{JoverJ}, that it is only the poles of the jet function.
As in the case of the form factor, the soft function $\tilde{S}_\sqcap$ resums the emission of gluons with vanishing momenta in the eikonal approximation. We shall shortly see however that while its ultraviolet behaviour is qualitatively the same as that of the form-factor soft function in eq.~\eqref{softFunction}, its infrared behaviour is qualitatively different, as it presents only single poles.

The function $\tilde{S}_\sqcap$ is defined by the Mellin transform of
the $x-$space soft function
\begin{align}
  S_\sqcap\left(x,\frac{\beta\cdot u\mu}{p\cdot u},\al_s(\mu^2),\eps\right)=(p\cdot u)\int\frac{dy}{2\pi}\,e^{iy(1-x)\,p\cdot u}\,W_\sqcap\left(\beta\cdot\, uy\mu,\alpha_s(\mu^2),\epsilon\right),
  \label{def:soft-x}
\end{align}
where $W_\sqcap$ is the Wilson loop with $\sqcap-$shaped contour, see figure~\ref{fig:sqcap} (in ref.~\cite{Korchemsky:1988si} it is defined in the axial gauge),
\begin{equation}
  W_\sqcap\left(\beta\cdot u \,y\mu,\alpha_s(\mu^2),\epsilon\right) \equiv \braket{0|T\left[W_{\beta}(+\infty,y)W_u(y,0)W_\beta(0,-\infty)\right]|0}.
  \label{WpiDef}
\end{equation}
Note that the time-ordering operation here acts on the product of the three Wilson lines together. The soft function can be written in this way, despite coming from a cross-section definition because of the particular relation between path-ordering and time-ordering~\cite{Korchemsky:1992xv}.

The definition~\eqref{def:soft-x} determines two important properties
concerning the analytic structure of $W_\sqcap$, as argued in~\cite{Korchemsky:1992xv}. First of all, the soft function has support
in the physical region with $x\leq 1$ only if the singularities of
$W_\sqcap$ are located on the positive imaginary axis in the complex
$y-$plane. Indeed, if this is the case, for $x>1$ we can close the
integration contour in $y$ in eq.~\eqref{def:soft-x} through the lower
half-plane getting a vanishing result. Furthermore, the reality of
soft function implies that $W_\sqcap$ is unchanged by the
transformation $y\rightarrow -y$ followed by complex conjugation. Both
these conditions are satisfied if $W_\sqcap$ is a holomorphic function
in the variable
\begin{equation}
  \rho(y) \equiv i(u\cdot\beta\,y - i0) = \left(\rho(-y)\right)^*.
  \label{def:rhovar}
\end{equation}
In section~\ref{sec:GammaPiCalc} we show that we can write the renormalised $W_\sqcap$ as,
\begin{align}
  \label{eq:renWPi}
  \log W_\sqcap &= -\frac{1}{2}\int_0^{\mu^2}\frac{d\lambda^2}{\lambda^2}\left\{2\gamma_{\text{cusp}}(\al_s\left(\lambda^2,\eps\right))\log\left(\frac{\rho(y)\mu}{\sqrt{2}}\right)+\Gamma_\sqcap(\al_s\left(\lambda^2,\eps\right))\right\},
\end{align}
where the factor $\sqrt{2}$ was introduced in order to identify $\mu$ as the $\mathrm{\overline{MS}}$ renormalisation scale. The quantity $\Gamma_\sqcap$ will admit Casimir scaling to three loops and the scaling is determined by the representation of the Wilson lines in eq.~\eqref{WpiDef}.
Following ref.~\cite{Korchemsky:1992xv}, the soft function $\tilde{S}$ in the limit of large $N$, which is conjugate to the behaviour of $W_\sqcap$ at large $y$ through the Fourier transform in eq.~\eqref{def:soft-x}, is obtained to leading power in $N$ by replacing $y\rightarrow -iN$ in eq.~\eqref{eq:renWPi}, which leads to
\begin{align}\label{eq:renSPi}
  \log \tilde{S}_\sqcap =\,& -\frac{1}{2}\int_0^{\mu^2}\frac{d\lambda^2}{\lambda^2}\left\{2\gamma_{\text{cusp}}(\al_s\left(\lambda^2,\eps\right))\log\left(\frac{N\mu\beta\cdot u}{\sqrt{2}p\cdot u}\right)+\Gamma_\sqcap(\al_s\left(\lambda^2,\eps\right))\right\}\nonumber\\
&+ \mathcal{O}\left(\frac{\log N}{N}\right)
\end{align}
so that $\tilde{S}_\sqcap$ admits the following evolution equation
\begin{align}
  \mu\frac{d}{d\mu}\log\tilde{S}_\sqcap=\,&-2\gamma_{\text{cusp}}\left(\al_s\left(\mu^2\right)\right)\log\left(N\mu\frac{\beta\cdot u}{\sqrt{2}p\cdot u}\right)-\Gamma_\sqcap\left(\al_s\left(\mu^2\right)\right)\nonumber\\& -\int_0^{\mu^2}\frac{d\lambda^2}{\lambda^2}\gamma_{\text{cusp}}\left(\al_s\left(\lambda^2,\eps\right)\right)+ \mathcal{O}\left(\frac{\log N}{N}\right).\label{eq:diffSPi}
\end{align}
Note that the UV behaviour of $\tilde{S}_\sqcap$ is \emph{double} logarithmic: the right-hand side of eq.~\eqref{eq:diffSPi} is dominated by $\gamma_{\text{cusp}}(\al_s(\mu^2))\log\mu^2$ and therefore it has the same UV behaviour as the one of the form-factor soft function $\mathcal{S}$ in eq.~\eqref{softFint}. However, in contrast with eq.~\eqref{softFint}, the argument of the logarithm in eq.~\eqref{eq:renSPi} is independent of $\lambda$ and thus the IR behaviour in eq.~\eqref{eq:diffSPi} is \emph{single} logarithmic. Of course, it must be so also in view of
eq.~\eqref{PDFfactorisePole}: there both the renormalised PDF on the left-hand side and the hard-collinear factor $J|_{\text{pole}}/\mathcal{J}$ feature single poles. The distinct UV and IR behaviour in $\tilde{S}_\sqcap$ is associated to the presence of a length scale $y$ in the definition of the soft function eq.~\eqref{def:soft-x}. The soft function of the form factor does not involve any scale and therefore eq.~\eqref{softFint} has double logarithmic behaviour both in the UV and in the IR\@.

As before with the form factor, we seek to isolate the hard-collinear and the purely soft contributions from the Mellin transform~\eqref{MellinSpace} of the splitting functions in eq.~\eqref{eqn:SplitFunc}, $\tilde{P}(N,\al_s)$. The following argument is in the spirit of~\cite{Dixon:2008gr}. As mentioned earlier, the bare PDFs $\tilde{f}^{\text{bare}}(N,\eps)$ formally vanish because they are scaleless in dimensional regularisation~\cite{Collins:1989gx}. They feature UV divergences which are renormalised by the splitting functions $\tilde{P}(N,\al_s)$ through $\tilde{Z}(N,\al_s,\eps)$, see eq.~\eqref{eqn:Zfactor}. They are also infrared divergent because there are massless on-shell incoming partons. The IR divergences are the same as in the renormalised PDFs described by eq.~\eqref{PDFfactorisePole}. In perturbation theory it must mean that in $\tilde{f}^{\text{bare}}(N,\eps)$ the IR poles match the UV poles. In a minimal subtraction scheme the factor $Z$ in eq.~\eqref{eqn:Zfactor} consists of only poles. We are then able to construct $\tilde{f}^{\text{bare}}(N,\eps)$ in a way that separates the UV from the IR,
\begin{align}\label{eqn:factorisedPole}
  \tilde{f}^{\text{bare}}(N)&=\underbrace{\tilde{Z}(N)^{-1}}_{\text{UV}}\,\,\underbrace{\left\{\left(\prod_{i=L,R}\frac{J_i|_{\text{pole}}}{\eik_i}\right)\tilde{S}_\sqcap(N)+\mathcal{O}\left(\frac{\log N}{N}\right)\right\}}_{\text{IR}},
\end{align}
where we have suppressed the dependence on $\al_s$, the renormalisation scale $\mu$, the kinematic dependence of the functions and $\eps$.
Let us now consider the logarithm of both sides of eq.~\eqref{eqn:factorisedPole} and compute the derivative with respect to $\log\left(\mu\right)$, using the evolution equation for the ratio of jet functions in eq.~\eqref{JJdef}. The terms of the form
\begin{equation}
\int_0^{\mu^2}\frac{d\lambda^2}{\lambda^2}\gamma_{\text{cusp}}\left(\al_s\left(\lambda^2,\eps\right)\right),
\end{equation}
cancel between $\mu\frac{d}{d\mu}\log\tilde{S}_\sqcap$ and $\mu\frac{d}{d\mu}\log\frac{J}{\eik}$. The bare PDFs do not depend on the renormalisation scale so by using eqs.~\eqref{eq:diffSPi}, \eqref{eqn:splitting}, and~\eqref{JJdef} we get,
\begin{align}
  0&=\mu\frac{d}{d\mu}\log \tilde{f}^{\text{bare}}(N)\nonumber\\
  &=-2\tilde{P}(N) + 2\gamma_{J/\eik} - 2\gamma_{\text{cusp}}\log\left(\frac{\sqrt{2}p\cdot n}{\beta\cdot n}\frac{\beta\cdot u}{\sqrt{2}p\cdot u}N\right)-\Gamma_\sqcap+ \mathcal{O}\left(\frac{\log N}{N}\right)\label{LogDerivativePDF}
\end{align}
The kinematic dependence in the argument of the logarithm cancels upon identifying $\frac{p\cdot n}{\beta\cdot n}=\frac{p\cdot u}{\beta\cdot u}=\frac{p^+}{\beta^+}$.  We now require the Mellin transform of eq.~\eqref{eq:splittinglarge-x} at large $N$~\cite{Korchemsky:1988si,Korchemsky:1992xv,Dokshitzer:2005bf,Belitsky:2008mg},
\begin{align}\label{splittingFunctionDiverge}
\tilde{P}(N)=-\gamma_{\text{cusp}}\log N + B_\delta + \mathcal{O}\left(\frac{\log N}{N}\right).
\end{align}
Substituting this into eq.~\eqref{LogDerivativePDF} the dependence on $\gamma_{\text{cusp}}$ drops. This shows that the factor $\sqrt{2}$ present in eq.~\eqref{eq:renWPi} is indeed necessary for $\mu$ to be identified as $\mathrm{\overline{\text{MS}}}$ scale. Comparing the non-logarithmic terms in eqs.~\eqref{LogDerivativePDF} and~\eqref{splittingFunctionDiverge} we finally arrive at the relation,
\begin{equation}
  2B_\delta = 2\gamma_{J/\eik} - \Gamma_\sqcap.
  \label{Bdeltarel}
\end{equation}

The above equation mirrors the form factor equation for $\gamma_G$ in eq.~\eqref{gammaJJrel}. In both equations the same hard-collinear anomalous dimension $\gamma_{J/\eik}$ is present. We now proceed to use its universality to extract $\Gamma_\sqcap$ at two loops from the above equation. As in the form factor case to specialise to quarks or gluons we simply add a superscript $i=q,g$. Up to two loops the expressions for $B_\delta$ may be read off the results in eq.~\eqref{app:Pqq} and eq.~\eqref{app:Pgg} of the calculation in the appendix, in agreement with refs.~\cite{Floratos:1977au,Floratos:1978ny,GonzalezArroyo:1979df,GonzalezArroyo:1979he,Curci:1980uw,Furmanski:1980cm,Floratos:1981hs,Moch:1999eb}. They read
\begin{align}
  B_\delta^{q}&=\bigg(\frac{\al_s}{\pi}\bigg)\frac{3}{4}C_F+\bigg(\frac{\al_s}{\pi}\bigg)^2\bigg\{C_A C_F\left(\frac{11 \zeta_2}{12}-\frac{3\zeta_3}{4}+\frac{17}{96}\right)
                \nonumber\\&\quad+C_F^2 \left(-\frac{3\zeta_2}{4}+\frac{3\zeta_3}{2}+\frac{3}{32}\right)-C_F T_f n_f\left(\frac{\zeta_2}{3}+\frac{1}{24}\right)\bigg\}+\mathcal{O}(\al_s^3),\nonumber\\
  B_\delta^{g}&=\bigg(\frac{\al_s}{\pi}\bigg)\hat{b}_0+\bigg(\frac{\al_s}{\pi}\bigg)^2\bigg\{C_A^2 \left(\frac{3\zeta_3}{4}+\frac{2}{3}\right)-\frac{C_A T_f n_f}{3}- \frac{C_F T_f n_f}{4}\bigg\}+\mathcal{O}(\al_s^3).\label{BdeltaExp}
\end{align}
Substituting these results into eq.~\eqref{Bdeltarel} along with the values of $\gamma_{J/\eik}^{i}$ calculated in eq.~\eqref{QuarkJJ} and eq.~\eqref{GluonJJ} we arrive at the \emph{same quantity} for $\Gamma_\sqcap$ for quarks and gluons up to an overall Casimir scaling:
\begin{equation}
  \Gamma_\sqcap=\bigg(\frac{\al_s}{\pi}\bigg)^2\left\{\frac{C_i}{2}\left(-2\hat{b}_0\zeta_2-\frac{56}{27}T_fn_f+C_A \left[\frac{202}{27}-4\zeta_3\right]\right)\right\}+\mathcal{O}(\al_s^3).
  \label{Gammasqcap}
\end{equation}
The fact that Casimir scaling is recovered is expected of course, as this quantity is defined by Wilson lines. Nevertheless, recovering it by subtracting non-eikonal quantities is a non-trivial consistency check. It is worthwhile noting that only the $\zeta_3$ term is different between $\frac{\Gamma_\sqcap}{2}$ and $\Gamma_\wedge$ in eq.~\eqref{GeikSE}. The factor of two is present because there are two cusp contributions for the $\sqcap$ contour as opposed to one for the $\wedge$ contour. The different coefficient in front of $\zeta_3$ will be discussed further in section~\ref{sec:finalRelation}.

We have found the anomalous dimension that controls the non-collinear soft divergences of the diagonal DGLAP kernels by separating it from the hard-collinear behaviour that is identical to that in the form factor. We shall now verify the above result, eq.~\eqref{Gammasqcap}, by computing it directly.


\section{Explicit calculation of \texorpdfstring{\boldmath $\Gamma_\sqcap$}{Gamma-cap}}
\label{sec:GammaPiCalc}
In this section we derive the integral representation~\eqref{eq:renWPi} of the renormalised Wilson loop $W_\sqcap$ defined
in eq.~\eqref{WpiDef} and we verify the two-loop result of
eq.~\eqref{Gammasqcap} for the anomalous dimension $\Gamma_\sqcap$
with a direct calculation. This provides a consistency check of
eq.~\eqref{Bdeltarel}, which follows from the all-order factorisation in eq.~\eqref{PDFfactorise}.

The derivation of eq.~\eqref{eq:renWPi} consists of two parts: firstly
we will compute the bare diagrams and the UV counterterms related to
the renormalisation of the QCD coupling constant, then we will
subtract the short-distance singularities associated with the
Wilson-line operators, thus completing the renormalisation of $\log
W_\sqcap$. The non-Abelian exponentiation theorem~\cite{Sterman:1981jc,Gatheral:1983cz,Frenkel:1984pz,Gardi:2010rn}
allows us to determine directly $\log W_\sqcap$ by computing only the
\emph{webs} that capture the maximally non-Abelian colour factors of
each Feynman diagram, as defined in~\cite{Gardi:2010rn}. Moreover,
$\log W_\sqcap$ has a simpler singularity structure compared to
$W_\sqcap$, which allows us to setup the renormalisation procedure
directly at the level of the webs.

\SetPFont{}{8}
We introduce the following parameterisation for the contour of the
Wilson loop $W_\sqcap$

\vspace{-8mm}

\begin{align}
  \label{eq:contour}
{  \begin{axopicture}(80,40)(0,20)
   \Line[double,sep=1,arrow,arrowscale=0.6](0,0)(20,40)
   \Line[double,sep=1,arrow,arrowscale=0.6](20,40)(60,40)
   \Line[double,sep=1,arrow,arrowscale=0.6](60,40)(80,0)
   \PText(20,42)(0)[b]{$0$}
   \PText(60,42)(0)[b]{$y$}
   \PText(5,20)(0)[r]{$\beta$}
   \PText(40,35)(0)[t]{$u$}
   \PText(75,20)(0)[l]{$\beta$}
  \end{axopicture}}
  \qquad
  x^\mu(t)=\left\{
    \begin{array}{l@{\quad}l}
      \beta^\mu t            & t\in(-\infty,0)\\[1mm]
      u^\mu t                & t\in(0,y)\\[1mm]
      yu^\mu+\beta^\mu (t-y)  & t\in(y,+\infty)
    \end{array}
    \right.
\end{align}
We use the following Feynman rules in configuration space for the
gluon propagator in Feynman gauge and for the gluon emission from the
eikonal lines, respectively \SetPFont{}{8}
\begin{align}
  \label{eq:prop}
  {\begin{axopicture}(20,10)
    \Gluon(0,5)(20,5){1}{6}
  \end{axopicture}}
  &= \frac{\mathcal{N}}{\left[-x^2+i0\right]^{1-\epsilon}}\,g_{\mu\nu}\,,\\
{  \begin{axopicture}(20,20)(0,5)
    \PText(0,0)(0)[rt]{$x_0$}
    \PText(3,17)(0)[r]{$v$}
    \Line[double,sep=1,arrow,arrowscale=0.7,arrowpos=0.8](0,0)(10,20)
    \Line[double,sep=1,dash,dsize=1](10,20)(13,26)
    \Vertex(0,0){1}
    \Gluon(5,10)(20,15){1}{4}
  \end{axopicture}}
  &=ig_s\,T^a\,v^\mu\,\int d^dz\,\int_0^\infty d\lambda \,\delta^d(z-x_0-\lambda v),
\end{align}
where $T^a$ is the $\mathrm{SU}(N)$ generator in the appropriate
representation and
$\mathcal{N}=-\frac{\Gamma(1-\epsilon)}{4\pi^{2-\epsilon}}$.

In section~\ref{sec:one-loopWPi} we consider the one-loop calculation of
$\log\left(W_\sqcap\right)$ and then establish its general form before
and after renormalisation. In section~\ref{sec:two-loopWPi} we perform
the calculation at two loops, verifying the general structure and
obtaining an explicit result for $\Gamma_\sqcap$ consistent with
eq.~\eqref{Gammasqcap}.

\subsection{One-loop calculation}
\label{sec:one-loopWPi}

As a direct consequence of the Feynman rules given above, all the
diagrams that feature a gluon exchange between two lines with the same
lightlike velocity $v$ are proportional to $v^2$ and therefore they
are automatically zero.  At one loop there will be only two
non-vanishing webs contributing to $\log\left(W_\sqcap\right)$
\SetPFont{}{8}
\begin{align*}
  d^{(1)}_{A}=
{  \begin{axopicture}(80,30)(0,15)
   \Line[double,sep=1,arrow,arrowscale=0.6](0,0)(20,40)
   \Line[double,sep=1,arrow,arrowscale=0.6](20,40)(60,40)
   \Line[double,sep=1,arrow,arrowscale=0.6](60,40)(80,0)
   \Gluon(5,10)(50,40){2}{9}
   \Vertex(5,10){1}
   \Vertex(50,40){1}
   \PText(20,42)(0)[b]{$0$}
   \PText(60,42)(0)[b]{$y$}
  \end{axopicture}}
  \;,\qquad\qquad
  d^{(1)}_{B}=
{  \begin{axopicture}(80,30)(0,15)
   \Line[double,sep=1,arrow,arrowscale=0.6](0,0)(20,40)
   \Line[double,sep=1,arrow,arrowscale=0.6](20,40)(60,40)
   \Line[double,sep=1,arrow,arrowscale=0.6](60,40)(80,0)
   \Gluon(75,10)(30,40){2}{9}
   \Vertex(75,10){1}
   \Vertex(30,40){1}
   \PText(20,42)(0)[b]{$0$}
   \PText(60,42)(0)[b]{$y$}
  \end{axopicture}  }\\[-3mm]
\end{align*}
which differ only by a translation and therefore yield the same result
\begin{align}
  d^{(1)}(\al_s,\beta\cdot uy,\eps)&=\frac{\al_s}{\pi}(\mu^2\pi)^{\eps}\left(u\cdot\beta\right)\,C_i\,\Gamma(1-\eps)\int_0^\infty dt\int_0^yds\,(-2\beta\cdot u \,ts +i0)^{-1+\eps},
\label{eq:d1loop}
\end{align}
where $C_i$, with $i=A,F$ is the quadratic Casimir in the adjoint or
in the fundamental representation.  We notice that the integral over
the parameter $t$ diverges both in the UV limit $t\rightarrow 0$ and
in the IR regime $t\rightarrow\infty$. This fact is a consequence of
the absence of any scale associated with the integration over an
infinite Wilson line and it implies that the bare diagram in
eq.~\eqref{eq:d1loop} yields a vanishing contribution. Nevertheless,
the diagram is non-trivial after the renormalisation procedure, which
subtracts the divergence for $t\rightarrow 0$ and allows us to define
uniquely the integrand in eq.~\eqref{eq:d1loop}.  In order to expose
the analytic structure of eq.~\eqref{eq:d1loop} in terms of the
variable $\rho$ defined in eq.~\eqref{def:rhovar}, we rotate the path
along the negative imaginary axis in the complex $t-$plane. Then we change variables $t=-i\,\sqrt{2}\lambda$,
$s=-i\,\sqrt{2}\frac{\sigma}{u\cdot\beta}$, obtaining
\begin{align}
  d^{(1)}(\al_s,\rho,\eps)&=-\frac{\al_s}{\pi}(4\pi\mu^2)^{\eps}C_i\frac{\Gamma(1-\eps)}{2}\int_0^\infty\frac{d\lambda}{\lambda^{1-\eps}}\int_0^{\frac{\rho}{\sqrt{2}}}\frac{d\sigma}{\sigma^{1-\eps}}.
  \label{case1Res}
\end{align}
The complete result for $\log\left(W_\sqcap\right)$ at one loop is
given by twice the contribution of eq.~\eqref{case1Res}. It is
convenient to write it with the factor $(4\pi e^{\gamma_E})^\epsilon$
absorbed into the $\overline{\text{MS}}$ running coupling as follows
\begin{align}
  \log W_\sqcap^{\text{bare}} &= -\frac{\al_s(\mu^2)}{\pi}e^{-\epsilon\gamma_E}\,C_i\Gamma(1-\eps)\int_0^\infty \frac{d\lambda}{\lambda}\int_0^{\frac{\rho}{\sqrt{2}}}\frac{d\sigma}{\sigma}\,\left(\lambda\sigma\mu^2\right)^\epsilon+\mathcal{O}(\al_s^2).
  \label{eq:w1bare}
\end{align}
The label ``bare'' reminds us that eq.~\eqref{eq:w1bare} still has the
UV divergences associated to the cusps of the Wilson loop in
eq.~\eqref{eq:contour}, which must be subtracted before IR
singularities can be identified. Indeed, it is convenient to show
explicitly that eq.~\eqref{eq:w1bare} is independent on the
renormalisation scale, by writing the running coupling as
\begin{equation}
  \al_s \left(\mu^2\right) = (\mu^2\lambda\sigma)^{-\eps}\,\al_s\left(\frac{1}{\lambda\sigma}\right)+\frac{\hat{b}_0}{\eps}(\mu^2\lambda\sigma)^{-2\eps}\left(1-(\mu^2\lambda\sigma)^{\eps}\right)\al_s\left(\frac{1}{\lambda\sigma}\right)^2+\mathcal{O}(\al_s^3),
  \label{eq:runcoup}
\end{equation}
which leads to the expression
\begin{align}
  \log W_\sqcap^{\text{bare}} &= -C_i\int_0^\infty \frac{d\lambda}{\lambda}\int_0^{\frac{\rho}{\sqrt{2}}}\frac{d\sigma}{\sigma}\,\frac{\alpha_s\left(\frac{1}{\lambda\sigma}\right)}{\pi}\,e^{-\epsilon\gamma_E}\Gamma(1-\eps)+\mathcal{O}(\al_s^2).
  \label{eq:w1baregen}
\end{align}

\subsection{Exponentiation and renormalisation}

The integrand in eq.~\eqref{eq:w1baregen} is finite in the limit
$\epsilon\rightarrow 0$ and the singularities of $\log W_\sqcap$ arise
only after the integration over $\lambda$, $\sigma$. In particular,
following the coordinate-space analysis of refs.~\cite{Sterman,Erdogan:2013bga,Erdogan:2014gha}, we distinguish three
possible types of singular behaviour: {\emph{cusp singularities}},
which are associated to the limit $\lambda\simeq\sigma\rightarrow 0$
in which all the vertices approach a cusp of the Wilson loop;
{\emph{collinear}} singularities, which arise if either $\lambda$ or
$\sigma$ approaches the cusp, while the other parameter stays finite;
finally, the large-distance region with $\lambda\rightarrow\infty$,
which determines the IR pole. At higher perturbative orders,
individual diagrams feature soft and collinear subdivergences when a
subset of the vertices approaches one of these limits, which give rise
to poles of higher order compared to those in
eq.~\eqref{eq:w1baregen}. However, owing to its exponentiation
property, upon considering the logarithm of the Wilson-line
correlator, all the subdivergences cancel in the sum of webs at each
perturbative order~\cite{Frenkel:1983di,Berger:2002sv,Berger:2003zh,Sterman,Erdogan:2014gha}. It
is always possible to organise the calculation of
$\log\left(W_\sqcap\right)$ such that the integral over the position
of the vertex that is located at the largest distance along the
infinite Wilson line is performed last. Thus, the single infrared pole
will be generated only in the final integration, while all the
subdivergences of individual diagrams cancel in the sum of webs. This
procedure, which follows the prescriptions of ref.~\cite{Sterman},
allows us to generalise the representation of eq.~\eqref{eq:w1baregen}
to all orders
\begin{equation}
  \log W_\sqcap^{\text{bare}}=\int_0^\infty\frac{d\lambda}{\lambda}\int_0^{\frac{\rho}{\sqrt{2}}}\frac{d\sigma}{\sigma} w\left(\al_s\left(\frac{1}{\lambda\sigma}\right),\epsilon\right),
  \label{eq:wbare}
\end{equation}
where the integrand $w$ has an expansion in $\epsilon$ that involves
only non-negative powers
\begin{equation}
  w\left(\al_s\left(\frac{1}{\lambda\sigma}\right),\epsilon\right)=\sum_{n=0}^\infty w_n\left(\al_s\left(\frac{1}{\lambda\sigma}\right)\right)\,\epsilon^n.
\end{equation}
The representation eq.~\eqref{eq:wbare} is analogous to the one
derived in~\cite{Sterman} for the soft function of the form factor,
defined in eq.~\eqref{softFunction}, with the difference that in the
latter case the integrals over both the parameters are unbounded. This
is consistent with the presence of a double pole of long-distance
origin in the form factor, as compared to the single pole of this type
arising in eq.~\eqref{eq:renWPi}.

\enlargethispage{\baselineskip}

We now proceed with the renormalisation of the singularities of
short-distance origin that are present in the bare expression of
eq.~\eqref{eq:wbare}. Following~\cite{Sterman}, we notice that the
integral of $w_0$ in eq.~\eqref{eq:wbare} generates double UV poles,
which are subtracted by cutting the integration domain with
$\lambda<\frac{1}{\mu}$, $\sigma<\frac{1}{\mu}$ in
eq.~\eqref{eq:wbare}, where $\mu$ defines the subtraction point. The
contributions of $w_i$ with $i\geq1$ generate at most one UV
singularity, which we subtract in the last integration. In conclusion
we derive the representation for the sum of renormalised webs in
configuration space
\begin{equation}
  \log\left(W_\sqcap^{\text{ren}}\right)=-\int_{\frac{1}{\mu}}^\infty\frac{d\lambda}{\lambda}\int_{\frac{1}{\mu}}^{\frac{\rho}{\sqrt{2}}}\frac{d\sigma}{\sigma}\,\gamma_{\text{cusp}}\left(\al_s\left(\frac{1}{\lambda\sigma}\right)\right)-\int_{\frac{1}{\mu}}^\infty\frac{d\lambda}{\lambda}\,\Gamma_\sqcap\left(\al_s\left(\frac{1}{\lambda^2}\right)\right),
  \label{eq:wren}
\end{equation}
where we performed the integral over $\sigma$ by expanding the
coupling constant $\al_s\left(\frac{1}{\lambda\sigma}\right)$ at the
scale $\frac{1}{\lambda^2}$, as in eq.~\eqref{eq:runcoup}
\begin{equation}
  \label{eq:aslambda2}
\al_s\left(\frac{1}{\lambda\sigma}\right)=\al_s\left(\frac{1}{\lambda^2}\right)\left(\frac{\lambda}{\sigma}\right)^{-\eps}+\left(\al_s\left(\frac{1}{\lambda^2}\right)\right)^2\frac{\hat{b}_0}{\eps}\left(\frac{\lambda}{\sigma}\right)^{-2\eps}\left[1-\left(\frac{\lambda}{\sigma}\right)^{\eps}\right]+\mathcal{O}(\al_s^3).
\end{equation}
Eq.~\eqref{eq:wren}
directly leads to the result eq.~\eqref{eq:renWPi} from the web
integrals in coordinate space and it allows us to extract the
coefficients $\gamma_{\text{cusp}}$ and $\Gamma_\sqcap$.  At one-loop
order, we expand the web in eq.~(\ref{eq:w1baregen}) and we get
\begin{equation}
w\left(\al_s\left(\frac{1}{\lambda\sigma}\right),\epsilon\right)=-\frac{\al_s\left(\frac{1}{\lambda\sigma}\right)}{\pi}\,C_i\left[1+{\cal{O}}\left(\epsilon^2\right)\right]+{\cal{O}}\left(\al_s^2\right).
\end{equation}
Applying the renormalisation procedure described above we find
\begin{align}
  \log W_\sqcap^{\text{ren}}&=-\frac{C_i}{\pi}\int_{1/\mu}^\infty \frac{d\lambda}{\lambda}\int_{1/\mu}^{\frac{\rho}{\sqrt{2}}}\frac{d\sigma}{\sigma}\al_s\left(\frac{1}{\lambda\sigma}\right)+\mathcal{O}(\al_s^2)\nonumber\\
  &=\frac{\al_s(\mu^2)}{\pi}\frac{C_i}{\eps}\log\left(\frac{\rho\mu}{\sqrt{2}}\right)+\mathcal{O}(\al_s^2),
 \label{w1ren}
\end{align}
where we have used the fact that $W_\sqcap$ consists of pure
poles. The pole is infrared and is exactly the one that replicates the
soft divergence of the PDF. We compare eq.~\eqref{w1ren} with the
poles of eq.~\eqref{eq:renWPi} getting
\begin{align}
  \begin{split}
    \gamma_{\text{cusp}}&=\frac{\alpha_s}{\pi}\,C_i+{\cal{O}}\left(\al_s^2\right),\\
    \Gamma_\sqcap&=0\cdot\al_s+{\cal{O}}\left(\al_s^2\right).
  \end{split}
\end{align}

\subsection{Two-loop calculation}
\label{sec:two-loopWPi}

We now apply the renormalisation procedure to the two-loop webs.  Only a
few diagrams contribute to this order and they are represented below

\vspace{-8mm}

\begin{align}
  d^{(2)}_{\mathrm{SE}}&=
  {\begin{axopicture}(80,40)(0,18)
   \Line[double,sep=1,arrow,arrowscale=0.6](0,0)(20,40)
   \Line[double,sep=1,arrow,arrowscale=0.6](20,40)(60,40)
   \Line[double,sep=1,arrow,arrowscale=0.6](60,40)(80,0)
   \Gluon(5,10)(50,40){2}{9}
   \GCirc(27.5,25){8}{0.6}
   \Vertex(5,10){1}
   \Vertex(50,40){1}
  \end{axopicture}},\,&
  d^{(2)}_{X_2}&=
 { \begin{axopicture}(80,40)(0,18)
   \Line[double,sep=1,arrow,arrowscale=0.6](0,0)(20,40)
   \Line[double,sep=1,arrow,arrowscale=0.6](20,40)(60,40)
   \Line[double,sep=1,arrow,arrowscale=0.6](60,40)(80,0)
   \Gluon(5,10)(30,40){-1}{6}
   \Gluon(15,30)(50,40){1}{6}
   \Vertex(5,10){1}
   \Vertex(30,40){1}
   \Vertex(15,30){1}
   \Vertex(50,40){1}
  \end{axopicture}},\,&
  d^{(2)}_{X_3}&=
{  \begin{axopicture}(80,40)(0,18)
   \Line[double,sep=1,arrow,arrowscale=0.6](0,0)(20,40)
   \Line[double,sep=1,arrow,arrowscale=0.6](20,40)(60,40)
   \Line[double,sep=1,arrow,arrowscale=0.6](60,40)(80,0)
   \Gluon(5,10)(50,40){-1}{8}
   \Gluon(30,40)(75,10){1}{8}
   \Vertex(5,10){1}
   \Vertex(50,40){1}
   \Vertex(30,40){1}
   \Vertex(75,10){1}
  \end{axopicture}},\nonumber\\
    d^{(2)}_{Y_s}&=
 { \begin{axopicture}(80,40)(0,18)
   \Line[double,sep=1,arrow,arrowscale=0.6](0,0)(20,40)
   \Line[double,sep=1,arrow,arrowscale=0.6](20,40)(60,40)
   \Line[double,sep=1,arrow,arrowscale=0.6](60,40)(80,0)
   \Gluon(5,10)(35,20){-1}{4}
   \Gluon(35,20)(25,40){1}{4}
   \Gluon(35,20)(45,40){1}{4}
   \Vertex(5,10){1}
   \Vertex(35,20){1}
   \Vertex(25,40){1}
   \Vertex(45,40){1}
  \end{axopicture}},&
    d^{(2)}_{Y_L}&=
{  \begin{axopicture}(80,40)(0,18)
   \Line[double,sep=1,arrow,arrowscale=0.6](0,0)(20,40)
   \Line[double,sep=1,arrow,arrowscale=0.6](20,40)(60,40)
   \Line[double,sep=1,arrow,arrowscale=0.6](60,40)(80,0)
   \Gluon(5,10)(35,20){-1}{4}
   \Gluon(15,30)(35,20){1}{4}
   \Gluon(35,20)(45,40){1}{4}
   \Vertex(5,10){1}
   \Vertex(15,30){1}
   \Vertex(35,20){1}
   \Vertex(45,40){1}
  \end{axopicture}},&
    d^{(2)}_{3s}&=
 { \begin{axopicture}(80,40)(0,18)
   \Line[double,sep=1,arrow,arrowscale=0.6](0,0)(20,40)
   \Line[double,sep=1,arrow,arrowscale=0.6](20,40)(60,40)
   \Line[double,sep=1,arrow,arrowscale=0.6](60,40)(80,0)
   \Gluon(5,10)(40,20){-1}{4}
   \Gluon(40,20)(40,40){1}{4}
   \Gluon(40,20)(75,10){1}{4}
   \Vertex(5,10){1}
   \Vertex(75,10){1}
   \Vertex(40,20){1}
   \Vertex(40,40){1}
  \end{axopicture}},
  \label{fig:dia2}\\[-2mm]\nonumber
\end{align}
where we omit the configurations that are simply obtained by mirror
symmetry.  The diagrams in the first row of eq.~(\ref{fig:dia2}) are
computed following the same steps as the one-loop case. As in the
one-loop case, we write the bare webs using the representation in
eq.~(\ref{eq:wbare}) and we
define the integrand $w^{(2)}_i$ of diagram $d^{(2)}_i$ as,
\begin{equation}
  d^{(2)}_i(\al_s,\rho,\eps) = \int_0^\infty\frac{d\lambda}{\lambda}\int_0^{\frac{\rho}{\sqrt{2}}}\frac{d\sigma}{\sigma}\left(\frac{\al_s\left(1/\lambda\sigma\right)}{\pi}\,e^{-\eps\gamma_E}\right)^2\,w^{(2)}_i(\eps).
  \label{eq:2looprep}
\end{equation}

From now on we drop the arguments on $d_i$ and $w_i$ which are understood to have the above arguments unless otherwise stated. The first diagram, $d^{(2)}_{\mathrm{SE}}$ is
obtained from eq.~(\ref{eq:d1loop}) by replacing the gluon propagator
eq.~(\ref{eq:prop}) with the one-loop expression
\begin{align}
  \begin{split}
    D_{\mu\nu}^{(1)}(x) =& -\frac{\alpha_s}{16\pi^3}\left(\pi^2\mu^2\right)^\epsilon\frac{\Gamma^2(1-\epsilon)}{\epsilon(1-2\epsilon)(3-2\epsilon)}\Big[C_A(5-3\epsilon)-4n_fT_f(1-\epsilon)\Big]\\
    &\hspace{8.5cm}\times\left[-x^2+i\,0\right]^{-1+2\epsilon}\,g_{\mu\nu},
  \end{split}
\end{align}
where we discarded the longitudinal components of the propagator, that
are proportional to $\partial_\mu\partial_\nu$, because they decouple
from the amplitude via Ward identities~\cite{Sterman}. The result is
\begin{align}
  w^{(2)}_{\mathrm{SE}}&=-
  C_i\frac{\Gamma^2(1-\epsilon)}{8\epsilon(1-2\epsilon)(3-2\epsilon)}\Big[C_A(5-3\epsilon)-4n_fT_f(1-\epsilon)\Big],
  \label{eq:f2SE}
\end{align}
in agreement with the results of~\cite{Korchemsky,Sterman}. We notice
immediately that, at two-loop level, the representation
eq.~\eqref{eq:2looprep} of the individual webs has subdivergences,
which are manifest as explicit poles in the integrand $w_i$. In this
case the subdivergence is cancelled by the coupling renormalisation in
the QCD Lagrangian that will be taken into account later in this
section. The double gluon exchange diagrams give
\begin{align}
  \label{eq:f2X2}
  w^{(2)}_{X_2}&=-
  C_i\,C_A\frac{\Gamma^2(1-\epsilon)}{8\epsilon^2}\,,\\
  \label{eq:f2X3}
  w^{(2)}_{X_3}&=-
  C_i\,C_A\frac{\Gamma^2(1-\epsilon)}{2\epsilon}\left[\frac{1}{\epsilon}-B(\epsilon,1+\epsilon)\right].
\end{align}
Both results are in agreement with the maximally non-Abelian
contributions of the diagrams $W_c$ and $W_d$ reported in~\cite{Korchemsky}. The integrand of the diagram $d^{(2)}_{X_2}$ in
eq.~\eqref{eq:f2X2} has poles of short-distance origin, associated to the configuration shown below, where the two innermost vertices on the Wilson lines are in proximity of the cusp

\vspace{-8mm}

\begin{align*}
  w^{(2),\text{subdiv}}_{X_2}=
  {  \begin{axopicture}(80,40)(0,10)
   \Line[double,sep=1,arrow,arrowscale=0.6,arrowpos=0.7](0,0)(20,40)
   \Line[double,sep=1,arrow,arrowscale=0.6](20,40)(60,40)
   \Line[double,sep=1,arrow,arrowscale=0.6,arrowpos=0.3](60,40)(80,0)
   \Gluon(5,10)(24,40){-1}{6}
   \Gluon(18,36)(50,40){-1}{6}
   \Vertex(5,10){1}
   \Vertex(24,40){1}
   \Vertex(18,36){1}
   \Vertex(50,40){1}
  \end{axopicture}}.\\[-4mm]
\end{align*}
These subdivergences are not related to QCD renormalisation and they
will cancel in the sum of all webs. Eq.~\eqref{eq:f2X3} is finite when
$\epsilon\rightarrow 0$ as we discuss more in detail in appendix~\ref{sec:appendixX3}.
The diagrams in the second row of eq.~(\ref{fig:dia2}) involve the
three-gluon vertex, whose Feynman rule in configuration space reads
\begin{align}
    V^{a_1a_2a_3}_{\mu_1\mu_2\mu_3}(x_1,x_2,x_3)=\,&g_sf^{a_1a_2a_3}\bigg[\left(-i\frac{\partial}{\partial x_1^{\mu_3}}+i\frac{\partial}{\partial x_2^{\mu_3}}\right)g_{\mu_1\mu_2}+\left(-i\frac{\partial}{\partial x_2^{\mu_1}}+i\frac{\partial}{\partial x_3^{\mu_1}}\right)g_{\mu_2\mu_3}\nonumber\\
      &\qquad\qquad\qquad+\left(-i\frac{\partial}{\partial x_3^{\mu_2}}+i\frac{\partial}{\partial x_1^{\mu_2}}\right)g_{\mu_3\mu_1}\bigg].
    \label{eq:tgv}
\end{align}
We notice that the diagrams $d^{(2)}_{Y_s}$ and $d^{(2)}_{Y_L}$ are
not related by symmetry transformations, because the former has two
gluon attachments on the segment of finite length $y$, while the
latter has two emissions from the semi-infinite line. We begin with the
calculation of $d^{(2)}_{Y_s}$
\begin{align}
  d^{(2)}_{Y_s}=\,&K_Y\int d^dz\int_{-\infty}^0dt_3\Bigg\{\int_0^yds_1\int_{s_1}^{y}ds_2\left(u\cdot\frac{\partial}{\partial s_2\,u}\right)-\int_0^yds_2\int_0^{s_2}ds_1\left(u\cdot\frac{\partial}{\partial s_1\,u}\right)\Bigg\}\nonumber\\
  &\times \left[-(u s_2-z)^2+i0\right]^{-1+\epsilon}\left[-(u s_1-z)^2+i0\right]^{-1+\eps}\left[-(\beta t_3-z)^2+i0\right]^{-1+\eps},\nonumber\\
\label{eq:intd2Ys}
\end{align}
where we introduced the normalisation factor
$K_Y=ig_s^4\frac{C_iC_A}{2}\mathcal{N}^3u\cdot\beta$. We write the differential operators in
eq.~(\ref{eq:intd2Ys}) in terms of total derivatives as follows
\begin{equation}
  u\cdot\frac{\partial}{\partial s_2\,u}\left[-z^2+2z\cdot u s_2+i0\right]^{-1+\eps} = \frac{d}{ds_2}\left[-z^2+2z\cdot u s_2+i0\right]^{-1+\eps},
\end{equation}
which allows us to perform immediately the integrals over $s_2$ and
over $s_1$, respectively in the first and in the second term in curly
brackets, by evaluating the appropriate propagator at the endpoints of
the integration interval. Eq.~(\ref{eq:intd2Ys}) becomes
\begin{align}
  \label{eq:d2Ys}
  d^{(2)}_{Y_s}&=d^{(2)}_{E}(y\,u,u\cdot\beta)+d^{(2)}_{E}(0,u\cdot\beta)-2d^{(2)}_{B}(u\cdot\beta),
\end{align}
in terms of the functions
\begin{align}
  \label{eq:d2E}
  d^{(2)}_E(v,u\cdot\beta) &= K_Y\int d^dz\int_{-\infty}^0dt\int_0^y ds \left[-(\beta t-z)^2\right]^{\eps-1}\left[-(u s-z)^2\right]^{\eps-1}\left[-(v-z)^2\right]^{\eps-1},\\
  \label{eq:d2B}
  d^{(2)}_B(u\cdot\beta) &= K_Y\int d^dz\int_{-\infty}^0dt\int_0^y ds \left[-(\beta t-z)^2\right]^{\eps-1}\left[-(u s-z)^2\right]^{2\eps-2},
\end{align}
where the prescription $+i0$ is understood in every factor appearing in
the integrals. Each function has a clear diagrammatic interpretation,
because the integrands are products of scalar propagators in
coordinate space. Thus, $d^{(2)}_{Y_s}$ is decomposed in a sum of
diagrams, as discussed in~\cite{Sterman}, giving in one-to-one
correspondence with the three terms in eq.~\eqref{eq:d2Ys}
\begin{align}
  \label{eq:d2YsDEC}
  d^{(2)}_{Y_s}=
  \SetPFont{}{8}
{  \begin{axopicture}(60,40)(0,10)
    \Line[double,sep=1,arrow,arrowscale=0.6](0,0)(20,40)
    \Line[double,sep=1,arrow,arrowscale=0.6](20,40)(60,40)
    \Line[dash](5,10)(35,20)
    \Line[dash](35,20)(60,40)
    \Line[dash](35,20)(25,40)
    \Vertex(5,10){1}
    \Vertex(35,20){1}
    \Vertex(25,40){1}
    \Vertex(60,40){2}
    \PText(5,10)(0)[tl]{$\beta t$}
    \PText(35,18)(0)[tc]{$z$}
    \PText(25,43)(0)[bc]{$u s$}
    \PText(60,42)(0)[bc]{$u y$}
  \end{axopicture}}\ \
  +\ \
{  \begin{axopicture}(60,40)(0,10)
    \Line[double,sep=1,arrow,arrowscale=0.6](0,0)(20,40)
    \Line[double,sep=1,arrow,arrowscale=0.6](20,40)(60,40)
    \Line[dash](5,10)(35,20)
    \Line[dash](35,20)(20,40)
    \Line[dash](35,20)(50,40)
    \Vertex(5,10){1}
    \Vertex(35,20){1}
    \Vertex(50,40){1}
    \Vertex(20,40){2}
    \PText(5,10)(0)[tl]{$\beta t$}
    \PText(35,18)(0)[tc]{$z$}
    \PText(50,43)(0)[bc]{$ u s$}
    \PText(20,43)(0)[bc]{$0$}
  \end{axopicture}}\ \
  -2\,
{  \begin{axopicture}(60,40)(0,10)
    \Line[double,sep=1,arrow,arrowscale=0.6](0,0)(20,40)
    \Line[double,sep=1,arrow,arrowscale=0.6,arrowpos=0.6](20,40)(60,40)
    \Line[dash](5,10)(35,20)
    \Bezier[dash](35,20)(30,25)(30,35)(35,40)
    \Bezier[dash](35,20)(40,25)(40,35)(35,40)
    \Vertex(5,10){1}
    \Vertex(35,20){1}
    \Vertex(35,40){1}
    \PText(5,10)(0)[tl]{$\beta t$}
    \PText(35,18)(0)[tc]{$z$}
    \PText(35,43)(0)[bc]{$u s$}
  \end{axopicture}}\;,\\[-2mm]\nonumber
\end{align}
where the dashed lines represent scalar propagators and dotted
vertices on the Wilson line indicate that the position of the
endpoint of the propagator is not be integrated over. We integrate
eqs.~(\ref{eq:d2E}) and~(\ref{eq:d2B}) over the position $z$ of the
three-gluon vertex and write the results in the two-dimensional
integral representation in eq.~(\ref{eq:2looprep}). We obtain
\begin{subequations}
\begin{alignat}{3}
  w^{(2)}_E(y\,u,u\cdot\beta) &= C_iC_A\frac{\Gamma(1-2\epsilon)}{16\epsilon}\Big[B(-\epsilon,1-\epsilon)-B(-\epsilon,1+\epsilon)\Big],\label{eq:dE}\\
  w^{(2)}_E(0,u\cdot\beta) &= C_iC_A\frac{\Gamma(1-\epsilon)\Gamma(1-2\epsilon)}{16\epsilon}\Gamma(\epsilon),\label{eq:d0}\\
  w^{(2)}_B(u\cdot\beta)& = C_iC_A\frac{\Gamma^2(1-\epsilon)}{16\epsilon(1-2\epsilon)}.\label{eq:dB}
\end{alignat}
\end{subequations}
Let us discuss the singularity structure of the separate integrals
above. The only one which is separately finite is eq.~\eqref{eq:dE},
which corresponds to the integrand of the first diagram in
eq.~\eqref{eq:d2YsDEC}. Eq.~\eqref{eq:d0} has single and double poles
that will cancel the corresponding singularities in
eq.~\eqref{eq:f2X2}. Indeed, the second diagram in
eq.~\eqref{eq:d2YsDEC}, which is associated to the integrand in
eq.~\eqref{eq:d0}, has subdivergences of short distance origin when
the three-gluon vertex approaches the cusp, similarly to the behaviour
shown by diagram $d^{(2)}_{X_2}$. The single pole in eq.~\eqref{eq:dB}
is entirely due to the presence of a one-particle-irreducible UV
divergent subgraph in the last diagram in
eq.~\eqref{eq:d2YsDEC}. Therefore, this singularity is removed by the
counterterms of the QCD Lagrangian. Using these results, the total
contribution of the diagram $d^{(2)}_{Y_s}$ in eq.~(\ref{eq:d2Ys})
agrees with the corresponding expression for diagram $W_e$ in~\cite{Korchemsky} and in the notation of eq.~\eqref{eq:2looprep} it
reads
\begin{align}
  w^{(2)}_{Y_s}&=-C_iC_A\frac{\Gamma(1-\epsilon)}{16\epsilon^2(1-2\epsilon)}\Big[\Gamma(1-\epsilon)-2\Gamma(2-2\epsilon)\Gamma(1+\epsilon)\Big].
  \label{eq:f2Ys}
\end{align}

The next diagram, $d^{(2)}_{Y_L}$, differs from eq.~(\ref{eq:intd2Ys})
only by the presence of the two gluon attachments on the semi-infinite
Wilson line rather than on the finite one. Once again, we write the
three gluon vertex in terms of total derivative and we decompose the
diagram as
\SetPFont{}{8}
\begin{align}
  \label{eq:d2YL}
  d^{(2)}_{Y_L}&=d^{(2)}_{E}(0,u\cdot\beta)-2d^{(2)}_{B}(u\cdot\beta)
  \equiv
 { \begin{axopicture}(60,28)(0,13)
    \Line[double,sep=1,arrow,arrowscale=0.6](0,0)(20,40)
    \Line[double,sep=1,arrow,arrowscale=0.6](20,40)(60,40)
    \Line[dash](5,10)(35,20)
    \Line[dash](35,20)(20,40)
    \Line[dash](35,20)(50,40)
    \Vertex(5,10){1}
    \Vertex(35,20){1}
    \Vertex(50,40){1}
    \Vertex(20,40){2}
  \end{axopicture}}\ \
  -2\
  {\begin{axopicture}(60,28)(0,13)
    \Line[double,sep=1,arrow,arrowscale=0.6](0,0)(20,40)
    \Line[double,sep=1,arrow,arrowscale=0.6,arrowpos=0.6](20,40)(60,40)
    \Line[dash](35,15)(55,40)
    \Bezier[dash](35,15)(25,20)(15,20)(4,8)
    \Bezier[dash](35,15)(25,5)(15,5)(4,8)
    \Vertex(4,8){1}
    \Vertex(35,15){1}
    \Vertex(55,40){1}
  \end{axopicture}}
  ,\\[-3mm]\nonumber
\end{align}
where we have used the functions defined in eqs.~(\ref{eq:d2E}) and~(\ref{eq:d2B}). The
comparison of eqs.~\eqref{eq:d2Ys} and~\eqref{eq:d2YL} shows that two
diagrams differ only by the term featuring a scalar propagator
connected to the endpoint of the Wilson line. In the case of
$d^{(2)}_{Y_L}$, the Wilson line is infinite and this term is
absent. This result was shown in~\cite{Sterman}, by introducing a
cutoff on the infinite line and carefully taking the limit to
infinity, which does not commute with the integration over $z$. The
same conclusion is found by computing $d^{(2)}_{Y_L}$ in momentum
space, as shown in appendix~\ref{sec:endpoint}. Using
eqs.~\eqref{eq:dE},~\eqref{eq:d0} and~\eqref{eq:dB} we get
\begin{align}
  \label{eq:f2YL}
  w^{(2)}_{Y_L}
  &=C_iC_A\frac{\Gamma(1-\epsilon)}{8\epsilon}\bigg[\frac{\Gamma(1-2\epsilon)\Gamma(\epsilon)}{2}-\frac{\Gamma(1-\epsilon)}{1-2\epsilon}\bigg].
\end{align}
By construction, the expression above has the same singularities as
$w^{(2)}_{Y_s}$, because the integrand of the diagram $d^{(2)}_{Y_L}$
differs from $d^{(2)}_{Y_s}$ only by the function in
eq.~\eqref{eq:dE}, which is~finite.

We compute the diagram $d^{(2)}_{3s}$ using the same procedure

\vspace{-8mm}

\begin{align}
  \label{eq:d3s}
  d^{(2)}_{3s}&=2\,d^{(2)}_E(yu,u\cdot\beta)
  \equiv
 { \begin{axopicture}(60,40)(0,15)
    \Line[double,sep=1,arrow,arrowscale=0.6](0,0)(20,40)
    \Line[double,sep=1,arrow,arrowscale=0.6](20,40)(60,40)
    \Line[dash](5,10)(35,20)
    \Line[dash](35,20)(60,40)
    \Line[dash](35,20)(30,40)
    \Vertex(5,10){1}
    \Vertex(35,20){1}
    \Vertex(30,40){1}
    \Vertex(60,40){2}
  \end{axopicture}}
\ \ +\ \
 { \begin{axopicture}(60,40)(0,15)
    \Line[double,sep=1,arrow,arrowscale=0.6](0,40)(40,40)
    \Line[double,sep=1,arrow,arrowscale=0.6,arrowpos=0.6](40,40)(60,0)
    \Line[dash](0,40)(25,20)
    \Line[dash](30,40)(25,20)
    \Line[dash](55,10)(25,20)
    \Vertex(55,10){1}
    \Vertex(30,40){1}
    \Vertex(25,20){1}
    \Vertex(0,40){2}
  \end{axopicture}},  \\[-3mm]\nonumber
\end{align}
getting
\begin{align}
  \label{eq:f23s}
  w^{(2)}_{3s}&=C_iC_A\frac{\Gamma(1-2\epsilon)}{8\epsilon}\Big[B(-\epsilon,1-\epsilon)-B(-\epsilon,1+\epsilon)\Big],
\end{align}
which is finite because it involves only the function in
eq.~\eqref{eq:dE}. We renormalise the UV divergences associated with
the QCD vertices and propagators by means of the one-loop counterterm
\begin{equation}
  \label{eq:d2ct}
  d^{(2)}_{\mathrm{ct}}=-\frac{\alpha_s}{4\pi\epsilon}\bigg[\frac{11}{3}C_A-\frac{4}{3}T_fn_f\bigg]\,d^{(1)},
\end{equation}
where $d^{(1)}$ is the result of the one-loop diagram
eq.~(\ref{case1Res}).

Finally, we sum all the diagrams depicted in eq.~(\ref{fig:dia2}),
including the symmetric configurations which are not shown there, getting
\begin{align}
\label{eq:wbarealldia}
\log W_\sqcap^{\text{bare}}&=2d^{(1)}+\Big(2d^{(2)}_{\text{SE}}+2d^{(2)}_{Y_s}+2d^{(2)}_{Y_L}+2d^{(2)}_{X_2}+d^{(2)}_{3s}+d^{(2)}_{X_3}+2d^{(2)}_{ct}\Big),\nonumber\\
&=2\Big\{d^{(1)}+d^{(2)}_{\text{SE}}+2d^{(2)}_{Y_L}+d^{(2)}_{X_2}+d^{(2)}_{ct}\Big\}+2d^{(2)}_{3s}+d^{(2)}_{X_3},
\end{align}
where to get to the second line we used the identity
$2d^{(2)}_{Y_s}=2d^{(2)}_{Y_L}+d^{(2)}_{3s}$ that is obtained by
comparing eqs.~\eqref{eq:d2YsDEC},~\eqref{eq:d2YL} and
eq.~\eqref{eq:d3s}. The terms in curly brackets in the final
expression are the same that appear in the calculation of the cusped
Wilson loop with two semi-infinite lightlike lines, discussed in~\cite{Sterman}. The last two contributions in
eq.~\eqref{eq:wbarealldia} are special to the configuration of
$W_\sqcap$, where the semi-infinite lines are connected by a finite
lightlike segment. The final expression in eq.~\eqref{eq:wbarealldia}
follows the decomposition of polygon-shaped Wilson loops presented in~\cite{Sterman}. The distinction between the terms inside and outside
the curly brackets in eq.~\eqref{eq:wbarealldia} stems from the
structure of their singularities.
The former ones give rise to cusp configurations characterised by
double UV poles and therefore they can be written in terms of the
representation in eq.~\eqref{eq:wbare} with a finite integrand. The
last contributions in eq~\eqref{eq:wbarealldia} generate at most a
single pole, associated to the configurations where all the vertices
simultaneously approach a lightlike segment, and therefore their
combination will give rise to an integrand of order $\eps$ in
eq.~\eqref{eq:wbare}, as we verify by expanding eqs.~\eqref{eq:f2X3}
and~\eqref{eq:f23s}
\begin{equation}
  \label{eq:endpointterms}
  2w^{(2)}_{3s}+w^{(2)}_{X_3} = \frac{3}{2}\eps\,C_iC_A\,\zeta_3.
\end{equation}

Substituting the results in eqs.~\eqref{eq:f2SE},~\eqref{eq:f2X2},~\eqref{eq:f2X3},~\eqref{eq:f2YL} and~\eqref{eq:f23s} into
eq.~\eqref{eq:wbarealldia}, the integral representation of the bare
diagrams reads
\begin{align}
  \log W_\sqcap^{\text{bare}} =\,&C_i\int_0^{\infty}\frac{d\lambda}{\lambda}\int_0^{\frac{\rho}{\sqrt{2}}}\frac{d\sigma}{\sigma}\bigg\{\frac{\al_s\left(\frac{1}{\lambda\sigma}\right)}{\pi}e^{-\eps\gamma_E}\Gamma(1-\eps)\bigg[-1+\frac{\al_s\left(\frac{1}{\lambda\sigma}\right)}{\pi}\frac{11C_A-4T_fn_f}{12\eps}\bigg]\nonumber\\
  &+\bigg(\frac{\al_s\left(\frac{1}{\lambda\sigma}\right)}{\pi}e^{-\eps\gamma_E}\bigg)^2\bigg[\frac{C_A}{4}\left(\frac{3(-4\!+\!3\eps)\Gamma(1\!-\!\eps)\Gamma(2\!-\!\eps)}{\eps^2(3-8\eps+4\eps^2)}-4\pi\Gamma(-2\eps)\cot\left(\frac{\pi\eps}{2}\right)\right)\nonumber\\
    &-T_fn_f\,\frac{\Gamma(2-\eps)\Gamma(-\eps)}{3-8\eps+4\eps^2}\bigg]\bigg\}.
\end{align}
By expanding the equation above in $\eps$ we get
\begin{align}
  \log W_\sqcap^{\text{bare}} &=
  -\int_0^{\infty}\frac{d\lambda}{\lambda}\int_0^{\frac{\rho}{\sqrt{2}}}\frac{d\sigma}{\sigma}\bigg\{\bigg(\frac{\al_s\left(\frac{1}{\lambda\sigma}\right)}{\pi}\bigg)\left[1+\frac{\eps^2}{2}\zeta_2+\mathcal{O}(\eps^3)\right]\nonumber\\
                &\hskip 90pt +\bigg(\frac{\al_s\left(\frac{1}{\lambda\sigma}\right)}{\pi}\bigg)^2\bigg[\gamma_{\text{cusp}}^{(2)}+\eps\bigg(\Gamma_\sqcap^{(2)}+\frac{3\hat{b}_0\zeta_2}{2}\bigg)+\mathcal{O}(\eps^2)\bigg]\bigg\},
\end{align}
where $\Gamma_\sqcap^{(2)}$ is the two-loop contribution to $\Gamma_\sqcap$ eq.~\eqref{Gammasqcap},
\begin{equation}
\Gamma_\sqcap^{(2)}=\frac{C_i}{2}\left(-2\hat{b}_0\zeta_2-\frac{56}{27}T_fn_f+C_A\left[\frac{202}{27}-4\zeta_3\right]\right).
\end{equation}

Now we renormalise using the procedure outlined in the one-loop case,
see eq.~\eqref{eq:wren}. For the terms of $\mathcal{O}(\eps^0)$, the
cusp terms, we integrate from $1/\mu$ on both integrals. For all the
subsequent terms, the $\sigma$ integral is performed first,
integrating from $0$ to $\frac{\rho}{\sqrt{2}}$. Then the parameter
$\lambda$ integrated from $\frac{1}{\mu}$. By doing this we get,
\begin{align}
\log W_\sqcap=\,&\al_s(\mu^2)\frac{1}{\eps}\log\left(\frac{\rho\mu}{\sqrt{2}}\right) \nonumber\\&+ \al_s(\mu^2)^2\bigg\{-\frac{\hat{b}_0}{2\eps^2}\log\left(\frac{\rho\mu}{\sqrt{2}}\right)+\frac{1}{\eps}\left(\frac{1}{4}\Gamma_\sqcap^{(2)}+\frac{1}{2}\gamma_{\text{cusp}}^{(2)}\log\left(\frac{\rho\mu}{\sqrt{2}}\right)\right)\bigg\}.
\end{align}
Again it is reminded that we have used the fact that $W_\sqcap$ consists of pure poles. The above is reproduced by eq.~\eqref{eq:renWPi}
\begin{equation}
\label{Wpi}
  \log W_\sqcap=-\frac{1}{2}\int_0^{\mu^2}\frac{d\xi^2}{\xi^2}\bigg\{2\gamma_{\text{cusp}}(\al_s(\xi^2,\eps))\log\left(\frac{\rho\mu}{\sqrt{2}}\right) + \Gamma_\sqcap(\al_s(\xi^2,\eps))\bigg\}.
\end{equation}
By this point we have determined the anomalous dimension
$\Gamma_\sqcap$ in two different ways, first by extracting it from the
evolution of PDFs using the universality of the hard-collinear poles
$J/\eik$ and now by a direct computation of the renormalisation of the
corresponding Wilson-line correlator.

\section{Relating Wilson-line geometries to physical quantities}
\label{sec:finalRelation}

In this section we establish a set of relations between different
physical quantities, based on the properties of the Wilson loops
discussed in section~\ref{sec:GammaPiCalc}. In section~\ref{sec:FFtoDGLAP}
we will show that the single infrared poles in the quark and in the
gluon form factors are related to the corresponding diagonal term in
the DGLAP kernels by a precise eikonal quantity that is associated to
the geometry of the Wilson loops with lightlike lines. The latter
emerges as the difference between the anomalous dimensions associated
with a wedged Wilson loop with two semi-infinite lines and a
$\sqcap-$shaped Wilson loop. This difference, in turn, can be
expressed as the anomalous dimension associated with a parallelogram
(or more generally) a polygon with lightlike segments. In
section~\ref{sec:highOrder} we use this relation to extract the anomalous
dimensions associated with a polygon Wilson loop to three loops, which
is related to the soft anomalous dimension appearing in the
resummation of threshold logarithms in the Drell-Yan process. Finally
we extract the fermionic components of the four-loop result in the planar limit.

\subsection{Relating the form factor with the DGLAP kernels}
\label{sec:FFtoDGLAP}

The direct calculation of the anomalous dimension $\Gamma_\sqcap$ in
section~\ref{sec:GammaPiCalc} confirms the identity in eq.~\eqref{Bdeltarel}
\begin{equation}
  2B_\delta = 2\gamma_{J/\eik} - \Gamma_\sqcap,
  \label{Bdeltarel2}
\end{equation}
which follows from the factorisation of the parton distribution
functions for large $x$. This identity is interpreted as a
decomposition of $B_\delta$, which was defined in
eq.~\eqref{splittingFunctionDiverge} as the coefficient of the delta
distribution in the splitting functions in the limit $x\rightarrow 1$,
into the contribution of the hard-collinear radiation,
$\gamma_{J/\eik}$, and the purely soft one, which is encoded by
$\Gamma_\sqcap$. In eq.~\eqref{Bdeltarel2} we suppressed the dependence on
the external parton: the relation holds for both quarks and
gluons. The hard-collinear contribution $\gamma_{J/\eik}$ is process
independent, as discussed in section~\ref{sec:FormFactorJJ} in the
context of the infrared factorisation of the form factor. Indeed,
eq.~\eqref{gammaJJrel} provides the analog of eq.~\eqref{Bdeltarel2}
\begin{equation}
  \gamma_G = 2\gamma_{J/\eik} - \Gamma_\wedge,
  \label{gammaJJrel2}
\end{equation}
where $\gamma_G$ is the anomalous dimension that determines the single
poles of the form factor. By comparing eq.~\eqref{Bdeltarel2} and
eq.~\eqref{gammaJJrel2} we derive the relation
\begin{equation}
 \label{MVVid2}
 \gamma_G - 2B_\delta = \Gamma_\sqcap - \Gamma_\wedge,
\end{equation}
which connects the single poles in the form factor with the diagonal
DGLAP kernels. The two quantities appearing on the left-hand side of
eq.~\eqref{MVVid2} depend on both the spin and the colour
representation of the external particles in a non-trivial way. In
contrast, the right-hand side involves the anomalous dimensions of two
eikonal quantities, which depend only on the colour representation of
the particles and obey Casimir scaling up to three loops.  Therefore,
eq.~\eqref{MVVid2} allows us to interpret the function
$f_{\text{eik}}$ eq.~\eqref{MVVidentities}, which was introduced in~\cite{Ravindran:2004mb,FormFactors} as the difference
$f_{\text{eik}}\equiv\gamma_G-2B_\delta$, in terms of the anomalous
dimensions of Wilson-line correlators.  By substituting the two-loop
expressions of $\Gamma_\sqcap$ and $\Gamma_\wedge$ from direct
calculations, respectively in eqs.~(\ref{Gammasqcap}) and~(\ref{GeikSE}), into the right-hand side of eq.~\eqref{MVVid2} we
reproduced the two-loop result obtained from the difference of
$\gamma_G$ and $B_\delta$ in ref.~\cite{FormFactors}, namely
\begin{align}
  \label{eq:feik2l}
  f_{\text{eik}}
  &=\left(\frac{\alpha_s}{\pi}\right)^2 C_i \bigg[C_A \left(-\frac{11 \zeta_2}{24}-\frac{7}{4} \zeta_3+\frac{101}{54}\right)+T_f n_f\left(\frac{\zeta_2}{6}-\frac{14}{27}\right)\bigg]+{\cal{O}}\left(\alpha_s^3\right),
\end{align}
thus verifying eq.~\eqref{MVVid2} through two loops.

The difference of anomalous dimensions appearing on the right-hand
side of eq.~\eqref{MVVid2} has also a geometric interpretation, which
suggests to define it as universal quantity. Following the analysis of
the singularities of the Wilson loops with lightlike lines detailed in
ref.~\cite{Sterman} and the calculation in section~\ref{sec:GammaPiCalc}
above, the anomalous dimensions $\Gamma_\sqcap$ and $\Gamma_\wedge$
receive contributions only from the singular configurations, in which
all the vertices approach one lightlike line. In this sense, these
anomalous dimensions depend only on the features of each lightlike
line separately and they are insensitive to the global shape of the
Wilson loop.  Both $\Gamma_\sqcap$ and $\Gamma_\wedge$ encode the
collinear singularities associated with the two semi-infinite lightlike
lines, but the former receives an additional contribution from the
configurations that are collinear to the finite segment. Such
singularities differ from the ones originating from infinite lines by
the presence of endpoint contributions, as we showed by computing the
diagrams $d^{(2)}_{Y_s}$ and $d^{(2)}_{Y_L}$ in
eqs.~\eqref{eq:d2YsDEC} and~\eqref{eq:d2YL}.  It is therefore useful
to define the difference of $\Gamma_\sqcap$ and $\Gamma_\wedge$ as the
anomalous dimension that captures the collinear singularities of a
finite lightlike segment. Similarly, we define also the collinear
anomalous dimension associated to infinite lines in terms of
$\Gamma_\wedge$ only
\begin{align}
  \label{eq:GammaCoF}
  \Gamma_{\text{co}}^{\text{fin}}&\equiv\Gamma_\sqcap-\Gamma_\wedge\,,\\
  \label{eq:GammaCoI}
  \Gamma_{\text{co}}^{\text{inf}}&\equiv\frac{\Gamma_\wedge}{2}\,.
\end{align}
The two-loop expression of $\Gamma_{\text{co}}^{\text{fin}}$ coincides
with the right-hand side of eq.~\eqref{eq:feik2l}, while
$\Gamma_{\text{co}}^{\text{inf}}$ to the same order is obtained from
eq.~\eqref{GeikSE}. Comparing the two expressions we get
\begin{equation}
 \Gamma_{\text{co}}^{\text{fin}} = 2\,\Gamma_{\text{co}}^{\text{inf}}-\frac{3}{2}\bigg(\frac{\alpha_s}{\pi}\bigg)^2\,C_iC_A\,\zeta_3 + {\cal{O}}\left(\alpha_s^3\right).
\end{equation}
The factor of two multiplying $\Gamma_{\text{co}}^{\text{inf}}$ is
consistent with the fact that the finite Wilson line is obtained as a
contour involving two semi-infinite lines. The remaining discrepancy
proportional to $\zeta_3$ is related to the endpoint contributions in
eq.~\eqref{eq:endpointterms}.

The geometric interpretation of $\Gamma_{\text{co}}^{\text{fin}}$ and
$\Gamma_{\text{co}}^{\text{inf}}$ allows one to derive the anomalous
dimensions of Wilson loops with the contour consisting of arbitrary,
possibly open, polygons with lightlike lines.  The first example is
the parallelogram-shaped Wilson loop $W_\Box$ that features four
lightlike segments of finite length (see figure~\ref{fig:box}), whose renormalisation was given
in~\cite{Korchemsky}
\begin{equation}
  \label{eq:WBoxren}
\mu\frac{d\log\left(W_\Box\right)}{d\mu}=-2\gamma_{\text{cusp}}\left[\log\left(\mu^2(x\cdot y+i\varepsilon)\right)+\log\left(\mu^2(-x\cdot y+i\varepsilon)\right)\right]-\Gamma_\Box,
\end{equation}
where $x$ and $y$ are the four-vectors that define the sides
of the parallelogram. $\Gamma_\Box$ receives contributions from the collinear divergences of four finite segments in lightlike directions, therefore it is
\begin{equation}
  \label{eq:gammaBox}
  \Gamma_\Box=4\,\Gamma_{\text{co}}^{\text{fin}}=4\left(\Gamma_\sqcap-\Gamma_\wedge\right).
\end{equation}
By replacing in the equation above the two-loop value of
$\Gamma_{\text{co}}^{\text{fin}}$ in eq.~\eqref{eq:feik2l}, we
reproduce the results of $\Gamma_\Box$ in ref.~\cite{Korchemsky}. In
the case of a generic polygonal Wilson loop $W_i$ with lightlike lines
the evolution equation in eq.~\eqref{eq:WBoxren} generalises~\cite{Korchemsky,Sterman}
\begin{equation}
  \label{eq:Wilsongenren}
  \mu\frac{d\log\,W_i}{d\mu}=-\sum_a\gamma_{\text{cusp}}\log\left(\mu^2 x_a\cdot x_{a-1}\right) - \Gamma_i,
\end{equation}
where the sum is extended over all the cusps in the contour and
$x_a$, $x_{a-1}$ define the sides adjacent to the cusp
$a$. The anomalous dimension $\Gamma_i$ collects all the collinear
contributions and it can be derived by summing the appropriate
multiples of $\Gamma_{\text{co}}^{\text{fin}}$ and
$\Gamma_{\text{co}}^{\text{inf}}$, given by the number of finite and
infinite sides, respectively.

Finally, having identified the difference
$\Gamma_\sqcap-\Gamma_\wedge=\frac{\Gamma_\Box}{4}$ in
eq.~\eqref{eq:gammaBox}, we may notice that eq.~\eqref{MVVid2} provides yet
another identity relating the form factor, the DGLAP kernel and the
Wilson loop $W_\Box$ computed in ref.~\cite{Korchemsky}, namely
\begin{equation}\label{eq:diffToBox}
  \gamma_G-2B_\delta = \frac{\Gamma_\Box}{4},
\end{equation}
thus explaining the numerical agreement of these two quantities
computed respectively in ref.~\cite{FormFactors} and in
ref.~\cite{Korchemsky}.

\subsection[The Drell-Yan soft function and \texorpdfstring{$\Gamma_\Box$}{Gamma-box} beyond two loops]{\boldmath The Drell-Yan soft function and $\Gamma_\Box$ beyond two loops}
\label{sec:highOrder}

We move to relate the abstract $W_\Box$ to a physical quantity relevant for soft-gluon resummation. It is known that the Drell-Yan \emph{cross-section} factorises near threshold~\cite{Sterman:1986aj,Catani:1989ne,Korchemsky:1993uz,Contopanagos:1996nh,Laenen:2005uz} (see also the more recent literature in the Soft Collinear Effective Theory~\cite{Becher:2007ty,IntroToSCET}). The hard-collinear region is described by the PDFs, the hard function by a squared timelike form factor and the soft region by Wilson lines in the DY configuration~\cite{Belitsky:1998tc}. This leads to the all-order relation $\gamma_G - 2 B_\delta=\Gamma_{\text{DY}}/2$, where $\Gamma_{\text{DY}}$ is the anomalous dimension associated to the DY configuration of Wilson lines (see e.g.~\cite{Becher:2007ty,Laenen:2005uz}). Using eq.~\eqref{eq:diffToBox} we have,
\begin{equation}\label{eq:DYequalsBox}
2\Gamma_{\text{DY}}=\Gamma_\Box .
\end{equation}
The ideas in section~\ref{sec:FFtoDGLAP} will allow us to test the identification in eq.~\eqref{eq:DYequalsBox}. The three-loop value for $\gamma_G-2B_\delta$ was first extracted in~\cite{FormFactors} using the three-loop results for $\gamma_G$ and $B_\delta$. If we expand $\Gamma_{\Box}$ as,
\begin{equation}
\Gamma_{\Box}=\sum_{n=0}^\infty\bigg(\frac{\al_s}{\pi}\bigg)^n\Gamma_{\Box}^{(n)},
\end{equation}
using the values in~\cite{FormFactors} we can then state
\begin{align}
  \label{eq:box1}
\Gamma_\Box^{(1)} =\,& 0\\ \label{eq:box2}
\Gamma_\Box^{(2)} =\,& C_i \bigg[C_A \left(-\frac{11 \zeta_2}{6}-7 \zeta_3+\frac{202}{27}\right)+T_f n_f\left(\frac{2\zeta_2}{3}-\frac{56}{27}\right)\bigg]\\
\Gamma_\Box^{(3)} =\,& C_i\bigg[C_A^2 \left(\frac{22 \zeta_2^2}{5}+\frac{11 \zeta_2 \zeta_3}{3}-\frac{6325 \zeta_2}{648}-\frac{329 \zeta_3}{12}+12 \zeta_5+\frac{136781}{11664}\right)\nonumber\\&\qquad+C_A n_f T_f \left(-\frac{12 \zeta_2^2}{5}+\frac{707 \zeta_2}{162}+\frac{91 \zeta_3}{27}-\frac{5921}{2916}\right)\nonumber\\
&\qquad+C_F n_f T_f \left(\frac{4 \zeta_2^2}{5}+\frac{\zeta_2}{2}+\frac{38 \zeta_3}{9}-\frac{1711}{216}\right)\nonumber\\&\qquad+n_f^2 T_f^2 \left(-\frac{10 \zeta_2}{27}+\frac{28 \zeta_3}{27}-\frac{520}{729}\right)\bigg].\label{eq:box3}
\end{align}
As mentioned in section~\ref{sec:FFtoDGLAP}, the two-loop $\Gamma_\Box^{(2)}$ was calculated explicitly using Wilson lines in~\cite{Korchemsky} and agrees with the extracted value in eq.~\eqref{eq:box2}. The three-loop $\Gamma_\Box^{(3)}$ displayed in eq.~\eqref{eq:box3} should be regarded as a prediction to be verified by direct calculation. For the Drell-Yan configuration of Wilson lines, $\Gamma_{\text{DY}}$ was computed at two loops in~\cite{Belitsky:1998tc} and three loops in~\cite{Li:2014afw}. The three-loop $\Gamma_{\text{DY}}$ coincides with eq.~\eqref{eq:box3}. This is a non-trivial three-loop test of the identification in eq.~\eqref{eq:DYequalsBox}, we arrive at the same value for $\Gamma_\Box^{(3)}$ by two different paths: the difference $\gamma_G-2B_\delta=\Gamma_\sqcap-\Gamma_\wedge=\Gamma_\Box/4$ and the explicit calculation of $\Gamma_{\text{DY}}$.

At four loops the complete picture in QCD for $\Gamma_\Box$ is unknown but in planar $\mathcal{N}=4$ super Yang-Mills the four-loop result for the difference $\gamma_G - 2 B_\delta$ was found in~\cite{Dixon:2017nat}. We identify this as $\Gamma_\Box^{\text{planar}\,\,\mathcal{N}=4}$ and quote the result here,
\begin{align}\label{eq:boxPlanar}
\Gamma_\Box^{\text{planar}\,\,\mathcal{N}=4}=&-\bigg(\frac{\al_s}{\pi}C_A\bigg)^27\zeta_3+\bigg(\frac{\al_s}{\pi}C_A\bigg)^3\left(12\zeta_5+\frac{11}{3}\zeta_2\zeta_3\right)\nonumber\\&-\bigg(\frac{\al_s}{\pi}C_A\bigg)^4\left(\frac{425}{16}\zeta_7+\frac{13}{2}\zeta_2\zeta_5+\frac{45}{4}\zeta_3\zeta_4\right)+\mathcal{O}(\al_s^5)
\end{align}
It is well-known that to reach the above result one can simply take the QCD result and take the limit $N_c\to\infty$ and the maximal transcendental weight term at each order in $\al_s$. We can do this at two and three loops by looking at eqs.~\eqref{eq:feik2l} and \eqref{eq:box3} respectively.

Above three loops, strict Casimir scaling has been proven to fail~\cite{Grozin:2017css,Boels:2017skl,Moch:2017uml}. As such, we need to distinguish between quarks and gluons or rather particles in the fundamental and adjoint representation. We focus on the quark case. To compute $\Gamma_\Box^{(4),q}$ we need $B_\delta^q$ and $\gamma_G^q$ at four loops. The state of the art is that some colour structures are known for both $B_\delta^q$~\cite{Moch:2017uml,Davies:2016jie} and $\gamma_G^q$~\cite{Henn:2016men} in the planar limit, $N_c\to\infty$. Using the values in~\cite{Moch:2017uml,Henn:2016men} we can extract the following terms in that limit for $\Gamma_\Box^{(4),q}$,
\begin{align}
  \Gamma_\Box^{(4),q}|_{N_c^3 n_f}=&-\frac{247315}{55296}+\frac{51529 \zeta_2}{11664}+\frac{102205 \zeta_3}{31104}-\frac{7589 \zeta_4}{768}\nonumber\\&-\frac{185 \zeta_5}{288}-\frac{103 \zeta_2 \zeta_3}{144}+\frac{15611 \zeta_6}{3456}+\frac{22 \zeta_3^2}{9}\label{fourLoopnfTerm}\\
  \Gamma_\Box^{(4),q}|_{N_c^2 n_f^2}=&\,\, \frac{329069}{2239488}-\frac{22447 \zeta_2}{93312}+\frac{6325 \zeta_3}{7776}+\frac{35 \zeta_4}{96}-\frac{107 \zeta_5}{144}-\frac{11 \zeta_2 \zeta_3}{72}\label{fourLoopnf2term}\\
  \Gamma_\Box^{(4),q}|_{N_c n_f^3}=&-\frac{505}{26244}-\frac{\zeta_2}{648}-\frac{25 \zeta_3}{1944}+\frac{\zeta_4}{27}\label{fourLoopnf3term}
\end{align}
where we have used $T_f=\frac{1}{2}$. We are unable to deduce the $N_c^4n_f^0$ term as it is unknown for $\gamma_G$ but it is known for $B_\delta$~\cite{Moch:2017uml}. In the planar limit $N_c\to\infty$ the quartic Casimirs $d_{FF}^{(4)}\equiv\left(d_F^{abcd}\right)^2$ contribute to the colour factor in eq.~\eqref{fourLoopnfTerm} since,
\begin{align}
  n_f\frac{d_{FF}^{(4)}}{N_F} &= n_fT_f^4\left(\frac{N_c^3}{6}-\frac{7 N_c}{6}+\frac{4}{N_c}-\frac{3}{N_c^3}\right).
\end{align}
It means that we are unable to fully construct the Casimir scaling of $N_c^3n_f$. The full (planar and non-planar) contribution of the quartic Casimir colour factor $d_{FF}^{(4)}$ to $\gamma_G$ is known~\cite{Lee:2019zop} but not to $B_\delta$. Only the low-$N$ values of the splitting functions or $\gamma_{\text{cusp}}$ is known~\cite{Moch:2018wjh,Bruser:2019auj}. In~\cite{Moch:2018wjh} it was also found that, within numerical errors, quartic Casimir contribution to the cusp anomalous dimension did not depend on the representation, i.e.\ it is the same for gluons and quarks. It was conjectured that although Casimir scaling is violated there is a generalised version where quartic factors are simply exchanged depending on gluon or~quarks,
\begin{align}
  \text{quarks} &\leftrightarrow \text{gluons}\nonumber\\[1mm]
  \frac{d_{FA}^{(4)}}{N_F} &\leftrightarrow \frac{d_{AA}^{(4)}}{N_A}\\
  n_f\frac{d_{FF}^{(4)}}{N_F} &\leftrightarrow n_f\frac{d_{FA}^{(4)}}{N_A}\nonumber
\end{align}
where $N_{F/A}$ denote the dimensions of the corresponding representations, namely $N_F=N_c=C_A$
and $N_A=N_c^2-1=2N_c C_F$.
The relation in eq.~\eqref{MVVid2} may be used as an interesting test for a generalised Casimir scaling extension to the anomalous dimension $\Gamma_\Box$.

However, the quartic Casimirs do not appear in the $n_f^2$ or $n_f^3$ terms of eqs.~\eqref{fourLoopnf2term} and~\eqref{fourLoopnf3term}. We are then able to use these terms for a leading-$N_c$, Casimir-scaling part of $\Gamma_\Box^{(4)}$. We put these terms together with the conjectured generalised scaling to create an ansatz for $\Gamma_\Box^{(4)}$,
\begin{align}
  \Gamma_\Box^{(4)}=& \,\,C_i\bigg[n_f^3\left(-\frac{505}{13122}-\frac{\zeta_2}{324}-\frac{25 \zeta_3}{972}+\frac{2\zeta_4}{27}\right)\nonumber\\ & \,\,\,\,\,\,\,\,\,\,                                                                                                                                                      +N_cn_f^2\left(\frac{329069}{1119744}-\frac{22447 \zeta_2}{46656}+\frac{6325 \zeta_3}{3888}+\frac{35 \zeta_4}{48}-\frac{107 \zeta_5}{72}-\frac{11 \zeta_2 \zeta_3}{36}\right)+\cdots\bigg] \nonumber\\&\,\,+n_f\frac{d_{Fi}^{(4)}}{N_i}\Gamma_\Box^{(4),dFi}+\frac{d_{Ai}^{(4)}}{N_i}\Gamma_\Box^{(4),dAi},
\end{align}
where the ellipsis represents all terms subleading in $N_c$, including the $n_f^1$ and $n_f^0$ terms, which are not found from the quartic Casimirs when they are expanded in $N_c$.


\section{Conclusions}
\label{sec:conc}

We have presented a detailed study of the infrared factorisation of form factors and PDFs at large $x$ using a common formalism.
By identifying the universal contributions from the hard-collinear region in both quantities, those controlled by the anomalous dimension $\gamma_{J/\eikJ}$, we were able to derive the relation in eq.~\eqref{MVVid2},
\begin{equation}\label{eq:bigRel}
{  \gamma_G - 2B_\delta = \Gamma_\sqcap - \Gamma_\wedge= \frac{\Gamma_\Box}{4}\,.}
\end{equation}
That is, the difference between anomalous dimension describing single poles in the on-shell form factor of quarks (gluons) and that associated with the $\delta(1-x)$ term in the large-$x$ limit of the quark (gluon) diagonal DGLAP splitting function, reduces to a difference between corresponding eikonal quantities, $\Gamma_\wedge$ and $\Gamma_\sqcap$ defined directly in terms of Wilson loops.
Furthermore, based on the configuration-space origin of the contributions to these two eikonal quantities we concluded that their difference simply corresponds to the anomalous dimension associated with a closed polygonal Wilson loop, such as the parallelogram analysed first in ref.~\cite{Korchemsky}. The contributions of the semi-infinite Wilson lines in $W_\sqcap$ and $W_\wedge$ cancel in the difference.
We emphasise that while each of the quantities on the left-hand side of eq.~(\ref{eq:bigRel}) depends in a non-trivial way on the spin of the partons, in addition to their colour representations, yielding very different results for quarks and for gluons, the eikonal quantities, by definition, depend only on the colour representation of these partons, and in particular admit Casimir scaling through three loops.
We stress that the relation in eq.~(\ref{eq:bigRel}) is expected to hold to all orders in perturbation theory.
An obvious next step is to compute $\Gamma_\Box$ to three loops in order to check it explicitly to this order.

In establishing the relation between $\Gamma_{\Box}$ and $\Gamma_{\sqcap}-\Gamma_{\wedge}$ we used the fact that singularities arise only from configurations where all the vertices approach a cusp or one where they all approach a particular lightlike segment~\cite{Sterman}. This underlies the cancellation of the two infinite segments, isolating a remaining finite segment. The very same logic may be applied to other, more complicated
 Wilson-line geometries involving both finite and semi-infinite lightlike segments. Specifically, the double pole is always governed by $\gamma_{\text{cusp}}$ while the single-pole anomalous dimension is written as a sum of terms, building blocks, each corresponding to either a finite or semi-infinite segment, which contribute
$\Gamma_{\text{co}}^{\text{fin}}$ and $\Gamma_{\text{co}}^{\text{inf}}$,
respectively.
An example of such a construction with only finite segments can be found in refs.~\cite{Drummond:2007cf,Drummond:2007au,Drummond:2008aq}, where polygons of up to six sides were computed to two loops.
Following our discussion in section~\ref{sec:FFtoDGLAP} it may be interesting to explicitly compute other Wilson-line configurations involving both finite and infinite segments. A simple example of direct relevance to physics is the non-forward amplitude, generalising the $\sqcap$ configuration.

One interesting aspect that we have encountered is that $W_\sqcap$
behaves very differently in the ultraviolet as compared to the infrared, as can be seen explicitly in eq.~(\ref{Wpi}). In the ultraviolet, one encounters a double logarithmic dependence on the scale $\mu^2$, originating from the cusp singularity, while in the infrared there is just a single pole. This stands in sharp contrast to the  $W_\wedge$, corresponding to the soft function of the form factor~(\ref{softFint}) (or more generally, in soft functions corresponding to multi-leg amplitudes) where the infrared behaviour entails a double pole, mirroring the ultraviolet. The absence of any distance scale in the relevant Wilson-line contour implies such mirroring. Indeed the symmetry between the ultraviolet and the infrared is broken in $W_\sqcap$ due to the presence of the scale $\beta\cdot y$.   The single-pole character of $W_\sqcap$ can be seen as intermediate in comparing $W_\Box$, which, lacking infinite rays, is infrared finite, to $W_\wedge$, which is double logarithmic.

The relation in eq.~(\ref{eq:DYequalsBox}) between the soft anomalous dimension in Drell-Yan production and the parallelogram  $W_\Box$ is interesting in its own right. The Drell-Yan soft function involves real gluon emission diagrams where the propagators connecting the amplitude side to the complex-conjugate one are cut, while in $W_\Box$ there are no cut propagators. A possible way to explain\footnote{We would like to thank Gregory Korchemsky for proposing this explanation.} this is to recall that a parallelogram made of four lightlike segments features two cusps where the exchanged gluons span timelike distances and two others where gluons span spacelike distances. The latter correspond to diagrams that feature in \emph{virtual} corrections to the Drell-Yan process (these propagators are not cut). In turn, the former are naturally time-ordered, because there path-ordering coincides with time-ordering (just as in the case of the $W_\sqcap$, discussed below eq.~(\ref{WpiDef})) and could be computed using either cut propagators or ordinary ones, giving the same answer. This way the calculation of the parallelogram can be mapped into that of the Drell-Yan soft function.
It would be interesting to turn this argument into a proof relating the two Wilson line configurations directly. It would also be interesting to explore in this context the conformal mapping techniques of refs.~\cite{Vladimirov:2017ksc,Vladimirov:2016dll}.

Another interesting direction to explore is the connection between partonic amplitudes in the Regge limit and anomalous dimensions of Wilson lines.
In particular, one would like to derive the relation between the Regge trajectory and $W_{\wedge}$ in eq.~(\ref{Regge}) and understand its generalisation to higher orders.


\acknowledgments

We would like to thank Andreas Vogt for a very useful discussion in March 2018 providing the impetus to revisit this topic; Lorenzo Magnea, Lance Dixon, and Vittorio Del Duca for a useful exchange early on; George Sterman for insightful discussions and encouragement, and Gregory Korchemsky for illuminating discussions at critical moments along the way.
GF's and EG's research is supported by the STFC Consolidated Grant `Particle Physics at the Higgs Centre'; CM's research is supported by a PhD scholarship from the Carnegie Trust for the Universities of Scotland. 
EG would like to thank the GGI and the Simons Foundation for an extended visit to the GGI, Florence in Autumn 2018, and the CERN theory department for hospitality as a Scientific Associate in 2019.
\appendix

\section{Direct calculation of the splitting functions at large \texorpdfstring{\boldmath $x$}{x}}
\label{subsec:pdf}

In this appendix we present a calculation for parton distribution splitting functions directly using the definitions \eqref{quarkPDF} and~\eqref{gluonPDF}. As explained in the main text we take incoming partons to be off shell $p^2\neq 0$ but with zero transverse momentum $p=(p_+,\frac{p^2}{2p_+},\mathbf{0}_{d-2})$. This regulates the infrared such that we are only exposed to UV poles.

To calculate a single diagram there is a general strategy talked through in the main text with a slight change for the off shell case:
\begin{itemize}
\item Write down the integral using Feynman rules
\item Integrate over the minus component of all loop momenta using Cauchy's residue theorem. This provides constraints on the plus component of loop momenta due to the location of the poles.
\item Integrate over the transverse component of all loop momenta. As we are only interested in the UV divergent terms, this can be simplified to just calculating iterated bubbles at two loops. Although we do need to calculate the finite terms of one loop graphs to perform the renormalisation.
\item Rescale the plus component to arrive at a general form such as,
  \begin{align}
    \text{Disc}\int_0^1dy dz \frac{N(x,y,z,\eps)}{(1-x+i\varepsilon)(1-x-y+i\varepsilon)(1-x-y z+i\varepsilon)}.
  \end{align}
  The denominators correspond to the Wilson line propagators.
\item Now we take the discontinuity in $x$ and perform the final integrations. Often these integrals evaluate to $_2F_1$ functions at two loops.
\item Finally we expand in $\eps$ using,
  \begin{align}
    (1-x)^{-1-m\eps} = -\frac{1}{n\eps}\delta(1-x) + \frac{1}{(1-x)_+} + \sum_{n=1}^\infty \frac{(-m\eps)^n}{n!}\left(\frac{\log(1-x)^n}{1-x}\right)_+.
  \end{align}
\end{itemize}
 For brevity of results we shall define,
\begin{align}
 \delta=\delta(1-x) && P = \frac{1}{(1-x)_+} && L^n = \frac{\log^n(1-x)}{(1-x)_+}.
\end{align}
Also, all the following expressions are valid up to but not including terms that diverge as slowly as $\log(1-x)$.

For PDFs we expand in powers of $\left(\frac{\al_s}{4\pi}\right)$,
\begin{equation}
f_{ii}=\sum_{n=0}^\infty\bigg(\frac{\al_s}{4\pi}\bigg)^nf_{ii}^{(n)}.
\end{equation}

\paragraph{Example.} Let us illustrate the above steps. For this we choose the two loop diagram in figure~\ref{qq5}. The Feynman rules for the diagram, in Feynman gauge, give,
\begin{align}
\text{Disc}\frac{ig_s^4}{2\pi}C_AC_F\int\frac{d^dq_1}{(2\pi)^d}\int\frac{d^dq_2}{(2\pi)^d}&\frac{p_+(p_+-q_{1+})(2q_{2+}-q_{1+})}{q_2^2q_1^2(q_1-q_2)^2(p-q_1)^2}\nonumber\\ &\times\frac{1}{(p-k)\cdot u\,\,\, (p-k-q_1)\cdot u\,\,\, (p-k-q_2)\cdot u},
\end{align}
where $k\cdot u=xp_+$ and the $+i\varepsilon$ prescription is implied. It is reminded that the Wilson line direction $u$ is in the (-) direction, $u=(0,1,\mathbf{0}_{d-2})$.

We shall define $f_{qq}^{(2),(e)}$ to be the contribution to $f_{qq}^{(2)}$ of diagram~\ref{qq5}. When integrating the $q_{1-}$ and $q_{2-}$ components using the residue theorem, constraints are placed on the plus~momenta,
\begin{equation}
p_+>q_{1+}>q_{2+}>0.
\end{equation}
The integrations over the transverse components are just iterated bubbles. Rescaling the plus component we arrive at,
\begin{align}
f_{qq}^{(2),(e)}=\text{Disc}\frac{i}{2\pi}C_AC_F\frac{\Gamma(\eps)\Gamma(2\eps)}{\Gamma(1+\eps)}\int_0^1&dydz y^{1-2\eps}(1-y)^{1-\eps}(1-z)^{-\eps}z^{-\eps}\nonumber\\&\times\frac{1-2z}{(1-x+i\varepsilon)(1-x-y+i\varepsilon)(1-x-yz+i\varepsilon)}.
\end{align}
Taking the discontinuity we use,
\begin{align}
\text{Disc}&\frac{1}{(1-x+i\varepsilon)(1-x-y+i\varepsilon)(1-x-yz+i\varepsilon)}\nonumber\\&=(-2\pi i)\left(\frac{\delta(1-x)}{(1-x-y)(1-x-yz)}+\frac{\delta(1-x-y)}{(1-x)(1-x-yz)}+\frac{\delta(1-x-yz)}{(1-x)(1-x-y)}\right).
\end{align}
There are three separate terms now to calculate. The first is the \emph{virtual} cut and evaluates~to,
\begin{equation}
f_{qq}^{(2),(e),(V)}=C_AC_F\left(\frac{1}{4 \eps^4}+\frac{1}{\eps^3}+\frac{1}{\eps^2}\left(\frac{7}{2}-\frac{\zeta_2}{4}\right)+\frac{1}{\eps}\left(-\zeta_2-\frac{8 \zeta_3}{3}+\frac{23}{2}\right)\right)\delta
\end{equation}
The second and third are \emph{real} cuts,
\begin{align}
  f_{qq}^{(2),(e),(R1)}&=C_AC_F\bigg(-\frac{\delta}{4 \eps^4}+\frac{P-\delta}{2\eps^3}+\frac{1}{\eps^2}\left(\left(-\frac{\zeta_2}{4}-1\right) \delta+P-L\right)\nonumber\\&\,\,\,\,\,\,+\frac{1}{\eps}\left(\delta \left(-\frac{\zeta_2}{2}+\frac{7 \zeta_3}{6}-2\right)+\left(\frac{\zeta_2}{2}+2\right) P-2 L+L^2\right)\bigg)\label{example:R1}\\
  f_{qq}^{(2),(e),(R2)}&=C_AC_F\bigg(\frac{-\delta-P}{2\eps^3}+\frac{1}{\eps^2}\left((\zeta_2-1) \delta+\frac{1}{2} P+\frac{3}{2} L\right)\nonumber\\&\,\,\,\,\,\,+\frac{1}{\eps}\left(\delta \left(-\frac{\zeta_2}{2}+\zeta_3-2\right)+\left(1-\frac{5 \zeta_2}{2}\right) P-\frac{1}{2} L-\frac{7}{4} L^2\right)\bigg)\label{example:R2}
\end{align}
In the sum we find,
\begin{align}
f_{qq}^{(2),(e)}=C_A C_F\bigg(&\frac{1}{\eps^2}\left(\left(\frac{\zeta_2}{2}+\frac{3}{2}\right) \delta+\frac{3}{2} P+\frac{1}{2} L\right)\nonumber\\&+\frac{1}{\eps}\left(\delta \left(-2 \zeta_2-\frac{\zeta_3}{2}+\frac{15}{2}\right)+(3-2 \zeta_2) P-\frac{5}{2} L-\frac{3}{4} L^2\right)\bigg)
\end{align}
Above we see a salient feature of two loop diagrams: individual cuts are $\eps^{-4}$ and $\eps^{-3}$ divergent. These are poles from when the emitted gluon goes soft and cancel in the sum of real and virtual cuts. The remaining divergences are UV, whose renormalisation gives the splitting functions. The $L$ and $L^2$ terms are present in individual diagrams but cancel in the combination such that the splitting functions diverge as in eq.~\eqref{splittingFunctionDiverge}.

Another feature is that we found that the real cuts, eqs.~\eqref{example:R1} and~\eqref{example:R2}, contribute to $B_\delta$. Rather than inferring the coefficient of $\delta$ from sum or momentum conservation rules, we are able to state that for the off-shell extraction of the splitting functions, real cuts contribute to $\delta(1-x)$.

We now calculate the two loop diagonal splitting functions at large $x$ for quarks \hbox{and~gluons.}

\subsection[Calculating \texorpdfstring{$P_{qq}$}{P(qq)}]{\boldmath Calculating $P_{qq}$}
\label{subsec:quark}

\newcommand\spit{\hspace{1cm}}

\begin{figure}
\centerline{  \begin{subfigure}[b]{0.3\textwidth}
            \caption{1}
        \includegraphics[width=.97\textwidth]{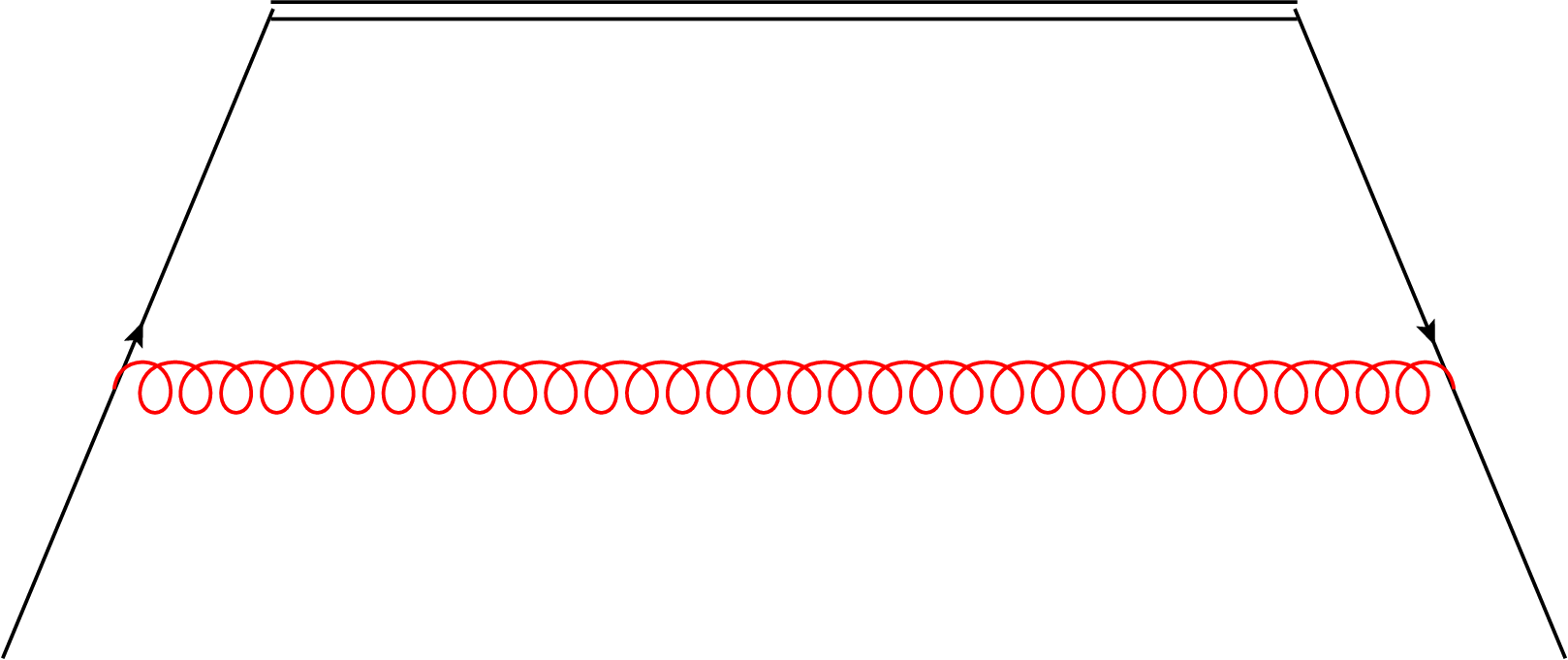}
        \label{qq1}
    \end{subfigure}\quad
    \begin{subfigure}[b]{0.3\textwidth}
      \caption{2}
        \includegraphics[width=.97\textwidth]{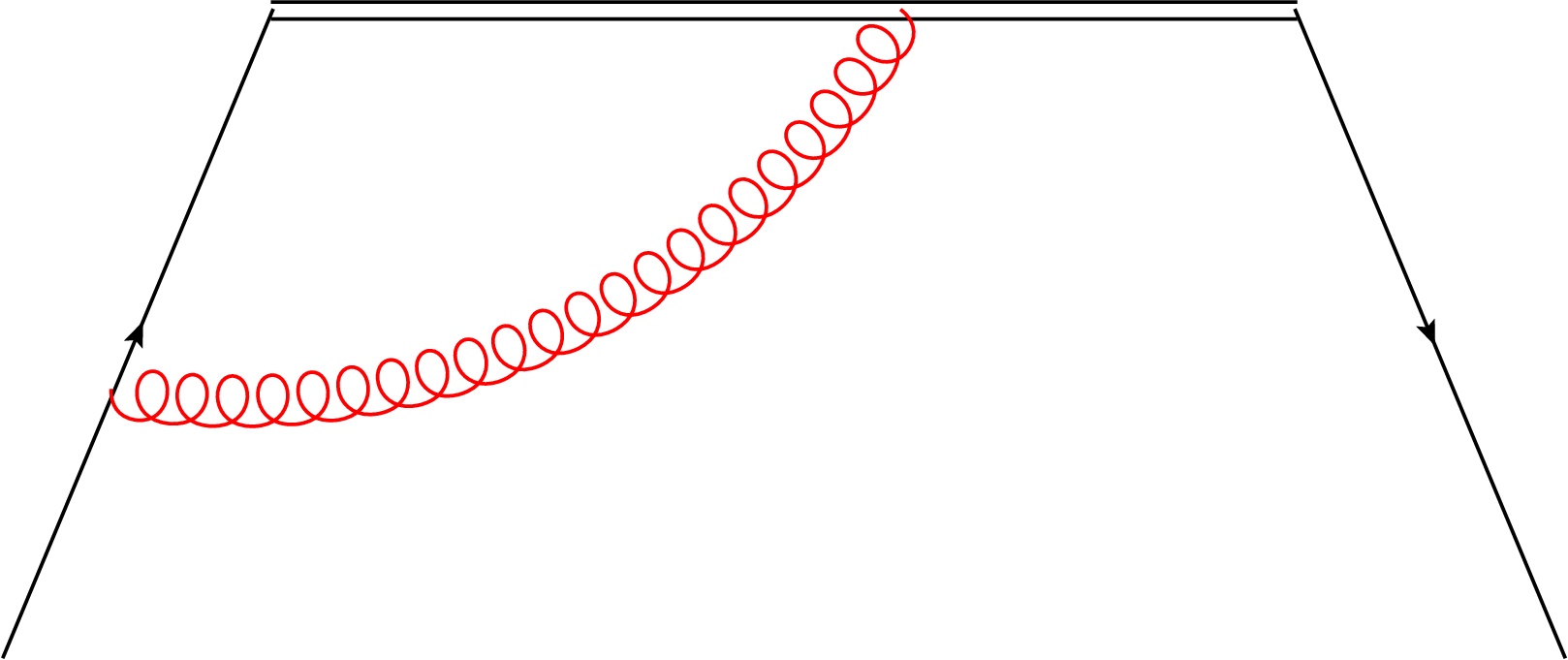}
        \label{qq2}
    \end{subfigure}\quad
    \begin{subfigure}[b]{0.3\textwidth}
      \caption{1}
        \includegraphics[width=.97\textwidth]{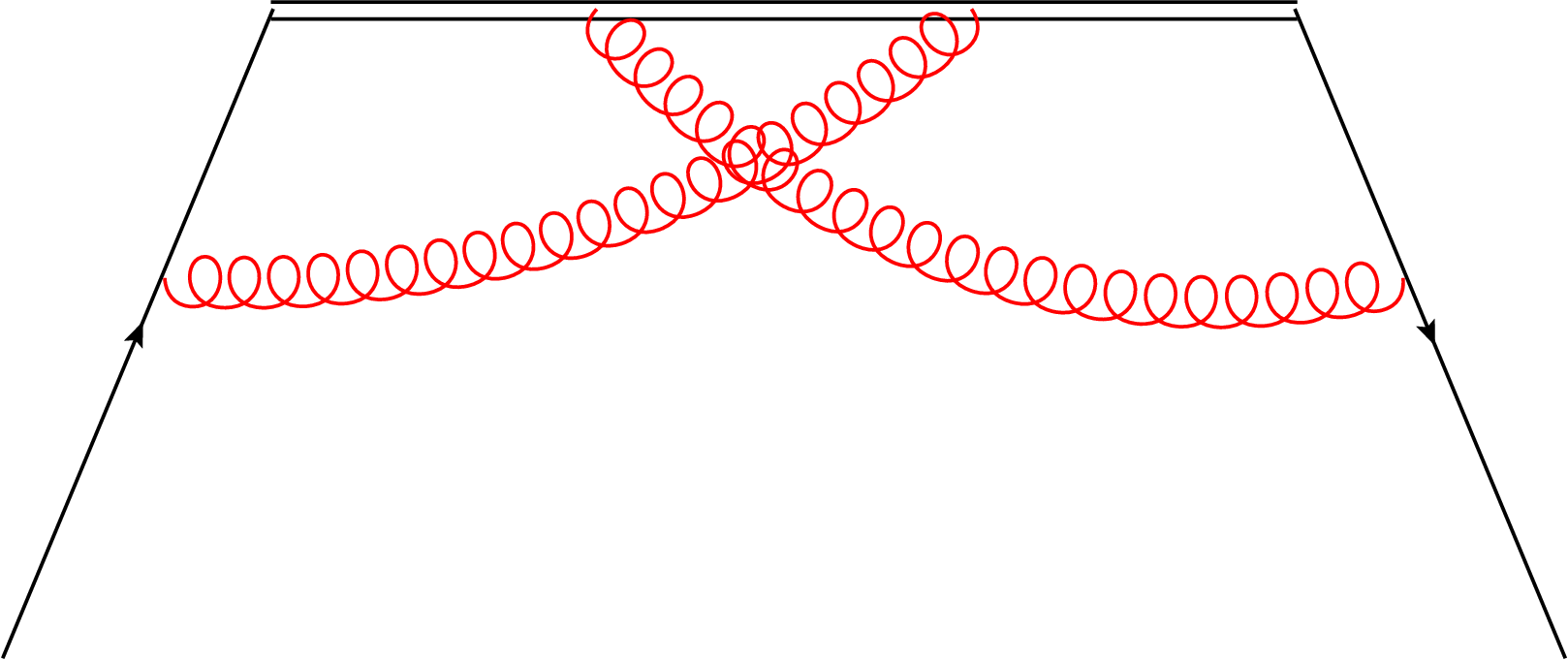}
        \label{qq3}
      \end{subfigure}
}
\vspace{2mm}
\centerline{
      \begin{subfigure}[b]{0.3\textwidth}
        \caption{1}
        \includegraphics[width=.97\textwidth]{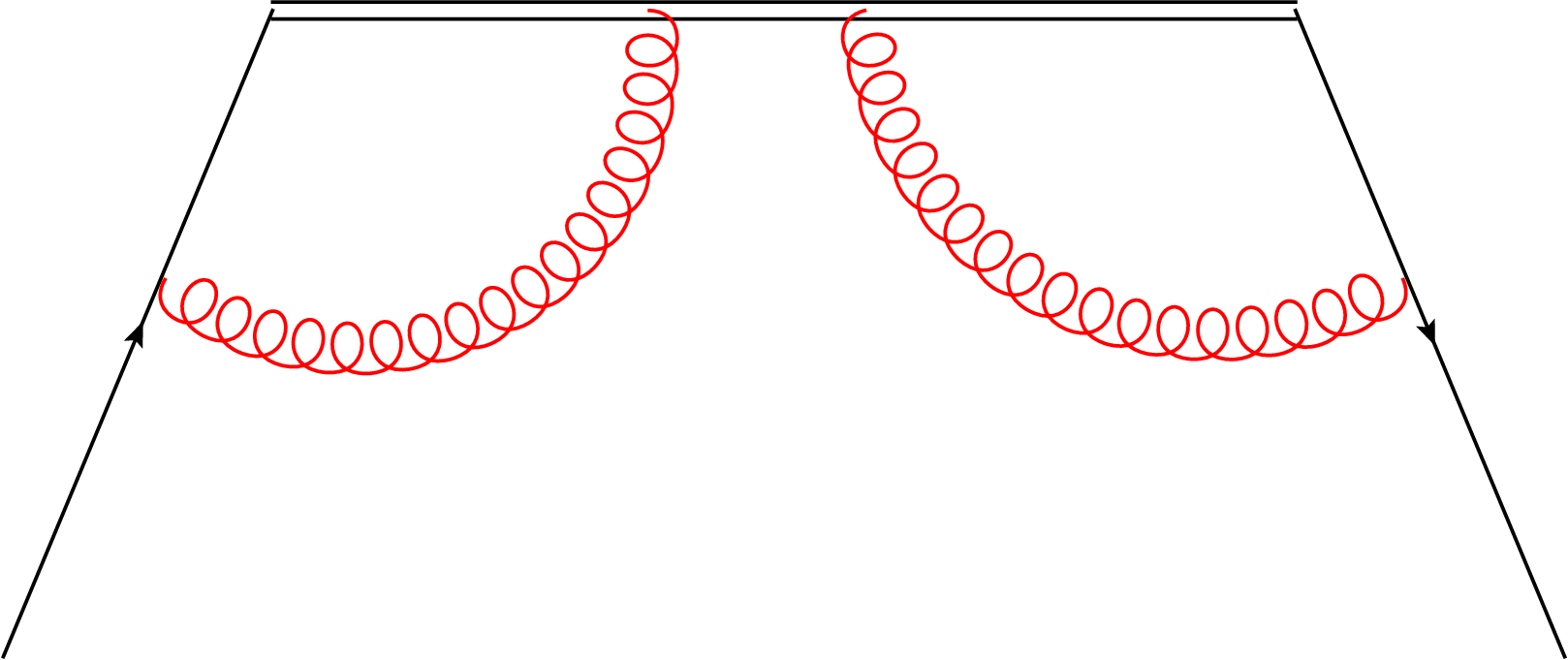}
        \label{qq4}
    \end{subfigure}\quad
    \begin{subfigure}[b]{0.3\textwidth}
      \caption{2}
        \includegraphics[width=.97\textwidth]{appendix/pdf_figs/quark_two_loop_12.ps}
        \label{qq5}
    \end{subfigure}\quad
    \begin{subfigure}[b]{0.3\textwidth}
      \caption{2}
        \includegraphics[width=.97\textwidth]{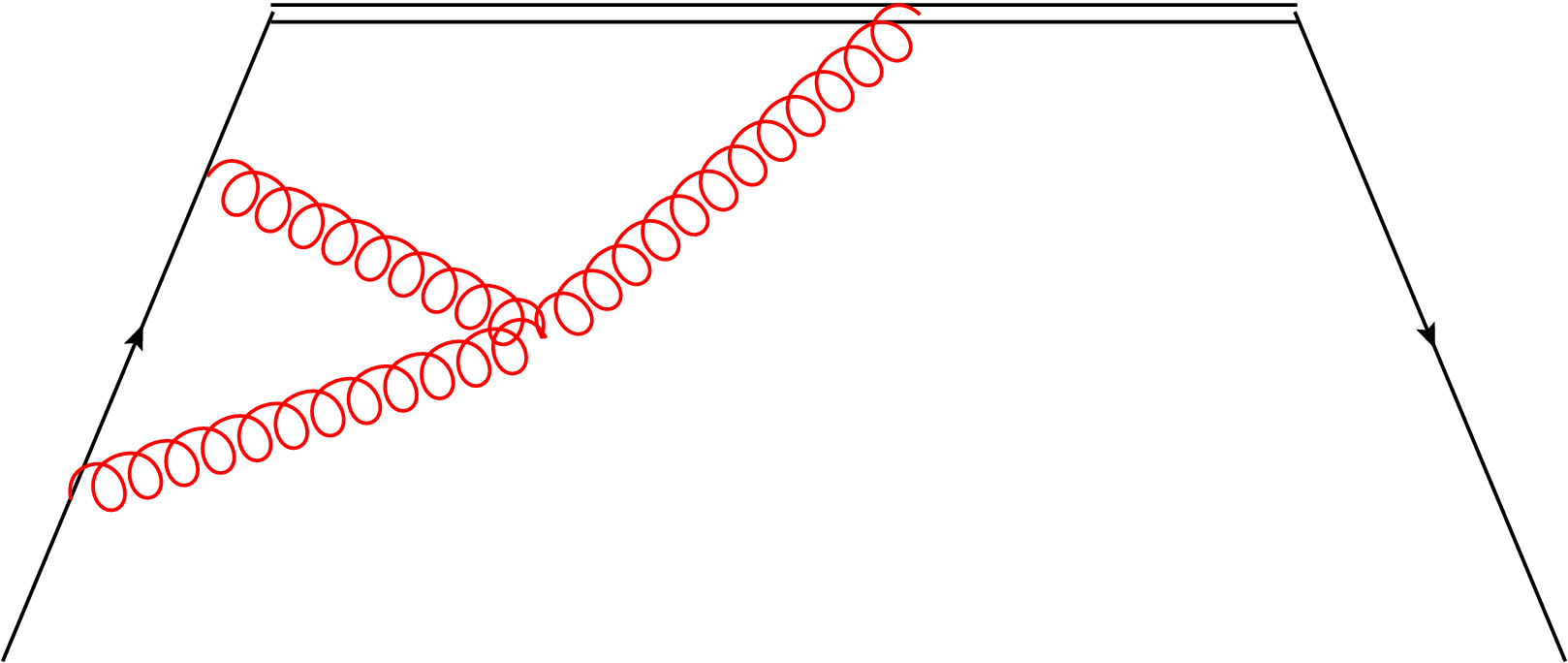}
        \label{qq6}
      \end{subfigure}
}
\vspace{2mm}
\centerline{
      \begin{subfigure}[b]{0.3\textwidth}
        \caption{2}
        \includegraphics[width=.97\textwidth]{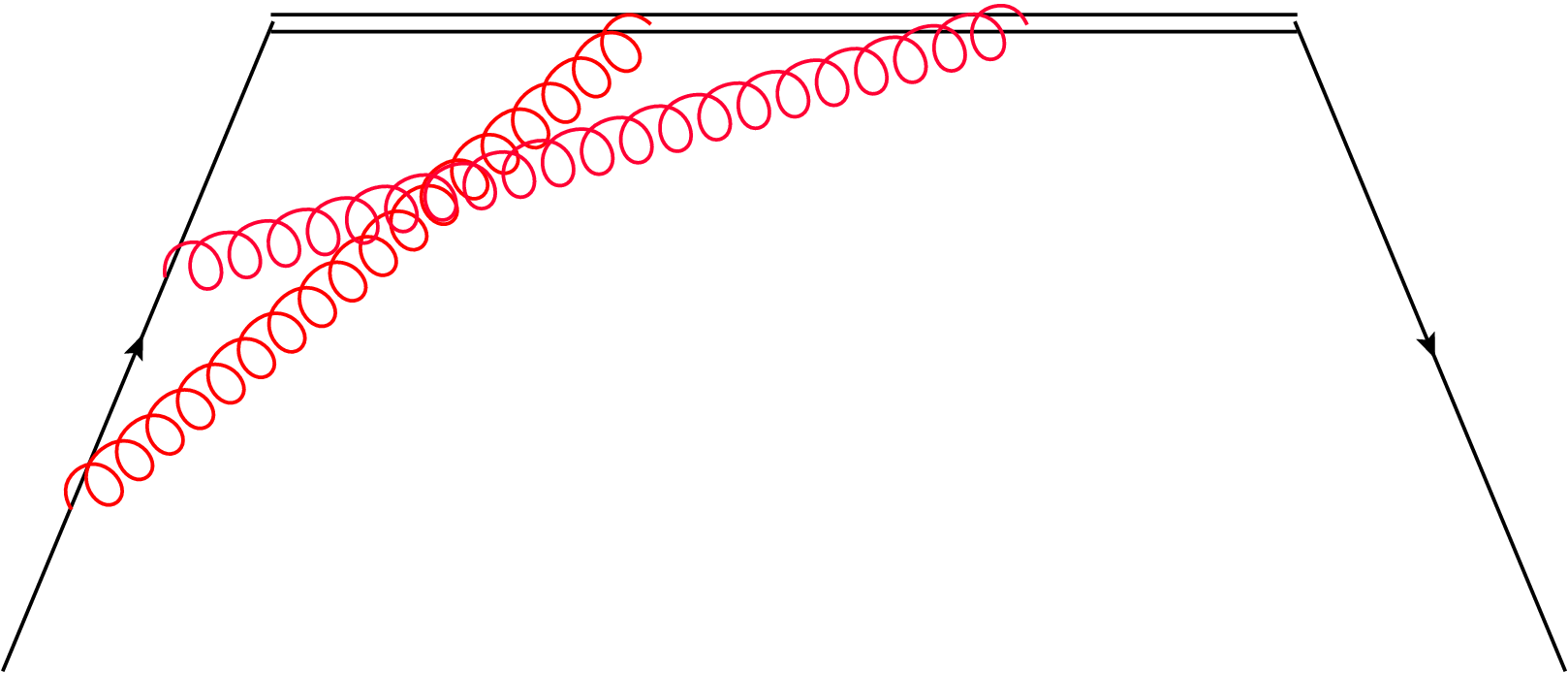}
        \label{qq7}
    \end{subfigure}\quad
    \begin{subfigure}[b]{0.3\textwidth}
      \caption{2}
        \includegraphics[width=.97\textwidth]{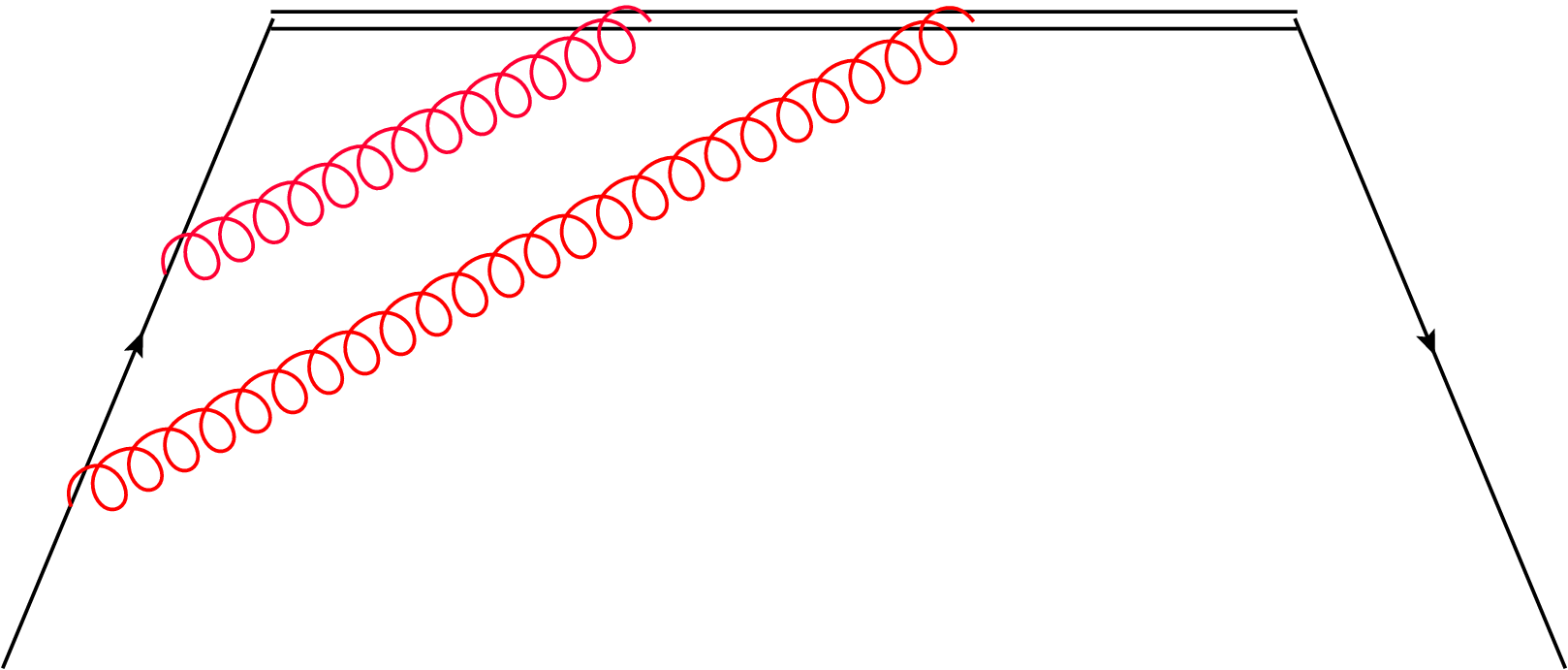}
        \label{qq8}
        \end{subfigure}\quad
        \begin{subfigure}[b]{0.3\textwidth}
          \caption{2}
        \includegraphics[width=.97\textwidth]{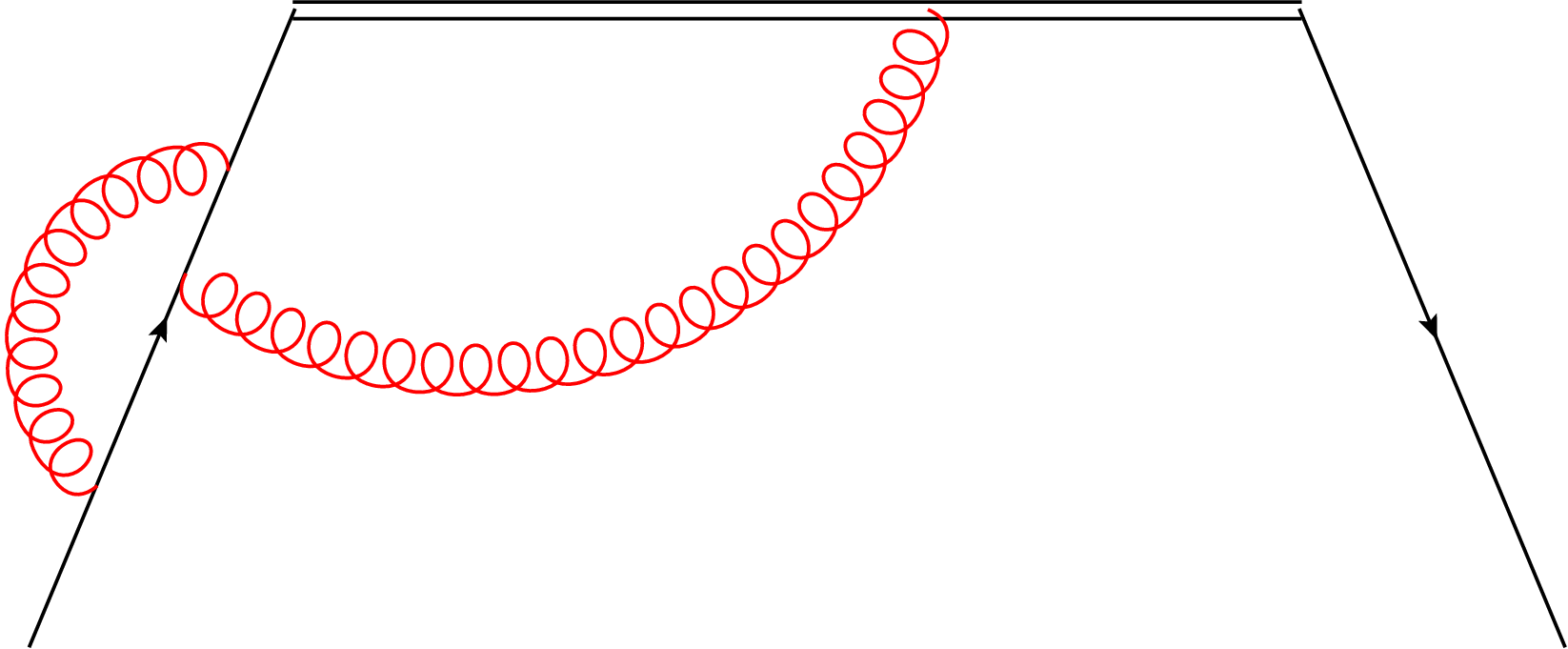}
        \label{qq9}
      \end{subfigure}
}
\vspace{2mm}
\centerline{
      \begin{subfigure}[b]{0.3\textwidth}
        \caption{1}
        \includegraphics[width=.97\textwidth]{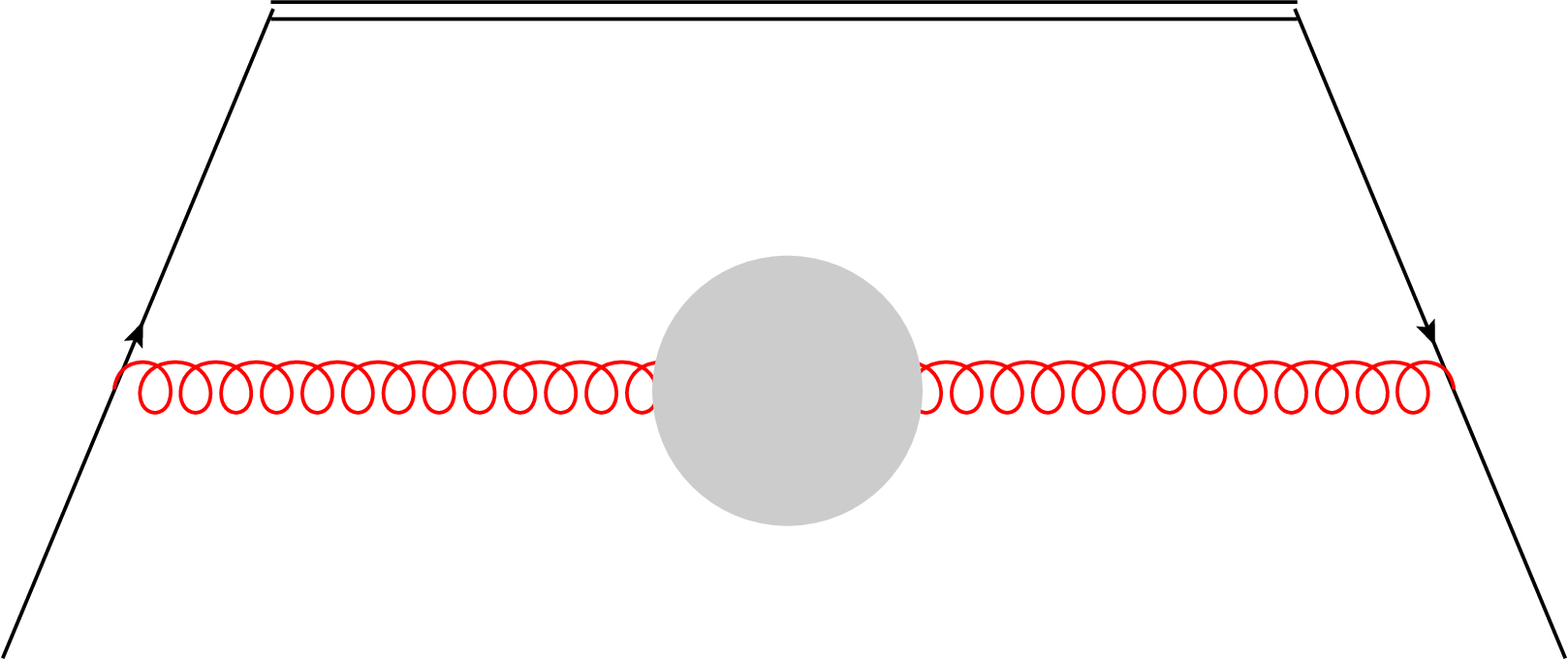}
        \label{qq10}
    \end{subfigure}\quad
    \begin{subfigure}[b]{0.3\textwidth}
      \caption{1}
        \includegraphics[width=.97\textwidth]{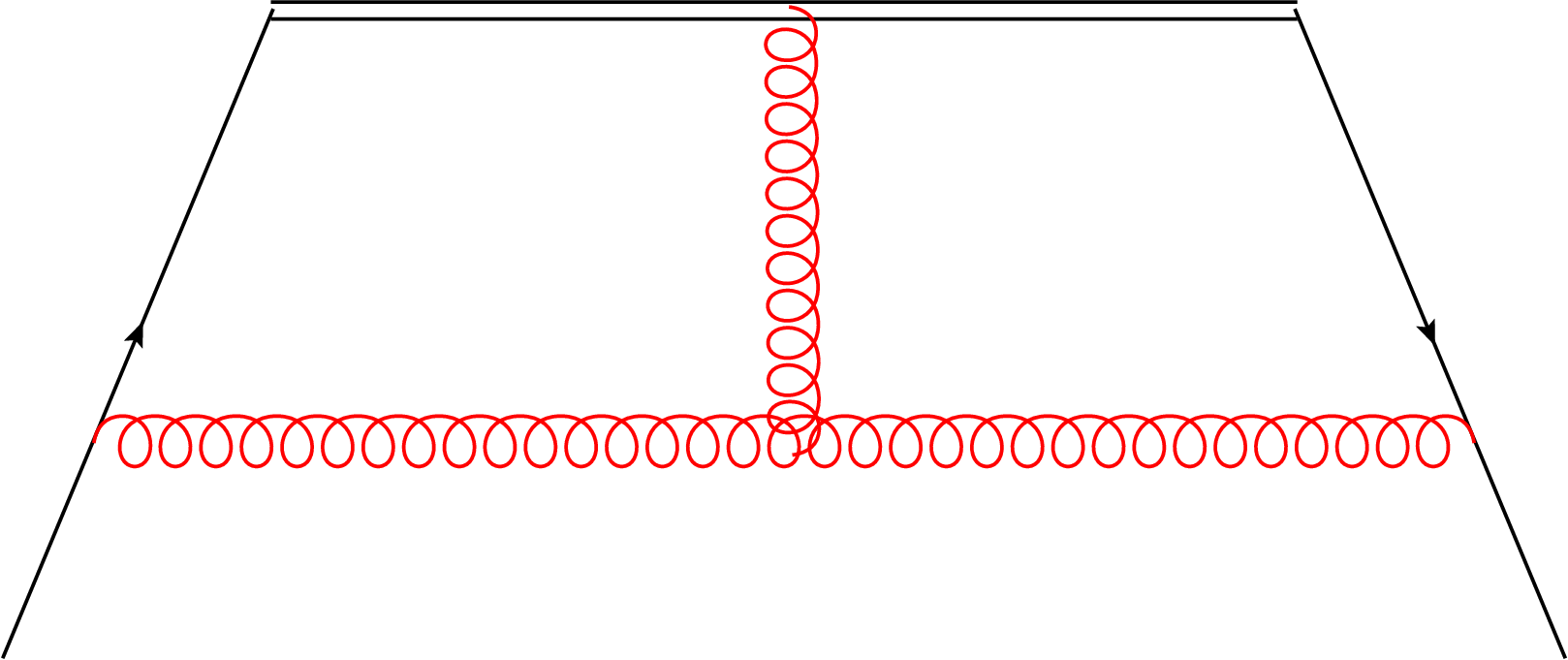}
        \label{qq11}
    \end{subfigure}\quad
    \begin{subfigure}[b]{0.3\textwidth}
      \caption{2}
        \includegraphics[width=.97\textwidth]{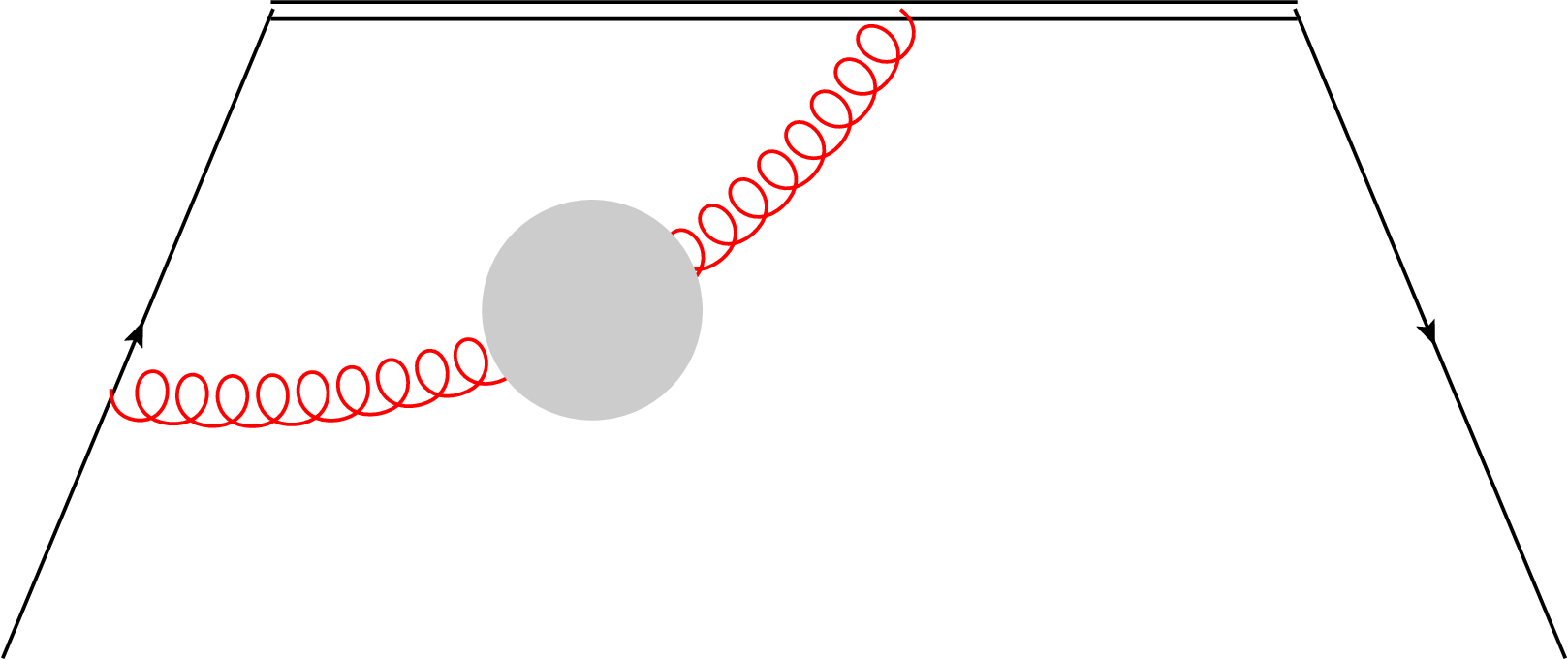}
        \label{qq12}
      \end{subfigure}
}
\vspace{2mm}
\centerline{
      \begin{subfigure}[b]{0.3\textwidth}
        \caption{2}
        \includegraphics[width=.97\textwidth]{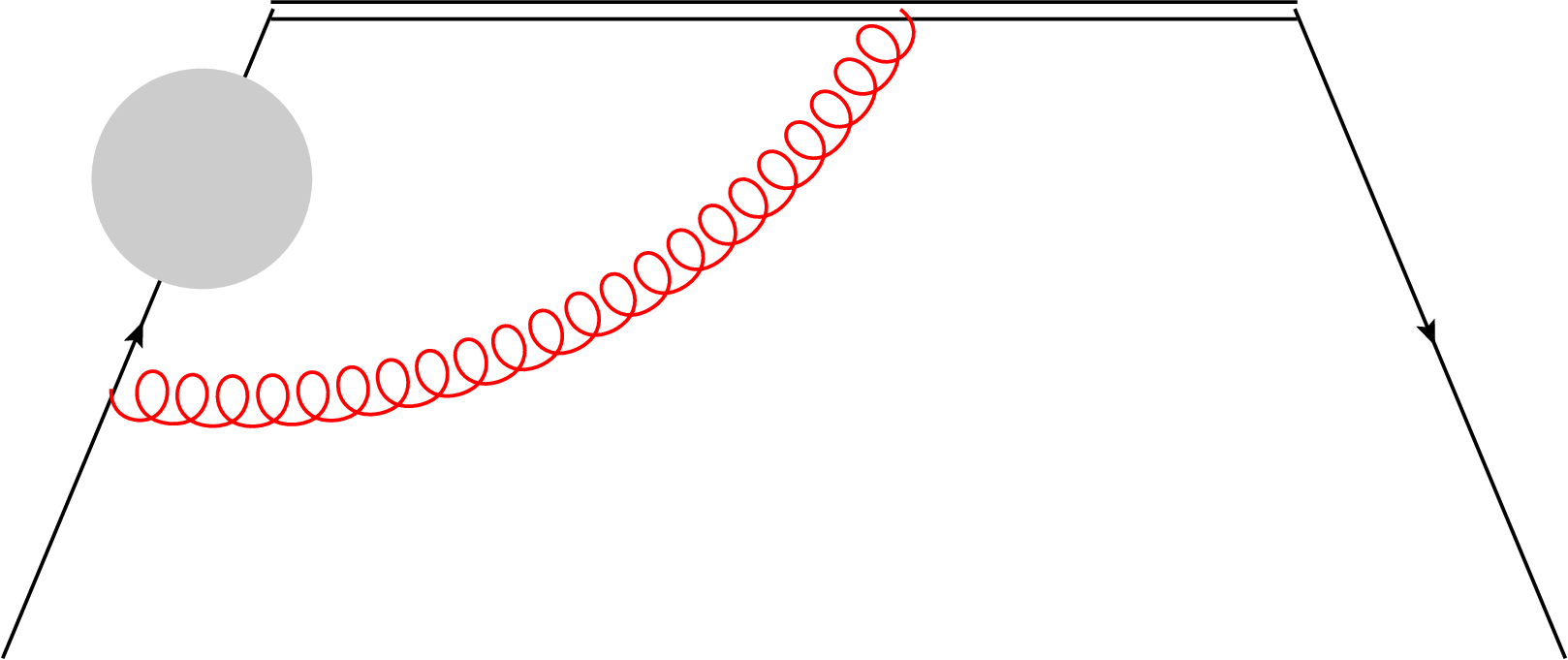}
        \label{qq13}
    \end{subfigure}
}
      \caption{Large-$x$ divergent contributions to the quark-quark parton distribution up to two loops. The grey blob represents a self energy insertion. Each diagram has a multiple factor displayed. Insertions on external legs are excluded.}\label{quarkPDFfigs}
    \end{figure}

    The one loop contributions are figures~\ref{qq1} and~\ref{qq2} and the self energy on each external leg. They sum to,
    \begin{align}
     f_{qq}^{(1)}&= f_{qq}^{(1),(a)}+2f_{qq}^{(1),(b)}+f_{qq}^{(1),\text{SE ext}} \nonumber\\&=C_F\left(\frac{1}{\eps}\left(4 P-(\xi -3) \delta\right)-\delta (\xi +4 \zeta_2-7)+(\xi -1) P-4 L\right),
    \end{align}
    where $\xi$ is the gauge parameter in a general covariant gauge. The two loop contributions are shown in figures~\ref{qq3}--\ref{qq13}. They exclude self energies on external legs. Their calculation was performed in Feynman gauge $\xi=1$ and the results are,
\allowdisplaybreaks
\begin{align*}
f_{qq}^{(2),(c)}&=C_F (C_A-2 C_F)\bigg[\frac{1}{\eps^2}\left(2 \zeta_2 \delta\right)+\frac{1}{\eps}\left(-4 \zeta_3 \delta-4 \zeta_2 P\right)\bigg]\\
f_{qq}^{(2),(d)}&=C_F^2\bigg[\frac{1}{\eps^2}\left((4-8 \zeta_2) \delta+8 P+8 L\right)+\frac{1}{\eps}\left((16-8 \zeta_2) \delta+(8 \zeta_2+16) P-8 L-12 L^2\right)\bigg]\\
f_{qq}^{(2),(e)}&=C_A C_F\bigg[\frac{1}{\eps^2}\left(\left(\frac{\zeta_2}{2}+\frac{3}{2}\right) \delta+\frac{3}{2} P+\frac{1}{2} L\right)\\&\spit  +\frac{1}{\eps}\left(\delta \left(-2 \zeta_2-\frac{\zeta_3}{2}+\frac{15}{2}\right)+(3-2 \zeta_2) P-\frac{5}{2} L-\frac{3}{4} L^2\right)\bigg]\\
f_{qq}^{(2),(f)}&=C_A C_F\bigg[\frac{1}{\eps^2}\left(\delta+2 P+\frac{1}{2} L\right)\\*&\spit+\frac{1}{\eps}\left(\delta \left(-\frac{5 \zeta_2}{2}-\zeta_3+3\right)+\left(\frac{7}{2}-\zeta_2\right) P-3 L-\frac{3}{4} L^2\right)\bigg]\\
f_{qq}^{(2),(g)}&=C_F (C_A-2 C_F)\bigg[\frac{1}{\eps^2}\left((1-\zeta_2) \delta+P+L\right)\\*&\spit  +\frac{1}{\eps}\left(\delta (-2 \zeta_2-3 \zeta_3+5)+2 P-L-\frac{3}{2} L^2\right)\bigg]\\
f_{qq}^{(2),(h)}&=C_F^2\bigg[\frac{1}{\eps^2}\left((2-2 \zeta_2) \delta+2 P+2 L\right)+\frac{1}{\eps}\left(\delta (-4 \zeta_2-6 \zeta_3+10)+4 P-2 L-3 L^2\right)\bigg]\\
  f_{qq}^{(2),(i)}&=C_F (C_A-2 C_F)\bigg[\frac{1}{\eps^2}\left(\left(\zeta_2-\frac{1}{2}\right) \delta+\frac{1}{2} P\right)\\& \spit +\frac{1}{\eps}\left(\delta \left(-\frac{\zeta_2}{2}+7 \zeta_3-4\right)+\frac{1}{2} P-\frac{1}{2} L\right)\bigg]\\
   f_{qq}^{(2),(j)}&=C_F\bigg[\frac{1}{\eps^2}\left(\delta \left(\frac{2 n_f T_f}{3}-\frac{5 C_A}{6}\right)\right)\\&
\spit +\frac{1}{\eps}\left(\delta \left(\frac{7 C_A}{9}-\frac{8n_f T_f}{9}\right)+P \left(\frac{5 C_A}{3}-\frac{4  n_f T_f}{3}\right)\right)\bigg]\\
  f_{qq}^{(2),(k)}&=C_A C_F\left(\frac{1}{\eps^2}\left(\zeta_2 \delta\right)+\frac{1}{\eps}\left(\zeta_3 \delta-2 \zeta_2 P\right)\right)\\
f_{qq}^{(2),(l)}&=C_F\bigg[\frac{1}{\eps^2}\left(\delta \left(\frac{5 C_A}{3}-\frac{4  n_f T_f}{3}\right)+P \left(\frac{5 C_A}{3}-\frac{4  n_f T_f}{3}\right)\right)\nonumber\\&\spit+\frac{1}{\eps}\bigg(\delta \left(-\frac{5 C_A  \zeta_2}{3}+\frac{61 C_A}{9}+\frac{4}{3}  n_f T_f \zeta_2-\frac{44 n_f T_f}{9}\right)\\&\spit+P \left(\frac{16 C_A }{9}-\frac{8  n_f T_f}{9}\right)+L \left(\frac{8  n_f T_f}{3}-\frac{10 C_A}{3}\right)\bigg)\bigg]\\
f_{qq}^{(2),(m)}&=C_F^2\left(-\frac{1}{\eps^2}\left( \delta + P\right)+\frac{1}{\eps}\left(\left(2 \zeta_2-4 \right) \delta -P + L\right)\right)\\
     f_{qq}^{(2),\text{SE ext}}&=\frac{1}{\eps^2}\left(-2 \left(C_A C_F+3 C_F^2\right) \delta-8 C_F^2 P\right)\\&\quad+\frac{1}{\eps}\left(\frac{1}{2} \delta \left(-25 C_A C_F+16 C_F^2 \zeta_2-37 C_F^2+4 C_F n_f T_f\right)-8 C_F^2 P+8 C_F^2 L\right)
\end{align*}%
Summing the two loop contributions with the factors shown in figure~\ref{quarkPDFfigs} we find,
\begin{align}
  f_{qq}^{(2)} =\,& \frac{C_F}{\eps^2}\bigg[\delta \left(\frac{9 C_A}{2}+C_F (2-8 \zeta_2)-2 n_f T_f\right)+P \left(\frac{22 C_A}{3}+8 C_F^2-\frac{8 n_f T_f}{3}\right)+16 C_F L\bigg]\nonumber\\&+\frac{C_F}{\eps}\bigg[\delta \bigg(C_A  \left(-\frac{22 \zeta_2}{3}-6 \zeta_3+\frac{125}{6}\right)\nonumber\\
&\qquad\qquad+C_F \left(-14 \zeta_2-4 \zeta_3+\frac{27}{2}\right)+n_f T_f \left(\frac{8 \zeta_2}{3}-\frac{26}{3}\right)\bigg)\nonumber\\&\qquad \qquad +P \left(C_A \left(\frac{119}{9}-4 \zeta_2\right)+24 C_F-\frac{28 n_f T_f}{9}\right)\nonumber\\&\qquad \qquad  +L \left(-\frac{44 C_A}{3}-8 C_F+\frac{16  n_f T_f}{3}\right) -24 C_FL^2\bigg]
\end{align}

At two loops we need to take into account the running from the one loop contribution, $\frac{\al_s}{4\pi}f_{qq}^{(1)}\to \frac{\al_s}{4\pi}f_{qq}^{(1),R}$. This is found by replacing $\xi\to\left(1+\frac{\al_s}{4\pi\eps}\left(\frac{10}{6}C_A-\frac{4}{3}T_fn_f\right)\right)\xi$ and $\al_s\to\big(1+\frac{\al_s\hat{b}_0}{\pi\eps}\big)\al_s$. We then specialise to Feynman gauge $\xi=1$.

We then find the $Z_{qq}$ that minimally subtracts the divergences in $\delta+\frac{\al_s}{4\pi}f_{qq}^{(1),R}+\left(\frac{\al_s}{4\pi}\right)^2f_{qq}^{(2)}$. As the renormalisation is multiplicative, convolutions need to be taken into account for one loop squared terms. For example,
\begin{equation}
P\otimes L = -\zeta_2P+\frac{3}{2}L^2+\zeta_3\delta.
\end{equation}
Equivalently the renormalisation can be transformed to Mellin space, eq.~\ref{MellinSpace}, where the convolutions become products ensuring that,
\begin{equation}
\tilde{Z}_{qq}\bigg(1+\frac{\al_s}{4\pi}\tilde{f}_{qq}^{(1),R}+\bigg(\frac{\al_s}{4\pi}\bigg)^2\tilde{f}_{qq}^{(2)}\bigg)
\end{equation}
is finite in $\eps$. We can then extract the splitting functions to two loops from,
\begin{equation}
\tilde{P}_{qq}=\bigg(-\eps\al_s-\al_s^2\frac{\hat{b}_0}{\pi}\bigg)\frac{d}{d\al_s}\log(\tilde{Z}_{qq}Z_q),
\end{equation}
where $Z_q$ is the wavefunction renormalisation in $\overline{\text{MS}}$ for the quark. Up to two loops,
\begin{align}
Z_q&=1-\left(\frac{\al_s}{4\pi}\right) \frac{C_F}{\eps}+\left(\frac{\al_s}{4\pi}\right)^2C_F \left(\frac{1}{\eps^2}\left(C_A +\frac{C_F}{2}\right)+\frac{1}{\eps}\left(-\frac{17 C_A}{4}+\frac{3 C_F}{4}+ T_fn_f\right)\right).
\end{align}

Converting back to $x$ space we find,
\begin{align}
  P_{qq}=\,&\frac{\al_s}{4\pi}C_F\left(3\delta+4P\right)\nonumber\\&+\bigg(\frac{\al_s}{4\pi}\bigg)^2\bigg[\delta \bigg(C_A C_F \left(\frac{44 \zeta_2}{3}-12 \zeta_3+\frac{17}{6}\right)+C_F^2 \left(-12 \zeta_2+24 \zeta_3+\frac{3}{2}\right)\nonumber\\& \spit  -C_FT_fn_f \left(\frac{16 \zeta_2}{3}+\frac{2}{3}\right)\bigg)+P\left(C_A C_F \left(\frac{268}{9}-8 \zeta_2\right)-\frac{80 C_F n_f T_f}{9}\right)\bigg].
  \label{app:Pqq}
\end{align}

Notice that we find that all $L^n$ terms cancel. This reproduces $B_\delta^{q}$, the coefficient of $\delta$ in eq.~\eqref{BdeltaExp}, and shows that the coefficient of $P$ is $\gamma_{\text{cusp}}$ as in eq.~\eqref{twoLoopCusp}.

\subsection[Calculating \texorpdfstring{$P_{gg}$}{P(gg)}]{\boldmath Calculating $P_{gg}$}
\label{subsec:gluongluon}

\begin{figure}
\centerline{
  \begin{subfigure}[b]{0.3\textwidth}
    \caption{1}
        \includegraphics[width=.97\textwidth]{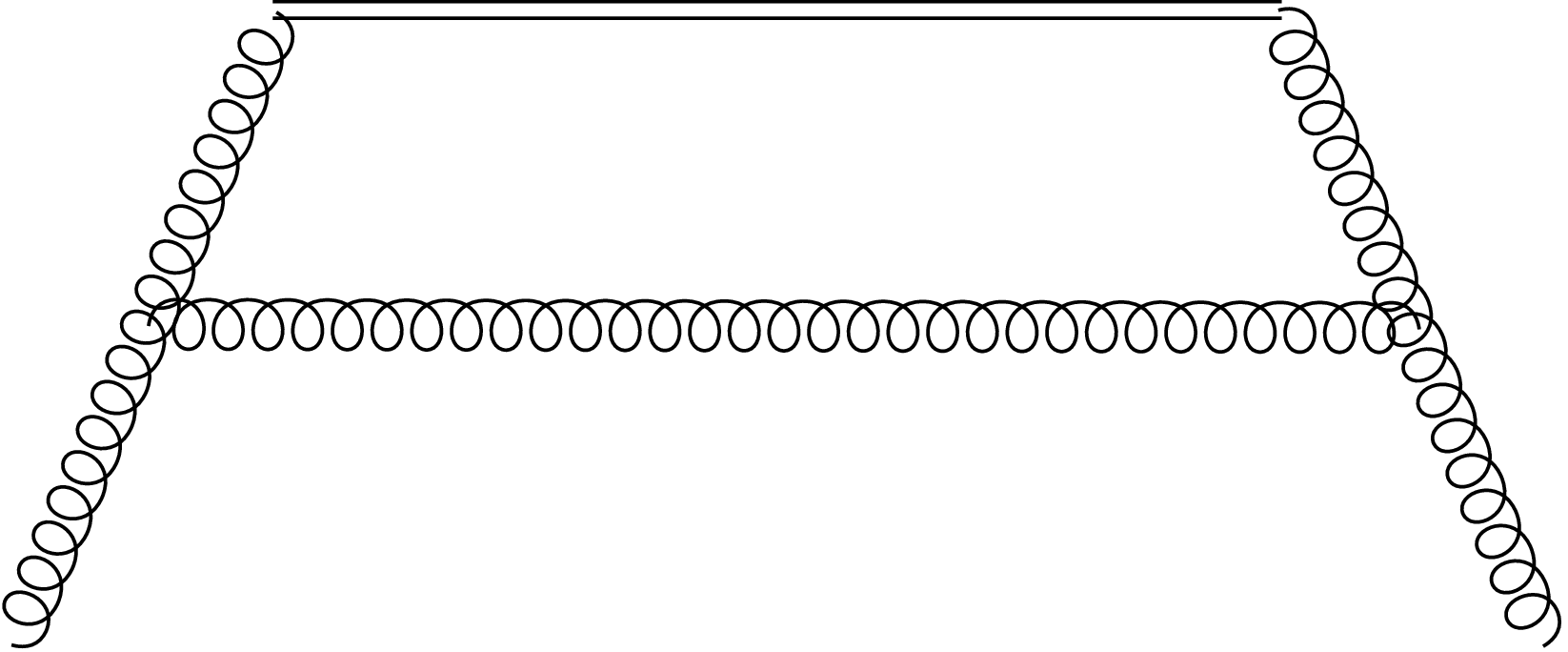}
        \label{gg1}
    \end{subfigure}\quad
    \begin{subfigure}[b]{0.3\textwidth}
      \caption{2}
        \includegraphics[width=.97\textwidth]{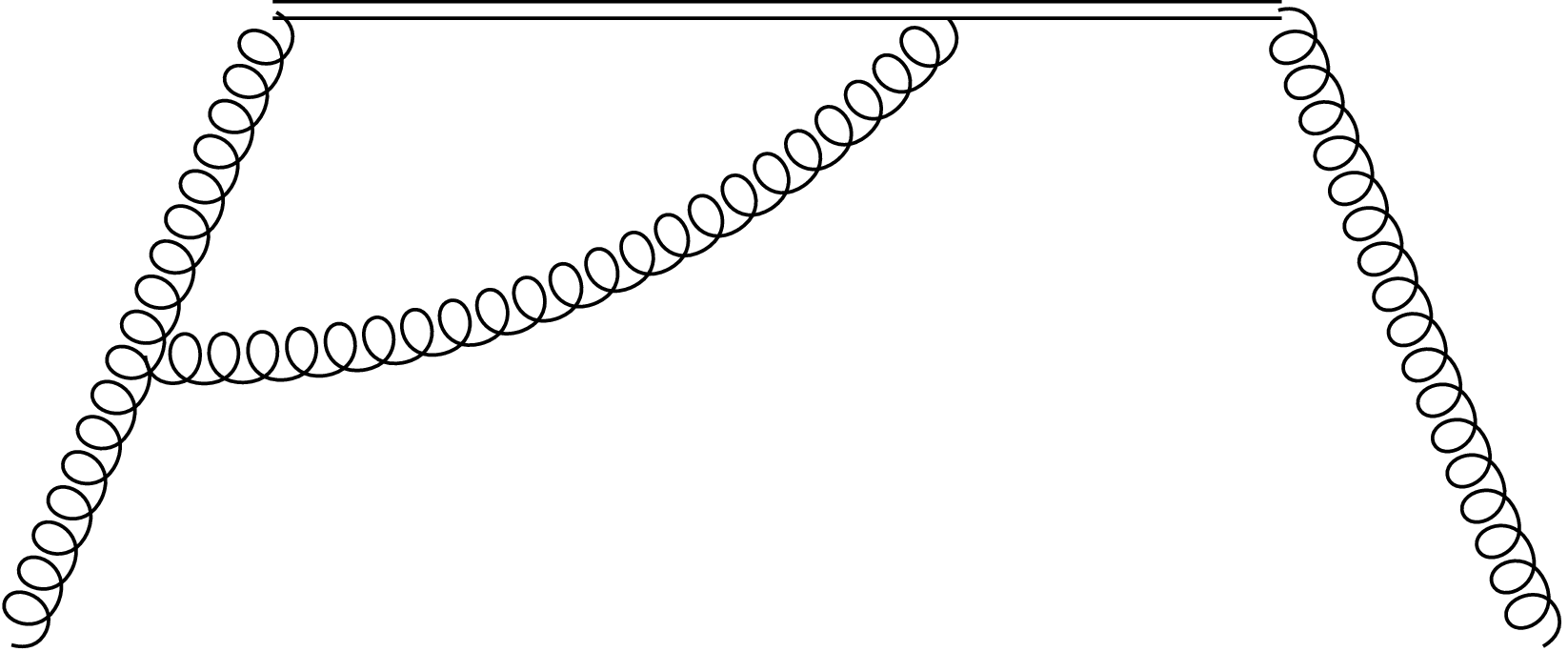}
        \label{gg2}
    \end{subfigure}\quad
    \begin{subfigure}[b]{0.3\textwidth}
      \caption{2}
        \includegraphics[width=.97\textwidth]{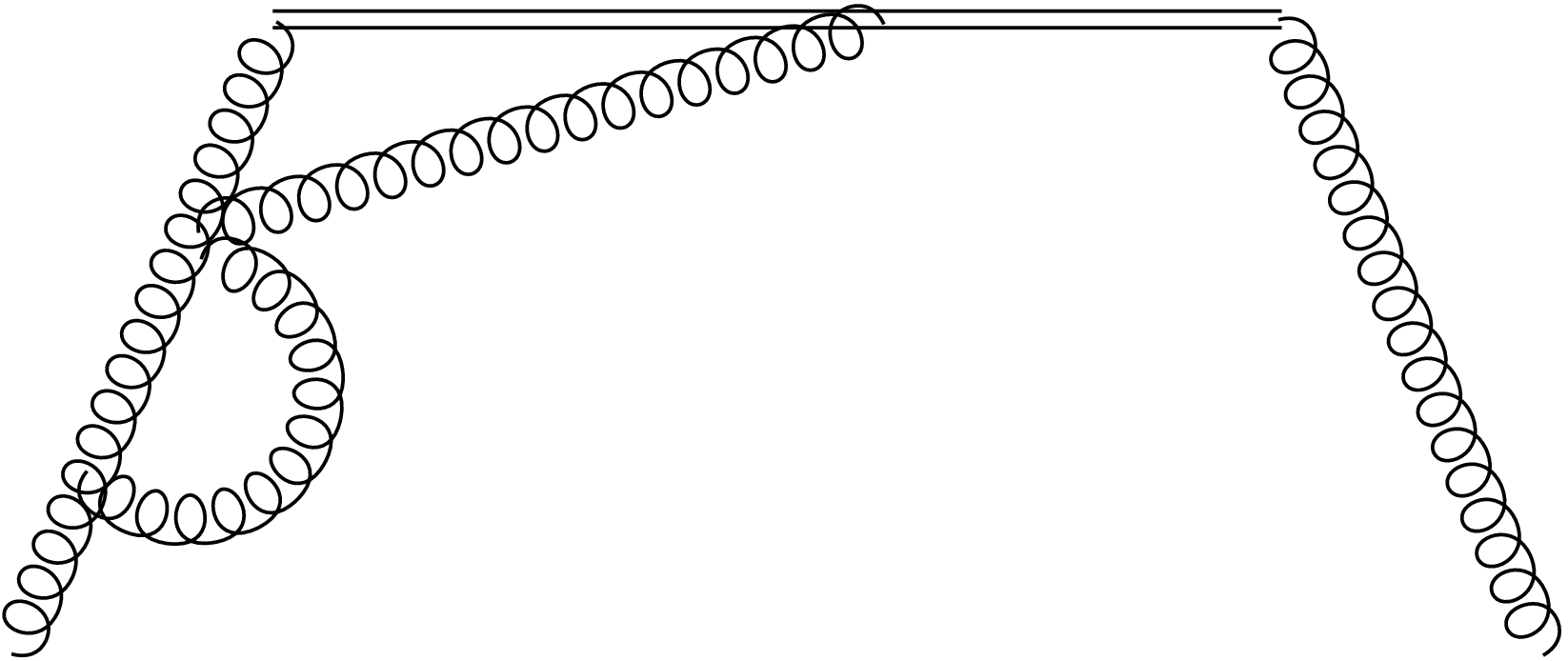}
        \label{gg3}
    \end{subfigure}
}
\vspace{3mm}
\centerline{
    \begin{subfigure}[b]{0.3\textwidth}
      \caption{1}
        \includegraphics[width=.97\textwidth]{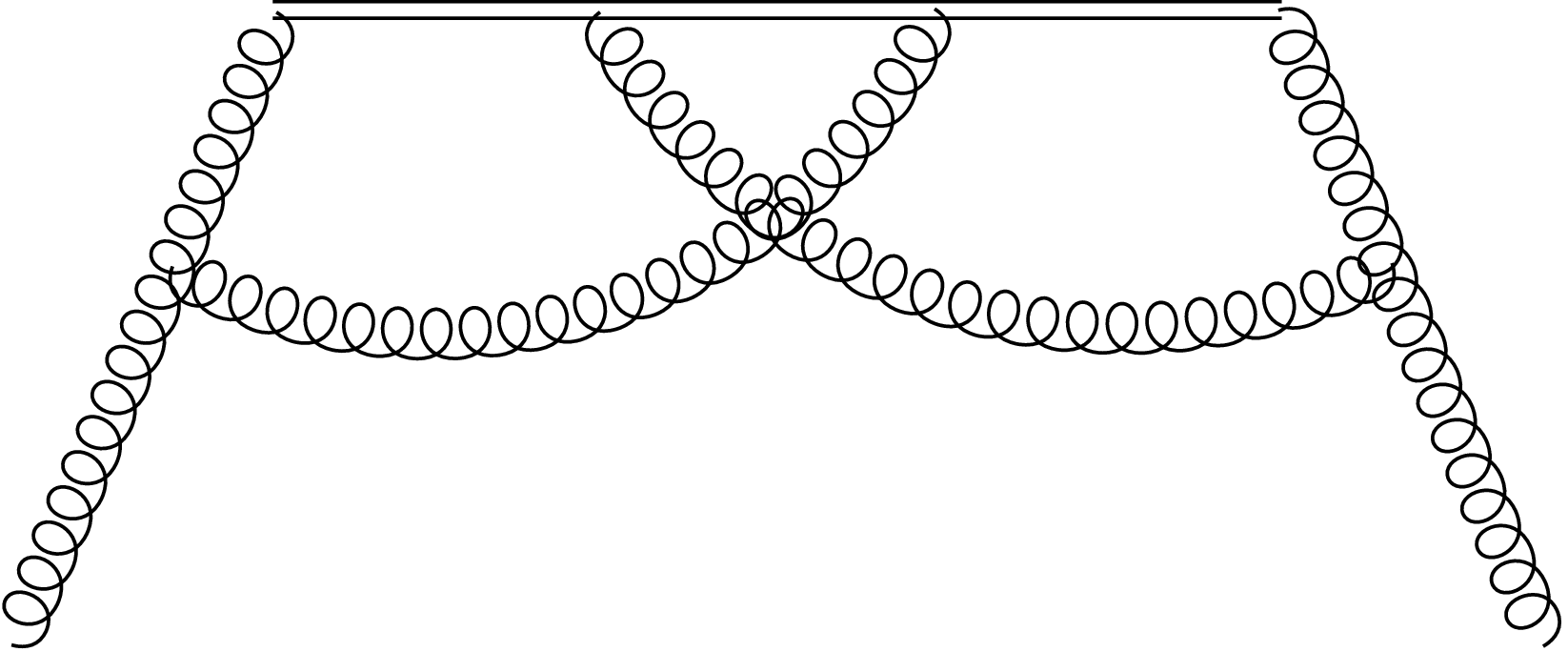}
        \label{gg4}
    \end{subfigure}\quad
    \begin{subfigure}[b]{0.3\textwidth}
      \caption{1}
        \includegraphics[width=.97\textwidth]{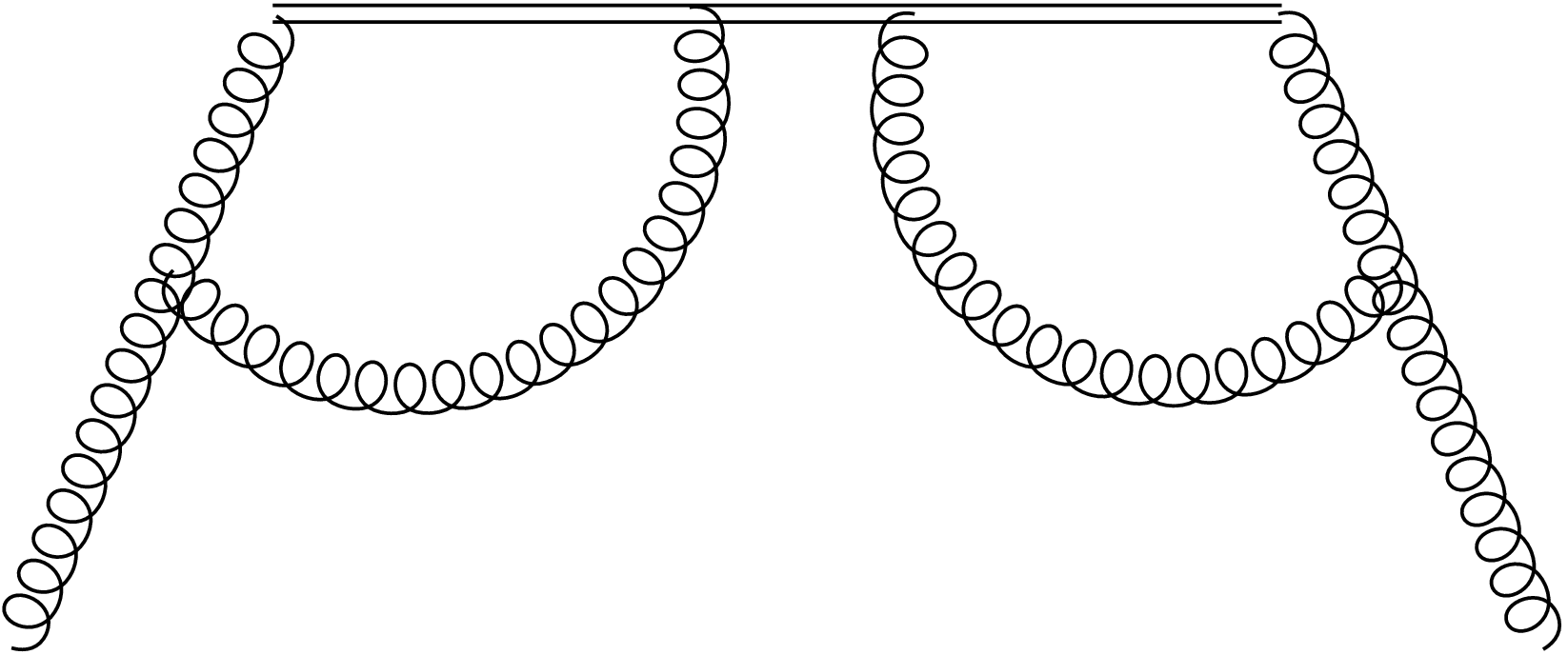}
        \label{gg5}
    \end{subfigure}\quad
    \begin{subfigure}[b]{0.3\textwidth}
      \caption{2}
        \includegraphics[width=.97\textwidth]{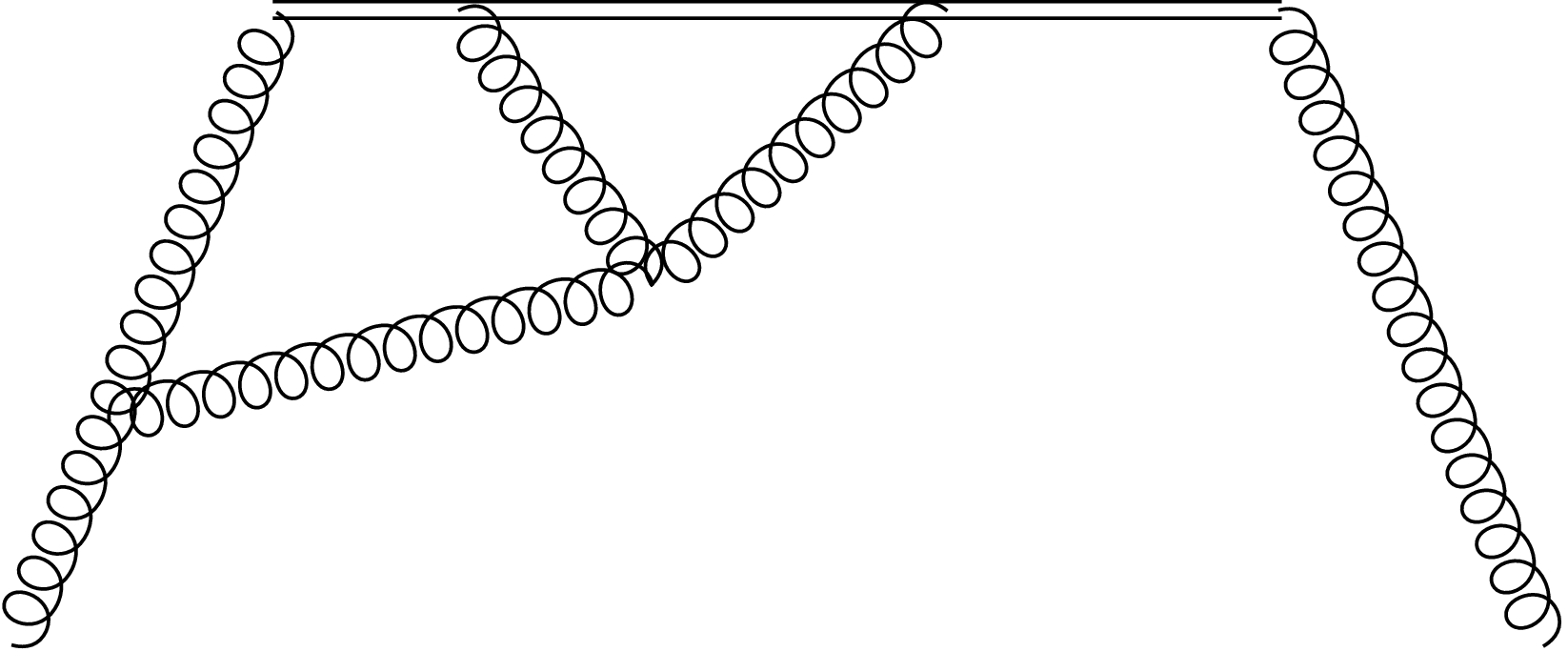}
        \label{gg6}
    \end{subfigure}
}
\vspace{3mm}
\centerline{
    \begin{subfigure}[b]{0.3\textwidth}
      \caption{2}
        \includegraphics[width=.97\textwidth]{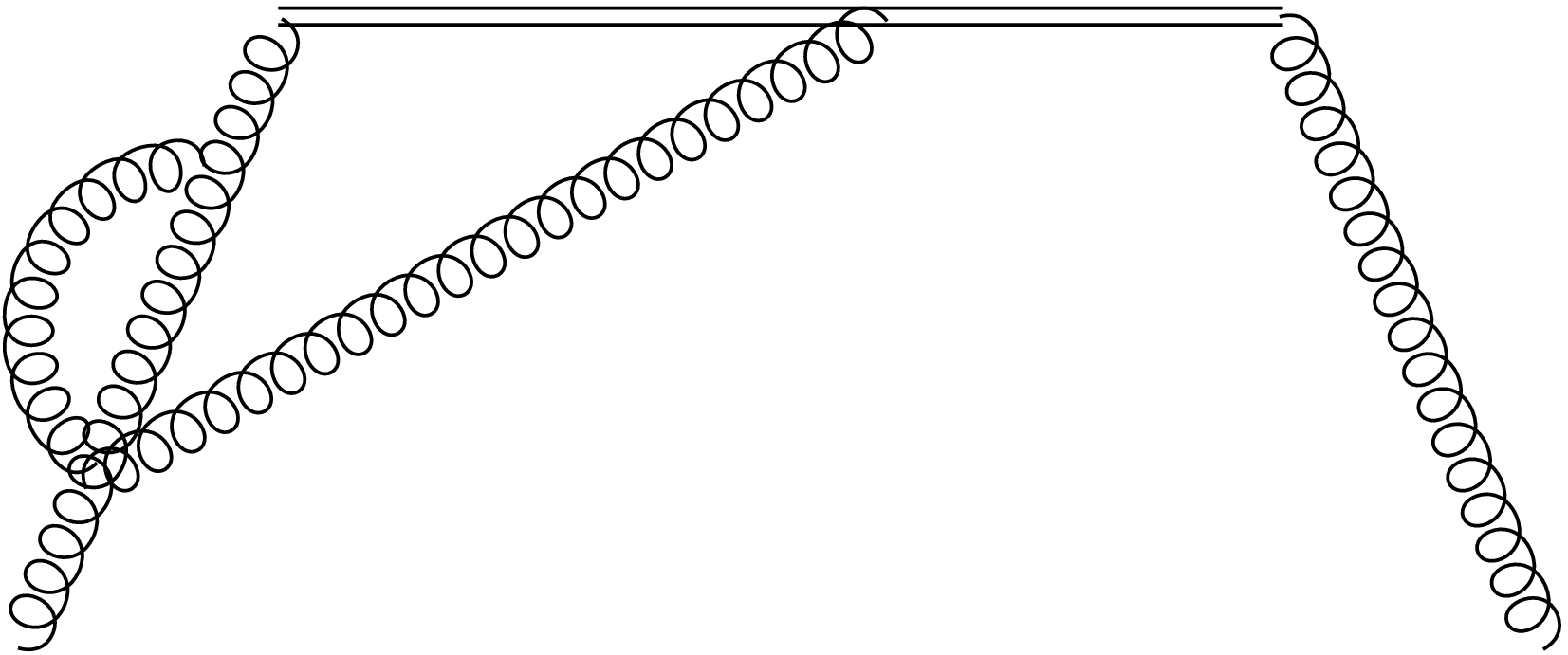}
        \label{gg7}
    \end{subfigure}\quad
    \begin{subfigure}[b]{0.3\textwidth}
      \caption{2}
        \includegraphics[width=.97\textwidth]{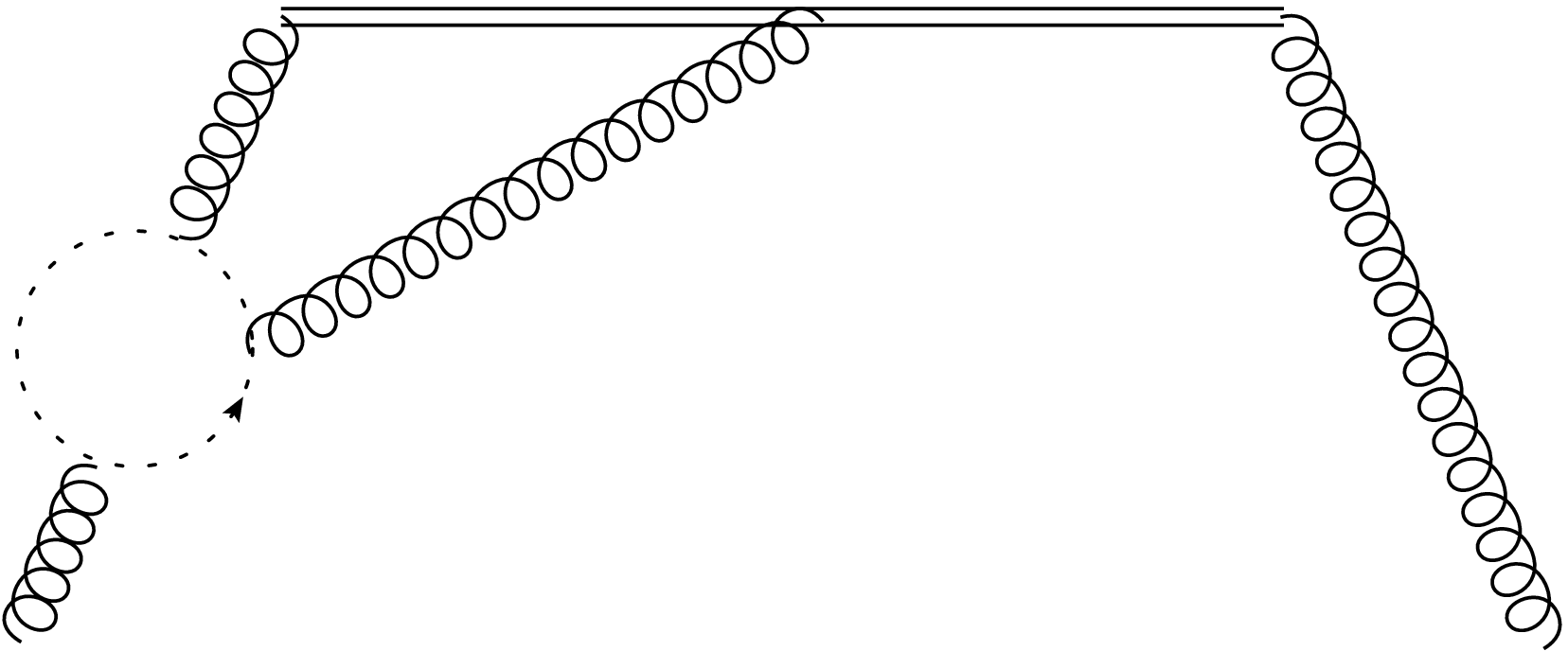}
        \label{gg8}
    \end{subfigure}\quad
    \begin{subfigure}[b]{0.3\textwidth}
      \caption{2}
        \includegraphics[width=.97\textwidth]{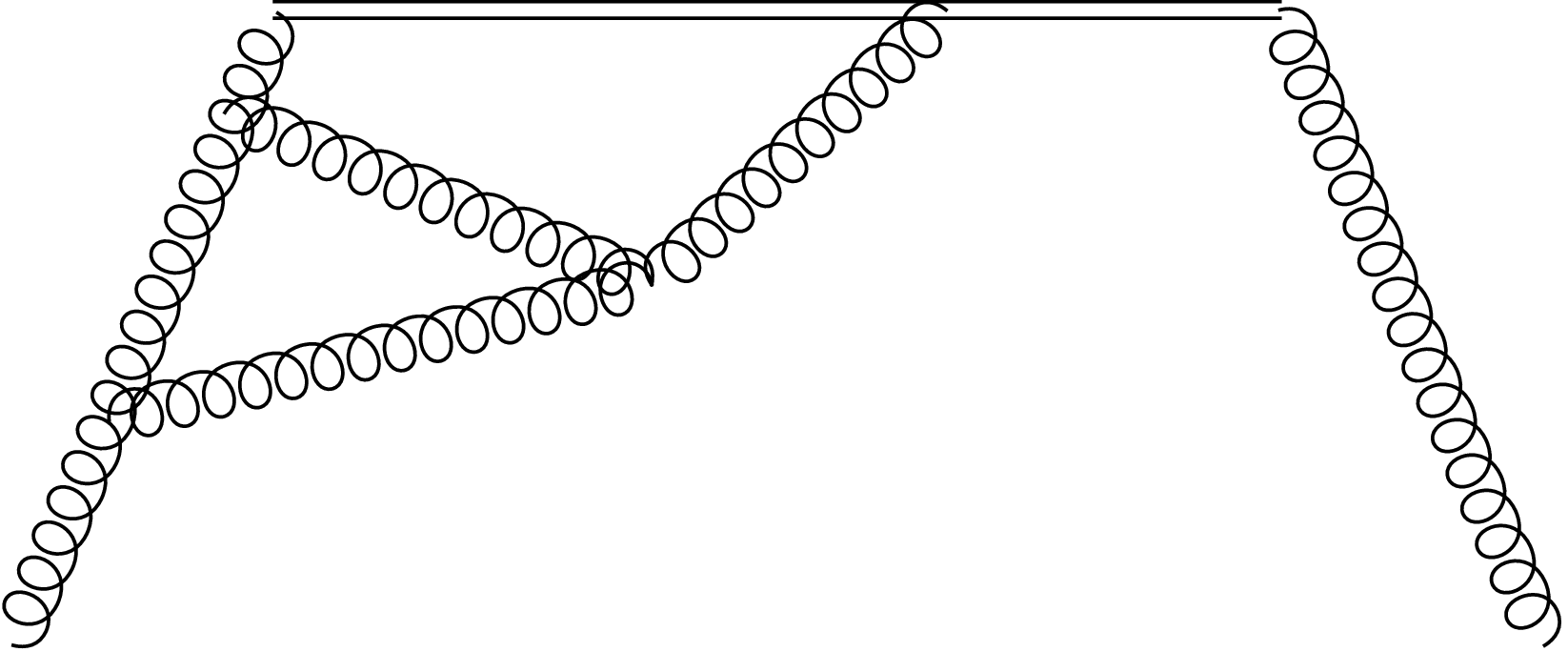}
        \label{gg9}
    \end{subfigure}
}
\vspace{3mm}
\centerline{
    \begin{subfigure}[b]{0.3\textwidth}
      \caption{4}
        \includegraphics[width=.97\textwidth]{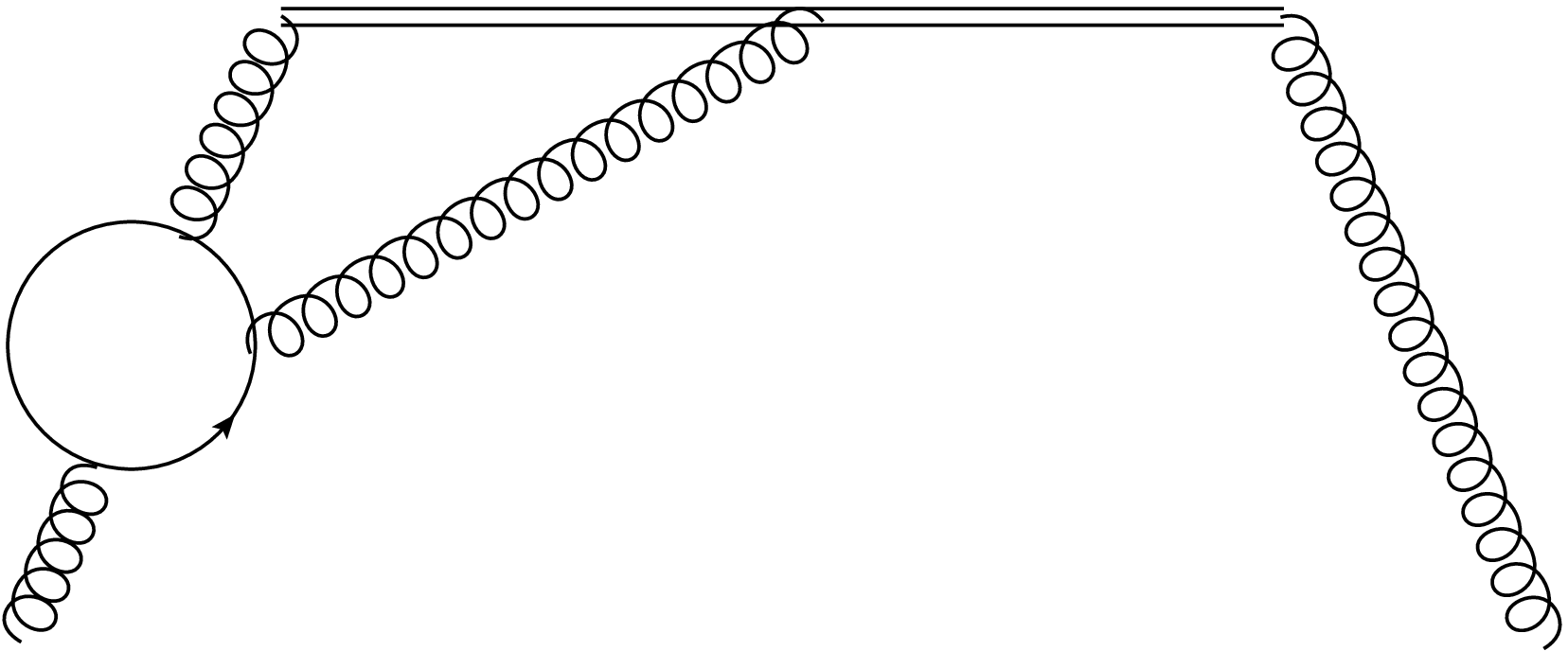}
        \label{gg10}
    \end{subfigure}\quad
    \begin{subfigure}[b]{0.3\textwidth}
      \caption{2}
        \includegraphics[width=.97\textwidth]{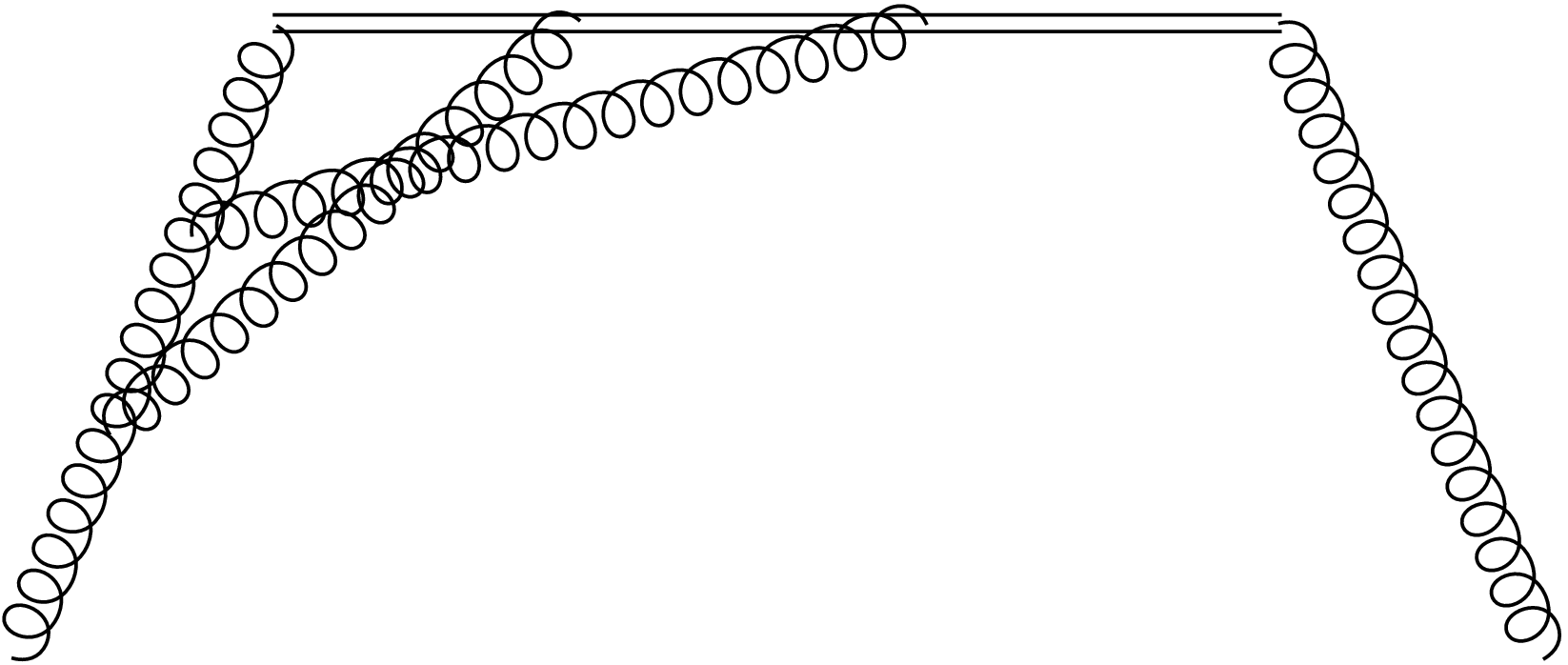}
        \label{gg11}
    \end{subfigure}\quad
    \begin{subfigure}[b]{0.3\textwidth}
      \caption{2}
        \includegraphics[width=.97\textwidth]{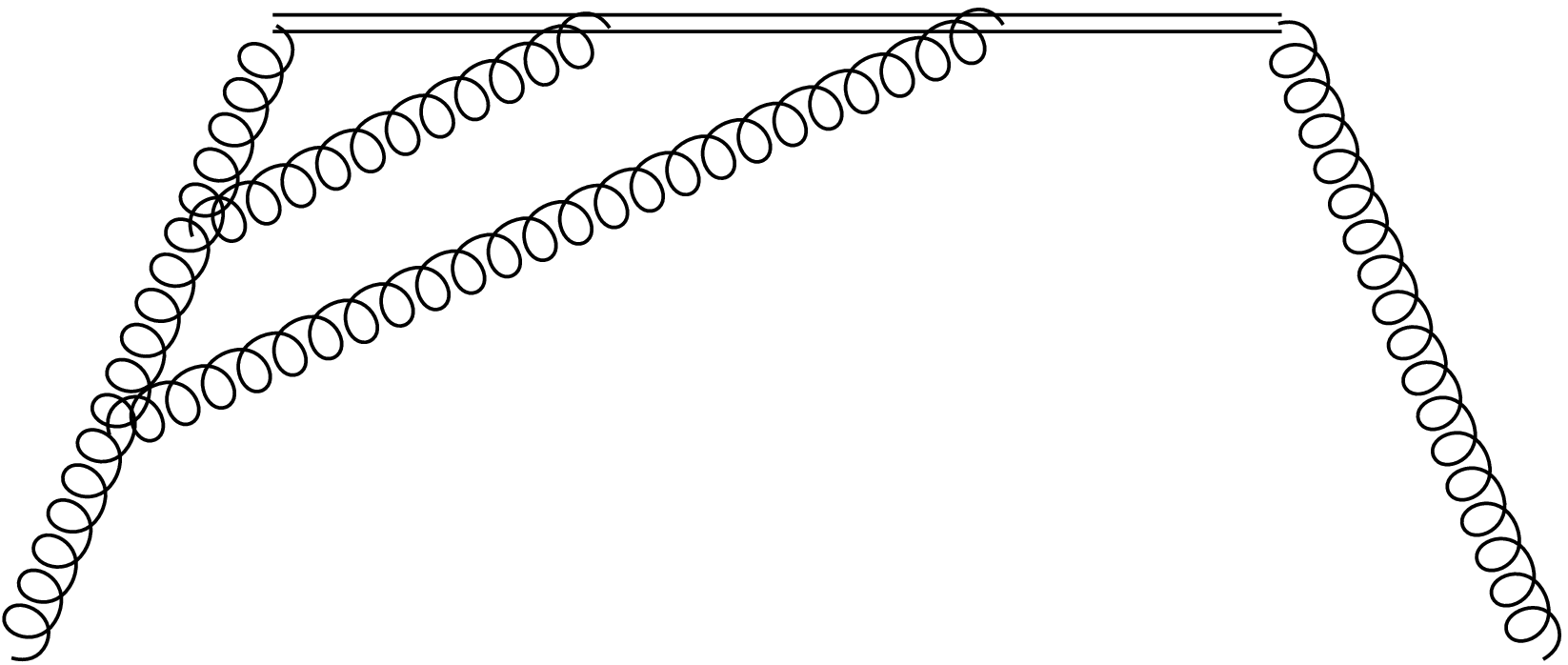}
        \label{gg12}
    \end{subfigure}
}
\vspace{3mm}
\centerline{
    \begin{subfigure}[b]{0.3\textwidth}
      \caption{1}
        \includegraphics[width=.97\textwidth]{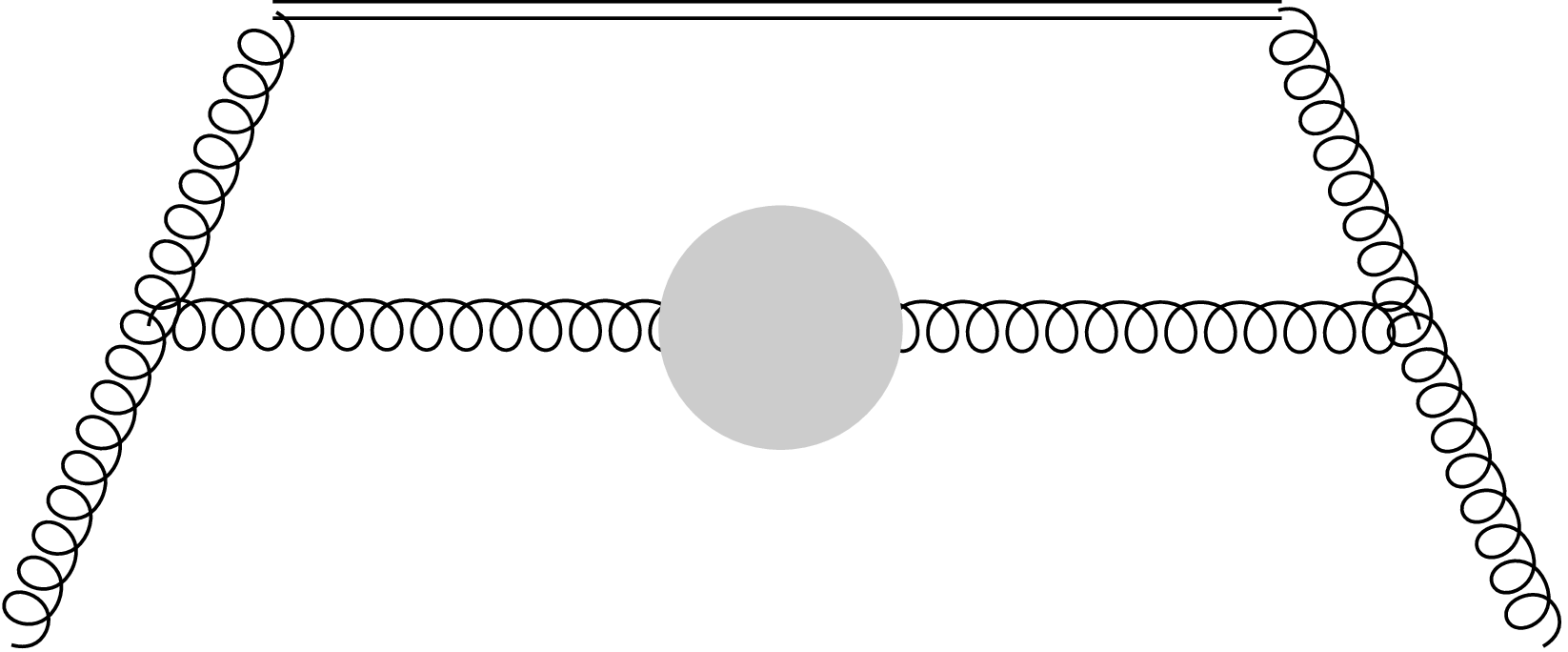}
        \label{gg13}
    \end{subfigure}\quad
    \begin{subfigure}[b]{0.3\textwidth}
      \caption{1}
        \includegraphics[width=.97\textwidth]{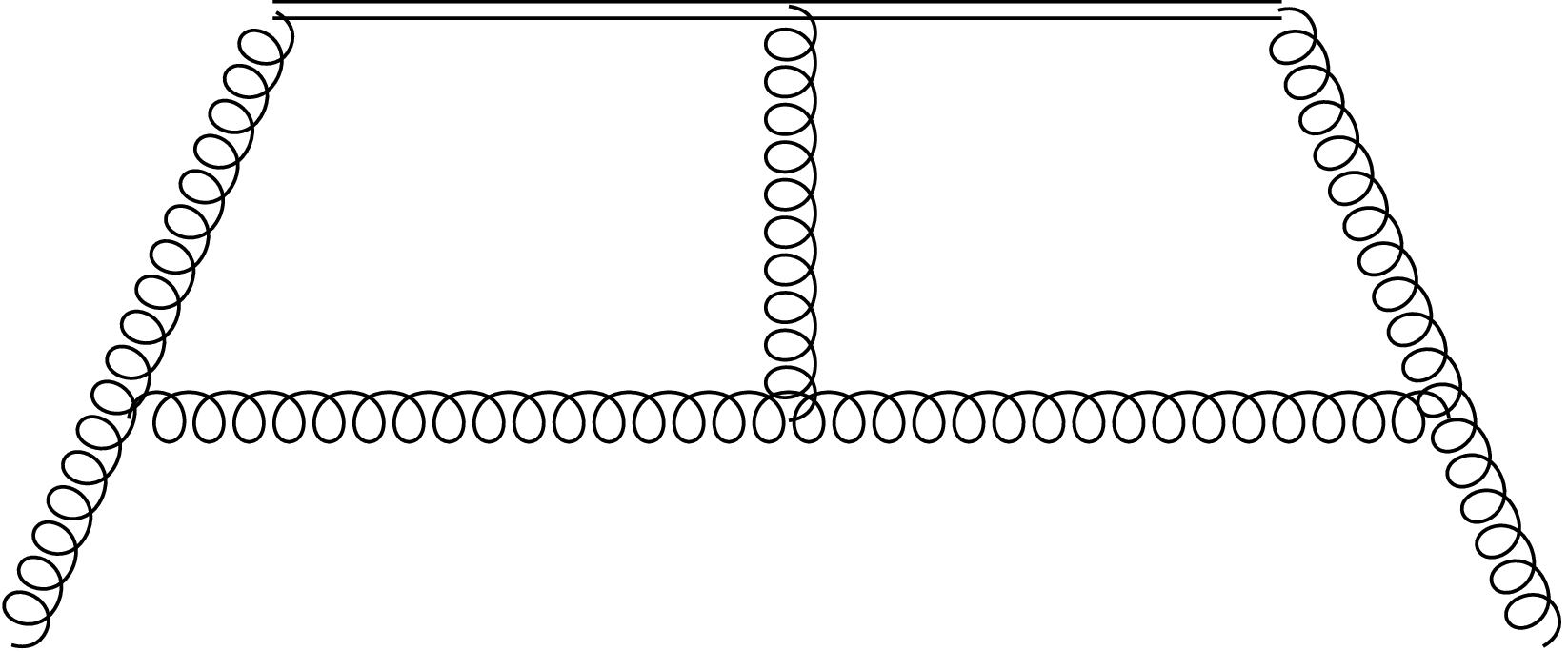}
        \label{gg14}
    \end{subfigure}\quad
    \begin{subfigure}[b]{0.3\textwidth}
      \caption{2}
        \includegraphics[width=.97\textwidth]{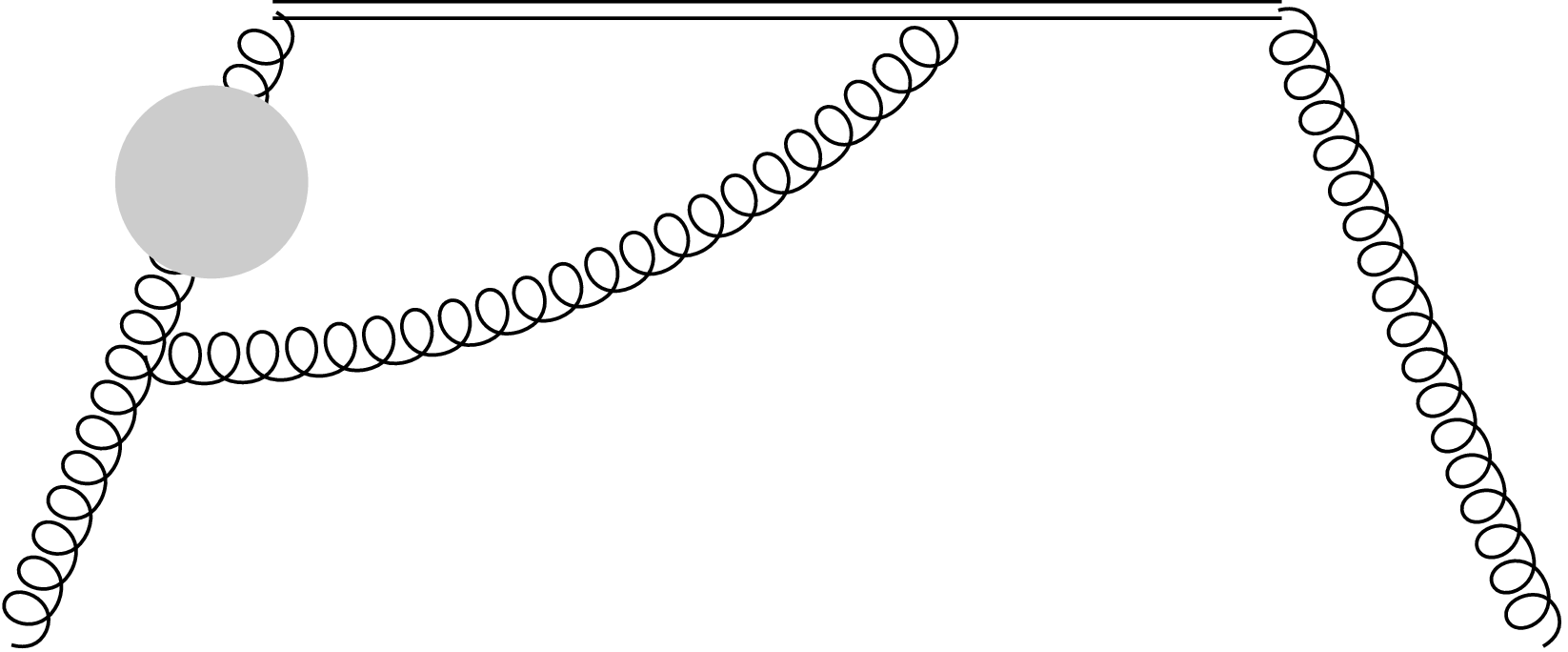}
        \label{gg15}
    \end{subfigure}
}
\vspace{3mm}
\centerline{
    \begin{subfigure}[b]{0.3\textwidth}
      \caption{2}
        \includegraphics[width=.97\textwidth]{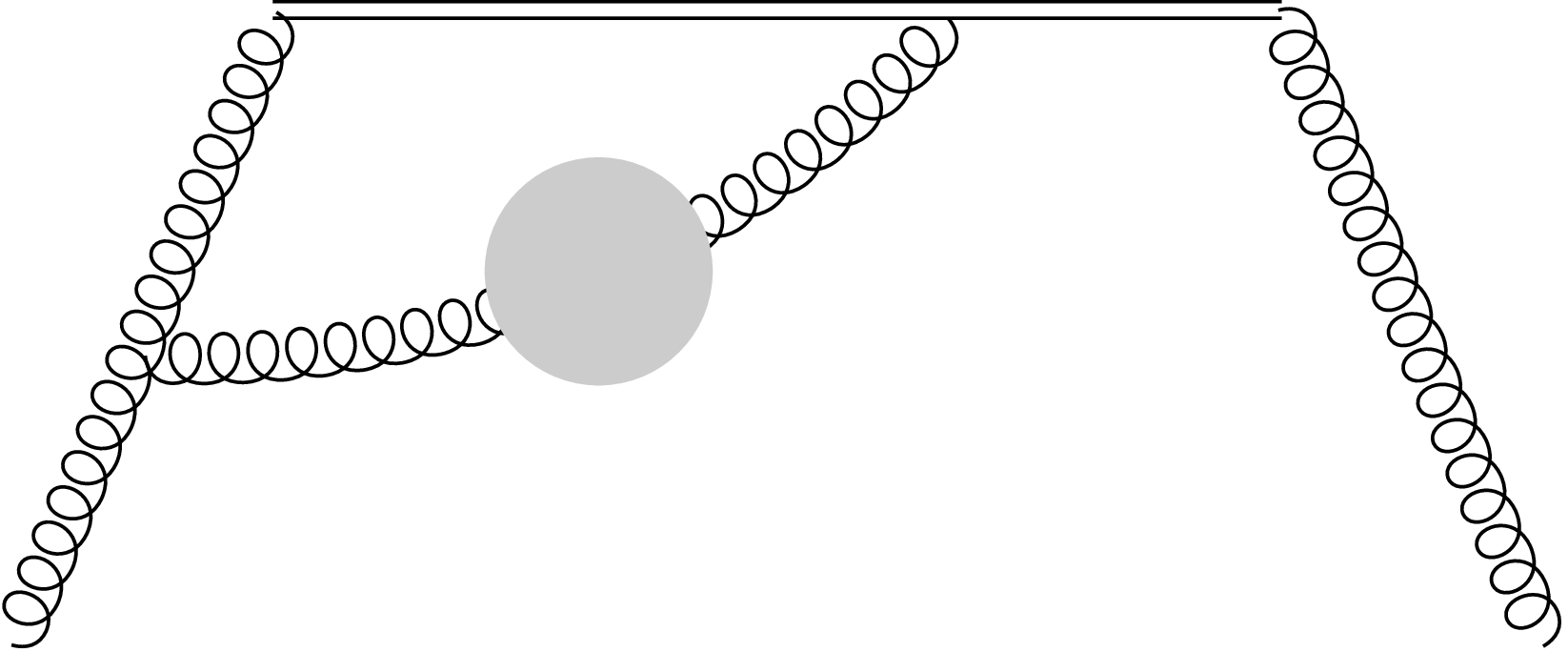}
        \label{gg16}
    \end{subfigure}}
     \caption{Large-$x$ divergent contributions to the gluon-gluon parton distribution up to two loops. The grey blob represents a self energy insertion. Insertions on external legs are excluded. The clockwise ghost is included in h).}
   \end{figure}

   The one loop contributions for the gluon gluon distribution function are shown in figures~\ref{gg1} and~\ref{gg2}. The total one loop contributions are,
   \begin{align}
     f_{gg}^{(1)}=\,&\frac{1}{\eps}\left[\delta \left(-\frac{C_A \xi }{2}+\frac{35 C_A}{6}-\frac{8 n_f T_f}{3}\right)+4 C_A P\right]\nonumber\\& +\delta \left(-4 C_A \zeta_2+\frac{98 C_A}{9}-\frac{40 n_f T_f}{9}\right)+(C_A \xi -C_A) P-4 C_A L.
   \end{align}
   The two loop contributions are shown in figures~\ref{gg3}--\ref{gg16}. The two loop contributions are,
   \allowdisplaybreaks
   \begin{align*}
     f_{gg}^{(2),(c)}=\,&C_A^2\left(\frac{1}{\eps^2}\left(-\frac{9}{4} P\right)+\frac{1}{\eps}\left(\frac{9}{4} \zeta_2 \delta-\frac{9}{2} P+\frac{9}{4} L\right)\right)\\
f_{gg}^{(2),(d)}=\,&C_A^2\left(\frac{1}{\eps^2}\left(2 \zeta_2 \delta\right)+\frac{1}{\eps}\left(-4 \zeta_3 \delta-4 \zeta_2 P\right)\right)\\
f_{gg}^{(2),(e)}=\,&C_A^2\left(\frac{1}{\eps^2}\left((1-8 \zeta_2) \delta+4 P+8 L\right)+\frac{1}{\eps}\left((4-4 \zeta_2) \delta+(8 \zeta_2+8) P-4 L-12 L^2\right)\right)\\
f_{gg}^{(2),(f)}=\,&C_A^2\bigg[\frac{1}{\eps^2}\left(\left(\frac{\zeta_2}{2}+\frac{3}{4}\right) \delta+\frac{5}{4} P+\frac{1}{2} L\right)\\&\spit+\frac{1}{\eps}\left(\delta \left(-\frac{3 \zeta_2}{2}-\frac{\zeta_3}{2}+\frac{15}{4}\right)+\left(\frac{5}{2}-2 \zeta_2\right) P-\frac{9}{4} L-\frac{3}{4} L^2\right)\bigg]\\
f_{gg}^{(2),(g)}=\,&C_A^2\left(\frac{1}{\eps^2}\left(-\frac{9}{8} \delta-\frac{9}{8} P\right)+\frac{1}{\eps}\left(\left(\frac{9 \zeta_2}{4}-\frac{45}{8}\right) \delta-\frac{9}{4} P+\frac{9}{8} L\right)\right)\\
f_{gg}^{(2),(h)}=\,&C_A^2\left(\frac{1}{\eps^2}\left(-\frac{1}{48} \delta-\frac{1}{24} P\right)+\frac{1}{\eps}\left(\left(\frac{\zeta_2}{24}-\frac{31}{288}\right) \delta-\frac{1}{9} P+\frac{1}{24} L\right)\right)\\
f_{gg}^{(2),(i)}=\,&C_A^2\bigg[\frac{1}{\eps^2}\left(\left(\frac{3 \zeta_2}{4}+\frac{5}{16}\right) \delta+3 P+\frac{1}{2} L\right)\nonumber\\*&\spit +\frac{1}{\eps}\left(\delta \left(-\frac{35 \zeta_2}{8}+\frac{17 \zeta_3}{4}+\frac{65}{96}\right)+\left(\frac{71}{12}-\zeta_2\right) P-4 L-\frac{3}{4} L^2\right)\bigg]\\
f_{gg}^{(2),(j)}=\,&C_A n_f T_f\left(\frac{1}{\eps^2}\left(\frac{1}{3} \delta+\frac{2}{3} P\right)+\frac{1}{\eps}\left(\left(\frac{59}{36}-\frac{2 \zeta_2}{3}\right) \delta+\frac{10}{9} P-\frac{2}{3} L\right)\right)\\
f_{gg}^{(2),(k)}=\,&C_A^2\bigg[\frac{1}{\eps^2}\left(\left(\frac{3}{4}-\frac{5 \zeta_2}{4}\right) \delta+\frac{1}{2} P+L\right)\\&\spit +\frac{1}{\eps}\left(\delta \left(-\zeta_2-\frac{19 \zeta_3}{4}+\frac{15}{4}\right)+P-\frac{1}{2} L-\frac{3}{2} L^2\right)\bigg]\\
     f_{gg}^{(2),(l)}=\,&C_A^2\left(\frac{1}{\eps^2}\left(-\frac{3}{2} \zeta_2 \delta+P+2 L\right)+\frac{1}{\eps}\left(\delta \left(-2 \zeta_2-\frac{5 \zeta_3}{2}\right)+2 P-L-3 L^2\right)\right)\\
     f_{gg}^{(2),(m)}=\,&\frac{1}{\eps^2}\left(\delta \left(\frac{2 C_A n_f T_f}{3}-\frac{5 C_A^2}{6}\right)\right)\\& +\frac{1}{\eps}\left(\delta \left(\frac{7 C_A^2}{9}-\frac{8 C_A n_f T_f}{9}\right)+P \left(\frac{5 C_A^2}{3}-\frac{4 C_A n_f T_f}{3}\right)\right)\\
     f_{gg}^{(2),(n)}=\,&C_A^2\left(\frac{1}{\eps^2}\left(\zeta_2 \delta\right)+\frac{1}{\eps}\left(-2 \zeta_2 P+\zeta_3 \delta\right)\right)\\
f_{gg}^{(2),(o)}=\,&\frac{1}{\eps^2}\left(\delta \left(-\frac{7 C_A n_f T_f}{9}+\frac{35 C_A^2}{36}\right)+P \left(-\frac{4 C_A n_f T_f}{3}+\frac{5 C_A^2}{3}\right)\right)\nonumber\\& \,\,\,\,\,\,\, +\frac{1}{\eps}\bigg[\delta \left(-\frac{10 C_A^2 \zeta_2}{3}+\frac{133 C_A^2}{27}+\frac{8}{3} C_A n_f T_f \zeta_2-\frac{98 C_A n_f T_f}{27}\right)\\&\,\,\,\,\,\,\,\qquad \qquad +P \left(-\frac{20 C_A n_f T_f}{9}+\frac{31 C_A^2}{9}\right)+L \left(-\frac{5 C_A^2}{3}+\frac{4 C_A n_f T_f}{3}\right)\bigg]\\
f_{gg}^{(2),(p)}=\,&\frac{1}{\eps^2}\left(\delta \left(-\frac{13 C_A n_f T_f}{18}+\frac{65 C_A^2}{72}\right)+P \left(-\frac{4 C_A n_f T_f}{3}+\frac{5 C_A^2}{3}\right)\right)\nonumber\\&\,\,\,\,\,\,\,  +\frac{1}{\eps}\bigg[\delta \left(-\frac{5 C_A^2 \zeta_2}{3}+\frac{1931 C_A^2}{432}+\frac{4}{3} C_A n_f T_f \zeta_2-\frac{355 C_A n_f T_f}{108}\right)\\&\qquad \qquad\,\,\,\,\,\,\,+P \left(-\frac{8 C_A n_f T_f}{9}+\frac{16 C_A^2}{9}\right)+L \left(-\frac{10 C_A^2}{3}+\frac{8 C_A n_f T_f}{3}\right)\bigg]\\
     f_{gg}^{(2)\text{SE ext}}=\,&\frac{1}{\eps^2}\left(\delta \left(\frac{95 C_A^2}{6}-\frac{58 C_A n_f T_f}{3}+\frac{16 n_f^2 T_f^2}{3}\right)+P \left(\frac{20 C_A^2}{3}-\frac{16 C_A n_f T_f}{3}\right)\right)\nonumber\\&+\frac{1}{\eps}\bigg[\delta \bigg(-\frac{1}{3} 20 C_A^2 \zeta_2+\frac{2311 C_A^2}{36}+\frac{16}{3} C_A n_f T_f \zeta_2-\frac{637 C_A n_f T_f}{9}-4 C_F n_f T_f\\&\spit +\frac{160 n_f^2 T_f^2}{9}\bigg)+P \left(\frac{124 C_A^2}{9}-\frac{80 C_A n_f T_f}{9}\right)+L \left(\frac{16 C_A n_f T_f}{3}-\frac{20 C_A^2}{3}\right)\bigg]
   \end{align*}%
   The total two loop contribution is,\enlargethispage{\baselineskip}
   \begin{align*}
     f^{(2)}=\,&\frac{1}{\eps}\bigg[\delta \bigg(C_A^2\left(-36 \zeta_2\!-\!10  \zeta_3\!+\!\frac{7591}{72}\right)\!+\!C_A n_f T_f\left(16 \zeta_2-\frac{527}{6}\right)\!-\!4 C_F n_f T_f+\frac{160 n_f^2 T_f^2}{9}\bigg)\\&\spit +P \left(C_A^2\left(-4\zeta_2\!+\!\frac{496}{9}\right)-\frac{176 C_A n_f T_f}{9}\right)\!+\!L \left(16 C_A n_f T_f-36 C_A^2\right)\!-\!24 C_A^2L^2\bigg]\\&  +\frac{1}{\eps^2}\bigg[\delta \left(C_A^2\left(-8 \zeta_2+\frac{101}{4}\right)-\frac{71 C_A n_f T_f}{3}+\frac{16 n_f^2 T_f^2}{3}\right)\\*&\spit+P \left(\frac{86 C_A^2}{3}-\frac{40 C_A n_f T_f}{3}\right)+16 C_A^2 L\bigg]
   \end{align*}
   The extraction of the splitting function from above is the same as in the quark case. Instead of $Z_q$ we use the gluon field renormalisation in $\overline{\text{MS}}$,
   \begin{align}
     Z_A=&1+\frac{\al_s}{4\pi}\frac{1}{\eps}\left(\frac{5 C_A}{3}-\frac{4 n_f T_f}{3}\right)\\& \nonumber +\bigg(\frac{\al_s}{4\pi}\bigg)^2 \bigg[\frac{1}{\eps}\left(\frac{23 C_A^2}{8}-\frac{5 C_A n_f T_f}{2}-2 C_F n_f T_f\right)+\frac{1}{\eps^2}\left(\frac{5 C_A n_f T_f}{3}-\frac{25 C_A^2}{12}\right)\bigg].
   \end{align}
   Performing those steps we find,
   \begin{align}     \label{app:Pgg}
     P_{gg}=\,&\frac{\al_s}{4\pi}\left(\delta  \left(\frac{11 C_A}{3}-\frac{4 n_f T_f}{3}\right)+4 C_A P\right)\nonumber\\& +\bigg(\frac{\al_s}{4\pi}\bigg)^2\bigg[\delta  \left(12 C_A^2 \zeta_3+\frac{32 C_A^2}{3}-\frac{16 C_A n_f T_f}{3}-4 C_F n_f T_f\right)\nonumber\\& \qquad\qquad\qquad    +P \left(-8 C_A^2 \zeta_2+\frac{268 C_A^2}{9}-\frac{80 C_A n_f T_f}{9}\right)\bigg].
   \end{align}
   Again this aligns with $B_\delta^g$ in eq.~\eqref{BdeltaExp} and $\gamma_{\text{cusp}}$ in eq.~\eqref{twoLoopCusp}.

   We have replicated previous splitting function calculations at large $x$ directly from the definitions~\eqref{quarkPDF} and~\eqref{gluonPDF} in a covariant gauge. By taking the incoming partons off shell, $p^2\neq 0$, we regulate the infrared divergences allowing the extraction of the UV poles of the PDFs. Although the divergent terms remain gauge independent the finite terms become gauge dependent. It means that we need to take into account the running of the gauge parameter $\xi\to Z_A\xi$ in finite terms, even when working in Feynman gauge.

\enlargethispage{1.8\baselineskip}

\section{Particular two-loop diagrams contributing to \texorpdfstring{\boldmath $W_\sqcap$}{W-cap}}
\label{sec:diagrams}

In this appendix we elaborate on aspects of the calculation of $W_\sqcap$ presented in section~\ref{sec:GammaPiCalc}.
We consider two specific diagrams where some subtle points arise.
In section~\ref{sec:endpoint} we discuss the endpoint contributions in diagram $d^{(2)}_{Y_L}$ in figure~\ref{fig:dia2} using momentum space, and in section~\ref{sec:appendixX3} we show the single IR divergent behaviour of $d^{(2)}_{X_3}$.

\subsection[Endpoint contribution in the diagram \texorpdfstring{$d^{(2)}_{Y_L}$}{d**(2)(Y(L))}]{\boldmath Endpoint contribution in the diagram $d^{(2)}_{Y_L}$}
\label{sec:endpoint}

In section~\ref{sec:GammaPiCalc} we revisited the analysis of non-Abelian contributions to the correlators of finite and semi-infinite Wilson lines~\cite{Sterman}. Specifically, we derived the representations of the two-loop diagrams that contain a three-gluon vertex and made a clear distinction between ones where two gluons are emitted from a finite Wilson-line segment as compared to the case where two emissions emerge from a semi-infinite line, corresponding respectively to diagrams $d^{(2)}_{Y_s}$ and $d^{(2)}_{Y_L}$ in~(\ref{fig:dia2}).
The difference is that in the former case both endpoint contributions appear, as in~(\ref{eq:d2YsDEC}), while in the latter case there is no endpoint contribution from infinity, so the representation of $d^{(2)}_{Y_L}$ simplifies to~(\ref{eq:d2YL}).
Let us now present this calculation in detail using momentum space and show explicitly that this endpoint contribution is indeed absent.

Using the
Feynman rules given in section~\ref{sec:GammaPiCalc}, diagram
$d^{(2)}_{Y_L}$ in~(\ref{fig:dia2}) reads
\begin{align}
  d^{(2)}_{Y_L}=\,&K_Y\int d^dz\int_{-\infty}^0ds_1\int_{s_1}^0 ds_2\int_0^ydt_3\,\left[\beta\cdot\frac{\partial}{\partial s_2\beta}-\beta\cdot\frac{\partial}{\partial s_1\beta}\right]\nonumber\\*
  &\times \left[-(s_1\beta-z)^2+i0\right]^{-1+\eps}\left[-(s_2\beta-z)^2+i0\right]^{-1+\eps}\left[-(ut_3-z)^2+i0\right]^{-1+\eps},
\end{align}
which is analogous to eq.~\eqref{eq:intd2Ys}. In the equation above, we
integrate over $z$ using the momentum-space representation of the propagators
\begin{equation}
  \label{eq:FTprop}
\mathcal{N}\left[-x^2+i0\right]^{-1+\epsilon} = -i\int\frac{d^dk}{(2\pi)^d}\frac{e^{-ik\cdot x}}{k^2+i0},
\end{equation}
obtaining
\begin{align}
d^{(2)}_{Y_L}=\,&ig_s^4\frac{C_iC_A}{2}u\cdot\beta\int\frac{d^dk_1d^dk_2}{(2\pi)^{2d}}\int_{-\infty}^0ds_1\int_{s_1}^0ds_2\int_0^ydt_3
\left[\beta\cdot\frac{\partial}{\partial
    s_2\beta}-\beta\cdot\frac{\partial}{\partial s_1\beta}\right]\nonumber\\
&\times(-i)^3\frac{e^{-ik_1\cdot\beta s_1}e^{-ik_2\cdot\beta s_2}e^{i(k_1+k_2)\cdot ut_3}}{k_1^2k_2^2(k_1+k_2)^2}.
\end{align}
After taking the derivatives with respect to $s_1\beta$, $s_2\beta$
and integrating over the infinite line we get
\begin{align}
  d^{(2)}_{Y_L}=\,&ig_s^4\frac{C_iC_A}{2}u\cdot\beta\int\frac{d^dk_1d^dk_2}{(2\pi)^{2d}}\int_0^ydt_3\frac{(-i)^3e^{i(k_1+k_2)\cdot ut_3}}{k_1^2k_2^2(k_1+k_2)^2}\frac{k_2\cdot\beta-k_1\cdot\beta}{k_2\cdot\beta+i0}\nonumber\\
  &\times\left\{\frac{1}{-i(k_1\cdot\beta+i0)}-\frac{1}{-i\left[(k_1+k_2)\cdot\beta+i0\right]}\right\},
\end{align}
where the prescription $+i0$ in the denominators ensures the
convergence of the integrals for $s_1\rightarrow -\infty$. The
expression above may be conveniently rewritten as
\begin{align}
  d^{(2)}_{Y_L}=\,&ig_s^4\frac{C_iC_A}{2}u\cdot\beta\int\frac{d^dk_1d^dk_2}{(2\pi)^{2d}}\int_0^ydt_3\,e^{i(k_1+k_2)\cdot ut_3}\frac{(-i)^3}{k_1^2k_2^2(k_1+k_2)^2}\nonumber\\
  &\times\left\{\frac{1}{-i(k_1\cdot\beta+i0)}-\frac{2}{-i\left[(k_1+k_2)\cdot\beta+i0\right]}\right\}\,.
\end{align}
This directly leads to the representation of eq.~\eqref{eq:d2YL}, as we now show. Upon introducing an auxiliary integration constrained by
momentum conservation we obtain:
\begin{align}
  d^{(2)}_{Y_L}=\,&ig_s^4\frac{C_iC_A}{2}u\cdot\beta\int\frac{d^dk_1d^dk_2d^dk_3}{(2\pi)^{3d}}\,(2\pi)^d\delta^d(k_1+k_2+k_3)\frac{(-i)^3}{k_1^2k_2^2k_3^2}\nonumber\\
  &\times\left\{\int_0^ydt_3\int_{-\infty}^0ds_1\,e^{-ik_3\cdot ut_3}\left[e^{-ik_1\cdot\beta s_1}-2 \,e^{-i(k_1+k_2)\cdot\beta s_1}\right]\right\}.
\end{align}
The representation of the delta function
$(2\pi)^d\delta^d(k_1+k_2+k_3)=\int d^dz\, e^{i(k_1+k_2+k_3)\cdot z}$
is interpreted as an integral over the position of the scalar ``three gluon''
vertex in eq.~\eqref{eq:d2YL}. Using eq.~\eqref{eq:FTprop} we recover the expression of the three gluon propagators in coordinate space, carrying momenta $k_1$, $k_2$ and $k_3$, obtaining
\begin{align}
  d^{(2)}_{Y_L}=\,&K_Y\,\int d^dz\int_0^ydt_3\int_{-\infty}^0ds_1\,[-(z-ut_3)^2+i0]^{\eps-1}\nonumber\\*
  &\times\left\{\left[-(z-\beta s_1)^2+i0\right]^{\eps-1}\left[-z^2+i0\right]^{\eps-1} - 2\,\left[-(z-\beta s_1)^2+i0\right]^{2\eps-2}\right\}.
\end{align}
Substituting the definitions in eqs.~\eqref{eq:dE},~\eqref{eq:d0} and~\eqref{eq:dB} we verify the result in eq.~\eqref{eq:d2YL}.

\subsection[The diagram \texorpdfstring{$d^{(2)}_{X_3}$}{d**(2)(X(3))} connecting three Wilson lines]{\boldmath The diagram $d^{(2)}_{X_3}$ connecting three Wilson lines}
\label{sec:appendixX3}

In this section we derive the representation of eq.~\eqref{eq:f2X3} of the
diagram $d^{(2)}_{X_3}$ that connects two cusps with a lightlike
segment of finite length. Following the discussion of
ref.~\cite{Sterman}, the singularities of the webs of this kind are
associated with the configuration where all the vertices approach the
lightlike segment of finite length. These webs do not contribute to
the cusp singularities because there is not any region of configuration
space where all the vertices are in proximity of the cusp. By using the
Feynman rules in eq.~\eqref{eq:prop}, the diagram $d^{(2)}_{X_3}$ reads
\begin{align}
  d^{(2)}_{X_3}=\,&-g_s^4\mu^{4\eps}\frac{C_AC_F}{2}\mathcal{N}^2(\beta\cdot u)^2\int_0^{+\infty} dt_1\int_0^{+\infty}dt_3\int_0^yds_1\int_{s_1}^y ds_2\nonumber\\&\times\left[-2 \beta\cdot u t_1s_2+i0\right]^{\eps-1}\left[-2\beta\cdot u t_3(s_1-y)+i0\right]^{\eps-1},
\end{align}
where the factor $-\frac{C_AC_F}{2}$ corresponds to the maximally
non-Abelian part of the colour factor of the diagram, which is exponentiated~\cite{Sterman:1981jc,Gatheral:1983cz,Frenkel:1984pz,Gardi:2010rn}.
We expose the overall infrared singularity in the last integration by rewriting the integration domain using
$\theta(t_1-t_3)+\theta(t_3-t_1)=1$ and changing the order of integrations. Thus we obtain
\begin{align}
  d^{(2)}_{X_3}=\,&-g_s^4\mu^{4\eps}\,C_AC_F\mathcal{N}^2(\beta\cdot u)^2\,\int_0^{+\infty} dt\int_0^{t}dt'(t\,t')^{-1+\eps}\int_0^yds_1\int_{s_1}^y ds_2\nonumber\\&\times\left[-2 \beta\cdot u s_2+i0\right]^{\eps-1}\left[-2\beta\cdot u (s_1-y)+i0\right]^{\eps-1}.
\end{align}
We stress that the expression above still has infrared
singularities from the limit $t\rightarrow\infty$ in the upper bound of the $t'$ integral. Therefore we decouple the infrared contributions by applying the changes of variables
\begin{align}
  t'&=t\left(\frac{s}{y}\right)^2,\quad s_1=y\,a_1,\quad s_2=y\,a_2,
\end{align}
which yields
\begin{align}
  d^{(2)}_{X_3}&=g_s^4\mu^{4\eps}\,C_AC_F\mathcal{N}^2(\beta\cdot u)\,\int_0^{+\infty} dt\int_0^{y}ds\left[-2 \beta\cdot u t s+i0\right]^{2\eps-1}\int_0^1\frac{da_1}{(1-a_1)^{1-\eps}}\int_{a_1}^1 \frac{da_2}{a_2^{1-\eps}}.
\end{align}
The parameters $a_1$ and $a_2$ are integrated immediately, leading to
\begin{align}
  d^{(2)}_{X_3}=\,&g_s^4\mu^{4\eps}\,C_AC_F\mathcal{N}^2(\beta\cdot u)\,\frac{1}{\eps}\left[\frac{1}{\eps}-B(\eps,1+\eps)\right]\nonumber\\
  &\times\int_0^{+\infty} dt\int_0^{y}d s \left[-2 \beta\cdot u t s+i0\right]^{2\eps-1}.
\end{align}
Thus we apply the change of variables introduced before
eq.~\eqref{case1Res} and we get
\begin{align}
  d^{(2)}_{X_3}=\,&-g_s^4\mu^{4\eps}\frac{C_AC_F}{2}\mathcal{N}^2\,\frac{1}{\eps}\left[\frac{1}{\eps}-B(\eps,1+\eps)\right]\nonumber\\*
  &\times\int_0^{+\infty} \frac{d\tau}{\tau}\int_0^{\frac{\rho}{\sqrt{2}}}\frac{d \sigma}{\sigma} \left(4\tau \sigma\mu^2\right)^{2\eps}.
\end{align}
By replacing the normalisation
$\mathcal{N}=-\frac{\Gamma(1-\eps)}{4\pi^{2-\eps}}$, we absorb the factor $(4\pi e^{\gamma_E})^\eps$ into the coupling constant, which we expand at the scale $\frac{1}{\tau\sigma}$ by means of eq.~\eqref{eq:runcoup}, thus getting
\begin{align}
  d^{(2)}_{X_3}=\,&-C_AC_F\,\frac{\Gamma^2(1-\eps)}{2\eps}\left[\frac{1}{\eps}-B(\eps,1+\eps)\right]\nonumber\\
  &\times\int_0^{+\infty} \frac{d\tau}{\tau}\int_0^{\frac{\rho}{\sqrt{2}}}\frac{d \sigma}{\sigma} \left(\frac{\al_s\left(\frac{1}{\tau \sigma}\right)}{\pi}e^{-\eps\gamma_E}\right)^{2},
\end{align}
which is written in terms of the representation in eq.~\eqref{eq:2looprep}
and reproduces the result of eq.~\eqref{eq:f2X3}. By expanding the
integrand for $\eps\rightarrow 0$ we get
\begin{equation}
  w^{(2)}_{X_3}=-C_AC_F\left[\frac{\zeta_2}{2}-\eps\zeta_3\right].
\end{equation}
We notice that the first term in the equation above, once integrated
with $\lambda$ and $\sigma\rightarrow0$, would yield a double UV pole,
which is expected to arise only from single cusp singularities. By
summing the contribution of $w^{(2)}_{X_3}$ above with the one
originated from the other webs connecting three lines, namely
$w^{(2)}_{3s}$, we verify that the cusp term cancels, leaving only the
subleading pole in eq.~\eqref{eq:endpointterms} associated with collinear
configurations around the finite~segment.


\bibliographystyle{JHEP}
\bibliography{biblio/bib}




\end{document}